\documentclass[modern]{aastex631}

 \usepackage[
   margin=1in
 ]{geometry}

\usepackage[T1]{fontenc}
\usepackage{libertinus}
\usepackage{ragged2e}
\usepackage{amssymb}
\usepackage{amsmath}
\usepackage{graphicx}
\usepackage{gensymb}
\usepackage{comment}
\usepackage{hyperref}
\hypersetup{
    colorlinks=true,
    linkcolor=black,
    citecolor=black,
    filecolor=black,      
    urlcolor=black}

\usepackage{tikz}
\usepackage{tcolorbox}
\usepackage{tabularx}
\usepackage{color}
\definecolor{cornflowerblue}{rgb}{0.39, 0.58, 0.93}
\graphicspath{{fig}}

\def\kms{\ {\rm km\, s}^{-1}}
\def\ms{\ {\rm m\, s}^{-1}}
\def\masyr{\ {\rm mas\, yr}^{-1}}
\def\teff{T_{\rm eff}}

\def\msun{\rm M_\odot}
\def\rsun{\rm R_\odot}

\newcommand{\gaia}{\textsl{Gaia}}
\newcommand{\MS}{\texttt{MINESweeper}}
\newcommand\ionn[2]{#1$\,${\scshape{#2}}}

\begin{document}

\pagestyle{empty}

\begin{tikzpicture}[remember picture, overlay]
    \node[anchor=north west,yshift=0.1in,xshift=-0.1in] 
        at (current page.north west)
        {\includegraphics[width=8.6in]{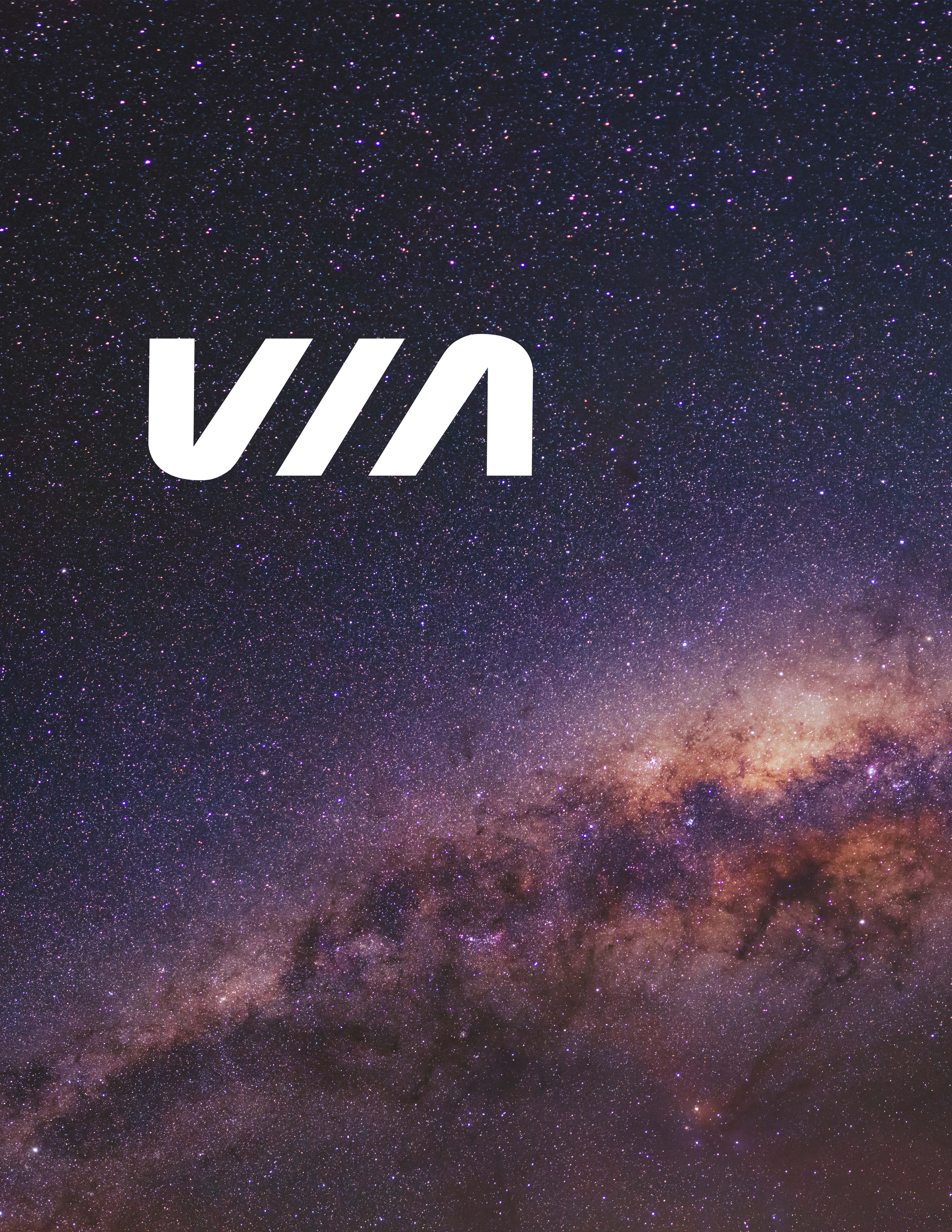}}; 
\end{tikzpicture}

\clearpage

\vspace*{\stretch{1}}
\begin{center}

{\bf {\Large The Via Project:}\\ \vspace{0.2cm} {\large Overview of the Science, Instrument, and Survey}}

\vspace{1cm}
v1.0

June 2026

\href{https://via-project.org}{via-project.org}

\vfill 
The front cover is a panorama of the southern sky  as seen from Las Campanas Observatory (courtesy of Igor Chilingarian). 

Via is supported in part by the Gordon \& Betty Moore Foundation, the Kavli Foundation, The John Templeton Foundation, the Ambrose Monell Foundation, Craig \& Barbara Barrett, Thomas Clay, an anonymous donor, and by the Center for Astrophysics $|$ Harvard \& Smithsonian, Carnegie Observatories, Stanford University, and Yale University.  

\end{center}

\clearpage

\noindent
{\bf Via Executive Committee}\\
\vspace{-0.15cm}
\hrule
\vspace{0.1cm}
\noindent
Ana Bonaca (Project Co-PI; Carnegie)\\
Charlie Conroy (Project Co-PI; CfA)\\
Daniel Fabricant (Instrument PI; CfA)\\
Marla Geha (Yale Lead Scientist; Yale)\\
Risa H. Wechsler (Stanford Lead Scientist; Stanford)\\

\noindent
{\bf Via Science Book Contributors}\\
\vspace{-0.15cm}
\hrule
\vspace{0.1cm}
\twocolumngrid
\noindent
\raggedright
Ebtihal M. Abdelaziz (Carnegie)\textsuperscript{1} \\
Christian Aganze (Stanford)\textsuperscript{1,3} \\
Dan Baldwin (CfA)\textsuperscript{2} \\
Carl Barcroft (CfA)\textsuperscript{2} \\
Earl P. Bellinger (Yale)\textsuperscript{1} \\
Andrew Benson (Carnegie)\textsuperscript{1} \\
Ana Bonaca (Carnegie)\textsuperscript{1,3} \\
Luke Bouma (Carnegie)\textsuperscript{1,3} \\
Warren R. Brown (CfA)\textsuperscript{1,2} \\
Nelson Caldwell (CfA)\textsuperscript{2} \\
Phillip A. Cargile (CfA)\textsuperscript{1,2,3} \\
Daniel Catropa (CfA)\textsuperscript{2} \\
William Cerny (Yale)\textsuperscript{1,3} \\
Vedant Chandra (CfA)\textsuperscript{1,2,3} \\
Susan E. Clark (Stanford)\textsuperscript{1,3} \\
Andrew Cline (CfA)\textsuperscript{2} \\
Liam Connor (CfA)\textsuperscript{1,3} \\
Charlie Conroy (CfA)\textsuperscript{1,2,3} \\
Jeffrey D. Crane (Carnegie)\textsuperscript{2} \\
David Cruz (Carnegie)\textsuperscript{2} \\
Tara Dacunha (Stanford)\textsuperscript{1} \\
M. Sten Delos (Carnegie)\textsuperscript{1} \\
Benjamin Dodge (Stanford)\textsuperscript{1,3} \\
Peter E. Doherty (CfA)\textsuperscript{2} \\
Yize Dong (CfA)\textsuperscript{1,3} \\
Daniel Durusky (CfA)\textsuperscript{2} \\
Daniel Fabricant (CfA)\textsuperscript{2,3} \\
Danielle Frostig (CfA)\textsuperscript{1,2,3} \\
Julian Garcia (Carnegie)\textsuperscript{2} \\
Marla Geha (Yale)\textsuperscript{1,3} \\
Matthew J. Holman (CfA)\textsuperscript{1} \\
Charlie Hull (Carnegie)\textsuperscript{2} \\
Benjamin D. Johnson (CfA)\textsuperscript{1,2,3} \\
Jan E. Kansky (CfA)\textsuperscript{2} \\
Vladimir Y. Kradinov (CfA)\textsuperscript{2} \\
Casey Lam (Carnegie)\textsuperscript{1,3} \\
Scott Lucchini (CfA)\textsuperscript{1} \\
Kyle MacKenzie (CfA)\textsuperscript{2} \\
Morgan MacLeod (CfA)\textsuperscript{1,3} \\
Philip Mansfield (Stanford)\textsuperscript{1} \\
Viraj Manwadkar (Stanford)\textsuperscript{1,3} \\
Alex McCarthy (CfA)\textsuperscript{2} \\
Brian A. McLeod (CfA)\textsuperscript{2} \\
Stephanie Melikian (CfA) \textsuperscript{2} \\
Catherine R. Miller (CfA)\textsuperscript{2} \\
Sean Moran (CfA)\textsuperscript{2} \\
Mark Mueller (CfA)\textsuperscript{2} \\
Antonio Braulio Neto (Steiner Institute)\textsuperscript{2} \\
Jacob Nibauer (Princeton)\textsuperscript{1,3} \\
Karin {\"O}berg (CfA)\textsuperscript{1,3} \\
Martin Paegert (CfA)\textsuperscript{2} \\
Nondh Panithanpaisal (Carnegie)\textsuperscript{1} \\
Anya Phillips (CfA)\textsuperscript{1} \\
John J. Piotrowski (Carnegie)\textsuperscript{2,3} \\
Anthony L. Piro (Carnegie)\textsuperscript{1,3} \\
William A. Podgorski (CfA)\textsuperscript{2} \\
Malena Rice (Yale) \textsuperscript{1,3} \\
Conor Sayres (University of Washington)\textsuperscript{2} \\
Andrew Schalk (CfA)\textsuperscript{2} \\
Joshua D. Simon (Carnegie)\textsuperscript{1} \\
Hunter Snell (CfA)\textsuperscript{2} \\
Jean J. Somalwar (Stanford; UC Berkeley)\textsuperscript{1} \\
Jessica J. Spake (Carnegie)\textsuperscript{1,3} \\
Zachary Svec (CfA)\textsuperscript{2} \\
David Theroux (CfA)\textsuperscript{2} \\
Andrew Vanderburg (CfA)\textsuperscript{1,3} \\
Pieter van Dokkum (Yale)\textsuperscript{1} \\
V. Ashley Villar (CfA)\textsuperscript{1,3} \\
Shreyas Vissapragada (Carnegie)\textsuperscript{1,3} \\
Newlin C. Weatherford (Carnegie)\textsuperscript{1} \\
Risa H. Wechsler (Stanford)\textsuperscript{1,3} \\
Matthew Werneken (CfA)\textsuperscript{1,2} \\
Abigail White (CfA)\textsuperscript{2,3} \\
Zijing (Julie) Xue (Carnegie)\textsuperscript{1} \\
Joseph Zajac (CfA)\textsuperscript{2} \\
Xinyue Alice Zhang (Stanford)\textsuperscript{1} 
\onecolumngrid
\justifying

\vspace{0.5cm}
\hrule

{\small
\vspace{0.5em}
\noindent\textsuperscript{1} Science Definition \& Planning;
\textsuperscript{2} Instrumentation \& Software;
\textsuperscript{3} Authorship of Science Book
}

\clearpage

\tableofcontents

\clearpage

\pagestyle{plain}
\pagenumbering{arabic}
\setcounter{page}{1}

\section*{Executive Summary}
\addcontentsline{toc}{section}{Executive Summary} 

\vspace{0.2cm}

Via is a forthcoming all-sky spectroscopic survey designed to achieve 100~m\,s$^{-1}$ radial velocity stability for millions of faint ($G \lesssim 21$) stars, opening a new observational regime at the intersection of dark matter physics, galaxy formation, and time-domain astrophysics. No existing or planned facility combines this velocity precision with the depth, multiplexing, and full-sky access that Via delivers. It  is timed to a singular convergence of wide-field facilities---the Vera C.\ Rubin Observatory's \textit{LSST}, \textit{Euclid}, \textit{Roman}, and \textit{Gaia}---supplying the spectroscopy needed to turn their imaging and astrometric discoveries into physical measurements.

Via will deploy identical instruments on the 6.5m MMT and Magellan/Clay telescopes. It will survey more than 2,000,000 stars over five years beginning in 2027, revisiting $\sim10^5$ stars five or more times to build multi-epoch velocities. Each instrument consists of 576 robotically positioned fibers patrolling a $1\degree$ diameter field of view, feeding two spectrographs. Viaspec, the primary instrument, is an $R \approx 15{,}000$ stabilized spectrograph covering 505--595~nm with 540 fibers, optimized for high-precision radial velocities and stellar chemical abundances. Boombox is an $R \approx 1{,}000$ spectrograph spanning 360--1010~nm with 36 fibers, designed to reach the single-epoch depth of LSST ($r \approx 24$) for transient classification and faint-object spectroscopy.

Four key science objectives motivate the survey. The \textit{Stream Perturbation Survey} will use stellar streams as gravitational detectors to constrain the abundance, mass, and density profiles of dark matter subhalos less massive than $\lesssim 10^7\,M_\odot$---below the threshold of galaxy formation, and into the regime where cold, warm, and self-interacting dark matter models make decisively different predictions. The \textit{Dwarf Galaxy Survey} will carry out a chemodynamical census of Milky Way satellite galaxies, measuring velocity dispersions and metallicities for systems down to the faintest galaxies known and newly discovered by LSST, \textit{Euclid}, and \textit{Roman}, directly probing the halo mass threshold at which galaxy formation ceases. The \textit{Cold Gas Survey} will use hundreds of thousands of halo stars as backlights to construct the first 3D tomographic maps of cold gas in the Milky Way's circumgalactic medium via \ion{Na}{1} absorption at 589~nm, leveraging the $R \approx 15{,}000$ resolution to separate interstellar from stellar features. Finally, the \textit{Transient Follow-up Survey} will spectroscopically characterize thousands of LSST-discovered transients---including superluminous supernovae, tidal disruption events, and gravitational-wave counterparts---with Via on-sky nearly one-third of the year. Beyond the core program, spare fibers in every pointing will support a rich ancillary science program---including the Ly$\alpha$ forest at $z \approx 3$--$4$, polluted white dwarfs, exoplanet host characterization, fast radio burst host galaxy redshifts, 
 and integrated kinematics of nearby low-mass galaxies.
 
Via's defining aim is a decisive test of the particle nature of dark matter: detecting or ruling out the gravitational imprints of dark, starless subhalos below the threshold of galaxy formation---while building a lasting, dual-hemisphere spectroscopic reference for near-field cosmology and time-domain astrophysics. The Via Project is a collaboration between the Center for Astrophysics $|$ Harvard \& Smithsonian, Carnegie Observatories, Stanford University, and Yale University.

\clearpage



\section{Motivation}
\label{motivate}

Over the next decade, wide-field 
imaging surveys---led by the NSF--DOE Vera C.\ Rubin Observatory's Legacy Survey of Space and Time (LSST)---will map stellar streams, discover ultra-faint dwarf galaxies, and detect millions of transients across the sky. Fully exploiting this wealth of discoveries requires a new kind of instrument: a massively multiplexed spectrograph capable of 100$\ms$ radial velocity stability for millions of faint targets, paired with an efficient low-resolution spectrograph that can reach the 
single-epoch depth of LSST for rapid transient characterization. 
Four key science drivers motivate the Via Project and define these instrumental requirements:  probing dark matter with Milky Way stellar streams (\S\,\ref{s:streams}), understanding the limit of galaxy formation with Milky Way satellite galaxies (\S\,\ref{s:dwarfs}), mapping cold gas in the Milky Way (\S\,\ref{s:coldgas}), and characterizing the transient universe (\S\,\ref{s:transients}).  We explain how the Via Project addresses these questions and provide a brief history of the collaboration in \S\,\ref{s:overview}.

\subsection{Stellar Streams and the Nature of Dark Matter}
\label{s:streams}

The nature of dark matter is one of the most fundamental unsolved problems in physics.
Almost a century of astronomical observations have revealed that all galaxies are embedded in massive halos of dark matter \citep[e.g.,][]{zwicky:1933,Rubin1970,Faber1979,bertone:2005,Bertone2018, Wechsler18}. 
In the hierarchical model of structure formation, these dark matter halos are assembled ``bottom up'' through the coalescence of smaller halos \citep[e.g.,][]{Press1974,Searle1978,white:1978,Davis1985,Lacey1993}.  
Plausible models that fit these observations span an enormous range from $\sim$ GeV-scale weakly interacting massive particles (WIMPs), to $\ll$ eV-scale axions, to asteroid-mass $\sim10^{-15}\,\msun$ primordial black holes \citep[e.g.,][]{bertone:2005,Bertone2018}.  
The models diverge in their predicted abundances of dark matter halos below the threshold of galaxy formation, i.e., below $\sim10^8\, \msun$.
Constraining the abundance of halos well below this scale, e.g., in the range $10^6$--$10^7\, \msun$ will decisively test a wide class of dark matter models.

\begin{figure*}[t!]
\centering
\includegraphics[width=\textwidth]{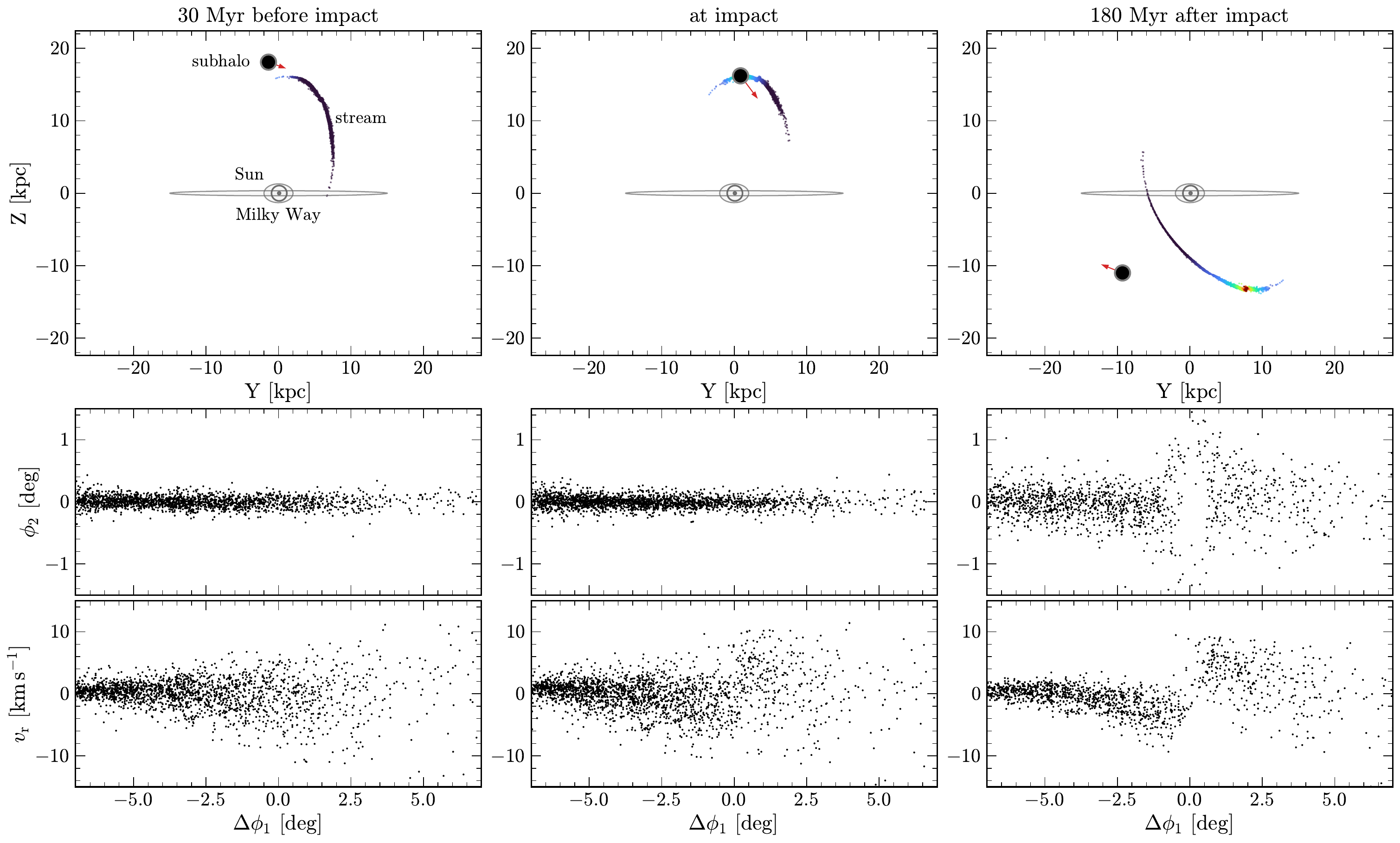}
\caption{A model of a $10^7\,M_\odot$ dark matter subhalo flying through a stellar stream, at three points of time from left to right. 
The top panels show the 3D distribution of stars around the Milky Way, with stars colored by the impulsive velocity kick they receive from the passing subhalo.
The central panels show the on-sky distribution of stars along the stream. 
The bottom panels show the radial velocity as a function of the angular coordinate along the stream, centered at the point of closest approach to the subhalo. 
{Left:} The un-perturbed stream 30~Myr prior to impact. 
{Middle:} At the time of impact, a velocity perturbation with a characteristic $1/\phi_1$ ``kick'' signature is created. 
{Right:} At longer times after impact, stars move from the impact site, forming a density caustic and/or gap. The dispersion in velocity will also be increased by the impact.  
}
\label{fig:gap_formation}
\end{figure*}

All past evidence for the existence of dark matter has been gravitational, so the most reliable approach to detecting truly dark halos is through their gravitational influence on luminous matter.
The Milky Way contains more than one hundred stellar streams---long, thin, and kinematically cold filaments of stars formed through tidal dissolution of dwarf galaxies and globular clusters.
Numerical simulations have shown that low-mass dark matter subhalos interact with stellar streams and leave observable signatures in their density and kinematic structure \citep{ibata:2002,johnston:2002,carlberg:2009,carlberg:2012,yoon:2011,erkal:2015,erkal:2015b,erkal:2016}.
A dark matter subhalo passing close to a stream imparts a velocity kick to stars at the impact site, changing their orbits, either (1) producing an underdensity or a gap in the stream (if the subhalo is sufficiently massive and close), or (2) dynamically heating the stream (if the subhalo is less massive or passes at a larger distance).
Figure \ref{fig:gap_formation} illustrates the velocity perturbation and gap formation after a stream directly encounters a $10^7\,\msun$ dark matter subhalo. 

So far, there have been more than 130 streams discovered in the Milky Way \citep[for a recent census, see][]{mateu:2023,bpw:2025}.  
Detailed maps of a handful of high-confidence streams show signatures of dynamical perturbation consistent with dark matter substructure: gaps and spurs in GD-1 \citep{pwb:2018,deboer:2018,deboer:2020}, density and width variations in Palomar~5 \citep{ibata:2016,erkal:2017,bonaca:2020a}, multiple components in Jhelum \citep{bonaca:2019b,awad:2024}, a break in ATLAS--Aliqa Uma \citep{li:2021}, and density variations and wiggles in Jet \citep{ferguson:2022} and Phoenix \citep{tavangar:2022}.  
These analyses have been limited to streams observed at a high contrast with respect to the field Milky Way stars, thanks either to a stream's retrograde orbit, which makes it stand out in \gaia\ proper motions, or to deep imaging, often targeted at individual streams.  
The LSST \citep[][]{Ivezic2019}, conducted at the Vera C. Rubin Observatory, will have the depth and spatial coverage to allow  uniform and systematic mapping of up to hundreds of stellar streams, and to detect gaps induced by subhalo perturbations \citep{lu:2025,  bpw:2025}. However, confirming the presence of dark subhalos is not possible from LSST data alone.

Despite the multitude of structural anomalies detected in stellar streams, an unambiguous identification of a truly dark subhalo is still missing.  
The main challenge has been distinguishing signatures of subhalo impacts from other processes. Streams preserve a record of their entire dynamical history because they are dynamically cold and evolve on long dynamical timescales.  
The dissolution process itself imprints a regular pattern of over- and underdensities onto globular cluster streams, as even stars escaping the cluster at a constant rate are offset from the cluster's orbit and, in projection, form density variations \citep{kupper:2008,just:2009}.  
The progenitor birth environment can also play a role. Globular clusters of extragalactic origin start to dissolve in their original host galaxy, and arrive at the Milky Way enshrouded in a cloud of already-stripped stars that expands into a complex, cocoon-like structure \citep[][]{malhan:2019,carlberg:2020}.  
Once in the Milky Way, streams can encounter compact objects other than dark matter subhalos, including giant molecular clouds \citep[GMCs,][]{amorisco:2016,banik:2019}, globular clusters \citep{doke:2022,ferrone:2025}, and dwarf galaxies \citep{woudenberg:2023,foote:2025}.
Finally, the large-scale gravitational potential may be complex.  Flattening gives rise to chaotic and/or resonantly trapped orbits \citep{price-whelan:2016,yavetz:2023}. A rotating bar \citep[][]{hattori:2016,pearson:2017} and active mergers \citep{erkal:2019,Nibauer:2024} also affect the structure of stellar streams.

To distinguish subhalo impacts from other mechanisms, accurately measuring the velocities of stars in stellar streams is essential.  Impulsive impacts produce a characteristic $1/\phi_1$ velocity signature across the perturbed region; the velocity amplitude and spatial extent are set by the impactor's mass, scale radius, and impact parameters (\citealt{erkal:2015}; see Figure~\ref{fig:gap_formation}).  
Therefore, it is possible to identify perturbers of different internal structures by modeling kinematic signatures of gaps \citep{erkal:2015b,bonaca:2019,hilmi:2024,lu:2025}.  
While there is a degree of overlap in the masses and sizes of possible stream impactors, at a fixed mass of $\approx10^6\,\msun$, globular clusters are more compact ($\approx1\,$pc) than dark matter subhalos ($\approx50\,$pc) and GMCs \citep[$\approx100\,$pc, see Figure 6 in][]{bonaca:2019}.  
Furthermore, the structure of subhalos varies for different dark matter models: while the cold dark matter model predicts cuspy halos \citep[e.g.,][]{navarro:1997,navarro:2010}, warm dark matter, made of lighter particles, forms lower-density, cored halos \citep[e.g.,][]{bode:2001,bose:2016}, and self-interacting dark matter halos can both form cores due to self-interactions, or undergo core collapse, achieving exceedingly high central densities \citep[e.g.,][]{spergel:2000,nadler:2023b}.

High-precision spectroscopy is needed to definitively identify dark-matter subhalo impacts on stellar streams.
The amplitude of velocity signatures imparted by $10^6$--$10^7\,\msun$ subhalos is $\approx0.1$--$1\kms$ \citep[see also \S\,\ref{s:sps}]{erkal:2015b}, which at a typical stream distance of 10\,kpc translates to a proper motion amplitude of $0.002$--$0.02\,\masyr$.
\gaia\ DR5 is expected to achieve this precision, but only for bright stars: $G<13$ and $G<18$ for $0.002\,\masyr$ and $0.02\,\masyr$, respectively.
For a typical stream, these magnitude limits impose a sampling too sparse to resolve signatures of subhalo impacts \citep[e.g.,][]{lu:2025}.
However, the necessary precision is within the reach of spectroscopy \citep[e.g.,][]{bonaca:2020}.
Medium-resolution, high-throughput, wide-field multi-object spectrographs are the best tool to detect low-mass subhalos impacting stellar streams.

\subsection{Milky Way Satellite Galaxies and the Threshold of Galaxy Formation}
\label{s:dwarfs}

Within the $\Lambda$CDM cosmological model, galaxies form from the cooling of gas within dark matter halos \citep[e.g.,][]{white:1978}.  However, as described above, not all halos host galaxies: galaxy formation only proceeds above a critical mass threshold. At lower masses, halos cannot provide the necessary gravitational potential wells to accumulate baryonic matter.   Identifying this halo mass threshold, and disentangling the complex interplay of baryonic and dark matter physics that governs it, is a central goal of near-field cosmology.

\begin{figure}[t!]
    \centering
    \includegraphics[width=1\textwidth]{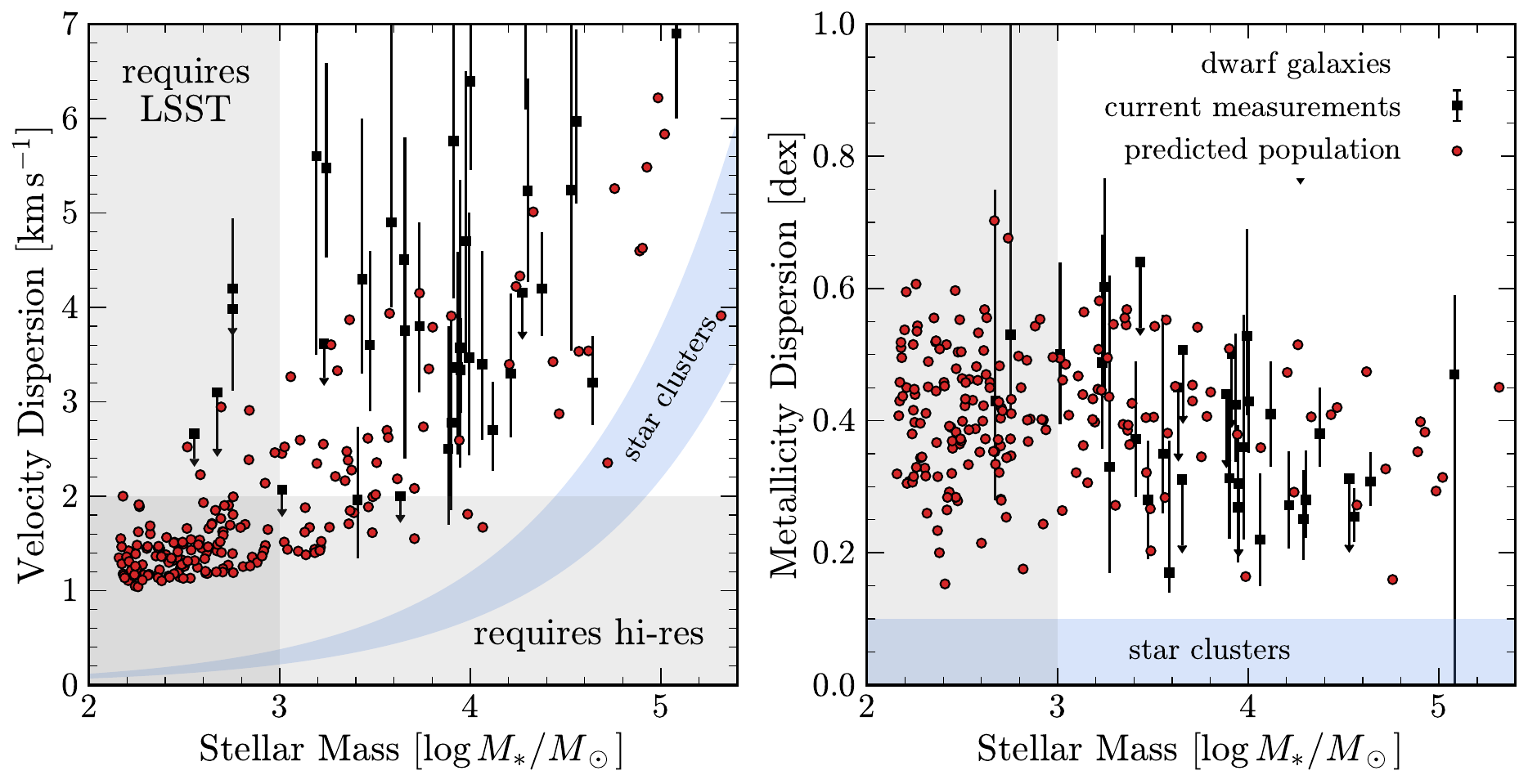}
    \caption{The landscape of spectroscopic measurements in dwarf satellite galaxies around the Milky Way. As a function of stellar mass, the velocity dispersion (left) and metallicity dispersion (right) are shown for existing measurements \citep{Pace2024, geha2026b}, and for a predicted population of LSST-discoverable dwarf galaxies \citep{Manwadkar2022}. 
    The locus of star clusters---whose kinematics and chemistry can be explained by baryonic physics alone---is shown in blue.
    Gray bands highlight current observational limitations; reliably measuring velocity dispersion in the $\lesssim 2~\kms{}$ regime requires stable high-resolution spectroscopy, and the faintest dwarf galaxies await discovery by surveys like LSST.}
    \label{fig:dwarfmotivate}
\end{figure}

Current models suggest the lowest-mass dark matter halos able to host a galaxy have virial masses of $M_{\rm peak}\approx 10^{7-8} M_\odot$ \citep[e.g.,][]{Jethwa2018, nadler:2020, Nadler2025}.    Below this mass, halo potential wells are too shallow to retain gas, particularly in the presence of cosmic reionization and energetic feedback processes that heat or expel gas from low-mass halos \citep{ReesOstriker1977,Efstathiou1992,Bullock2000,Somerville2002}.    As a result, only sufficiently massive halos can cool their gas efficiently and initiate star formation.  Many details of this transition remain uncertain: (1) the relative importance of different gas cooling channels (\ionn{H}{i} vs.~H$_2$ vs.~metal-line) and H$_2$ self-shielding \citep{Tegmark1997, Greif2008, Munshi19, Nadler2025}, (2) the speed, timing, and patchiness of cosmic reionization and its overall ability to suppress star formation \citep[e.g.,][]{Graus2019}, (3) the impact of dark matter--baryon streaming \citep[e.g.,][]{TseliakhovichHirata10,BovyDvorkin13},  and (4) the strength of feedback from massive stars and early supernova explosions \citep[e.g.,][]{DekelSilk1986, Chen2022}. These unsettled questions have direct bearing on the galaxy occupation fraction at low halo masses (the fraction of halos that host a galaxy of any mass) and more broadly, observable effects on the properties of the faintest galaxies.

Faint dwarf galaxies are the best observational laboratories for investigating the threshold of galaxy formation: they occupy the smallest dark matter halos known to have formed stars. Because galaxy formation within $\Lambda$CDM proceeds hierarchically, these faint dwarf galaxies are abundant satellites orbiting more massive hosts.  The Milky Way, in particular, is predicted to host $\sim 150$--$300$ such satellite galaxies \citep[e.g.,][]{Tollerud:2008, hargis:2014,nadler:2020,Manwadkar2022,Tan2026}.  Only $\sim70$ of these are known to date (\citealt{Simon2019,Pace2024}). The ensemble properties of this satellite population---the raw number of dwarfs, their luminosity function, and their radial distribution---are key constraints on the limit of galaxy formation and models of particle dark matter \citep[e.g.,][]{Kennedy2014, Newton2021, Nadler21}.  The masses, density profiles, metallicity distributions, and chemical abundances of the very faintest galaxies, ultra-faint dwarf (UFD) galaxies, are stringent probes of the fundamental nature of dark matter and the physics of galaxy formation at the low-mass end.

LSST \citep{Ivezic2019}, {\it Euclid} \citep{EuclidI}, and the {\it Nancy Grace Roman Telescope} (Roman; \citealt{spergel:2015}), will discover a deluge of new ultra-faint dwarf galaxy candidates in the local universe \citep{Manwadkar2022, Tsiane2025}, and help place transformative constraints on low-mass galaxy formation physics and the nature of dark matter.  Spectroscopic characterization of imaging survey ultra-faint dwarf galaxy \textit{candidates} is required to confirm that they are bona fide dwarf galaxies \citep{WillmanStrader2012} and to connect their luminous components to their host dark matter halos.  The velocity dispersions of the least-massive UFDs are expected to be $\mathcal{O}(1)\kms$ \citep[e.g.,][]{simon:2007, Simon2019}.

Figure~\ref{fig:dwarfmotivate} illustrates the current landscape of velocity and metallicity dispersion measurements in MW satellite galaxies. 
Current spectroscopic studies of ultra-faint dwarf galaxies rely on single-object or slit-based spectrographs on large telescopes, observing one system at a time with velocity precisions of 1--2~km~s$^{-1}$ \citep[e.g.,][]{Simon2019, Geha2026}. This precision is sufficient for the more massive classical dwarfs but inadequate for the faintest systems, where dispersions approach 1~km~s$^{-1}$ and reliable mass estimates require per-star uncertainties several times smaller. At these low dispersions, even confirming that a candidate is a dark matter-dominated galaxy rather than a  star cluster depends on resolving the velocity dispersion with high confidence \citep{WillmanStrader2012, Cerny2026}. Moreover, the anticipated flood of new candidates from LSST, \textit{Euclid}, and \textit{Roman} will require confirmation and characterization of dozens to hundreds of systems, each needing $\gg10$ member stars observed to $G \gtrsim 23$ (see Figure~\ref{fig:dwarfmotivate}). Meeting this challenge requires a combination of capabilities that no current facility provides: radial velocity stability at the 100~m~s$^{-1}$ level, depth to $G \gtrsim 23$ on a large-aperture telescope, and multi-object multiplexing over a wide field to efficiently survey extended, sparse stellar systems.

\subsection{Cold Gas Tomography in the Milky Way}
\label{s:coldgas}

\begin{figure}[t!]
    \centering
    \includegraphics[width=0.95\textwidth]{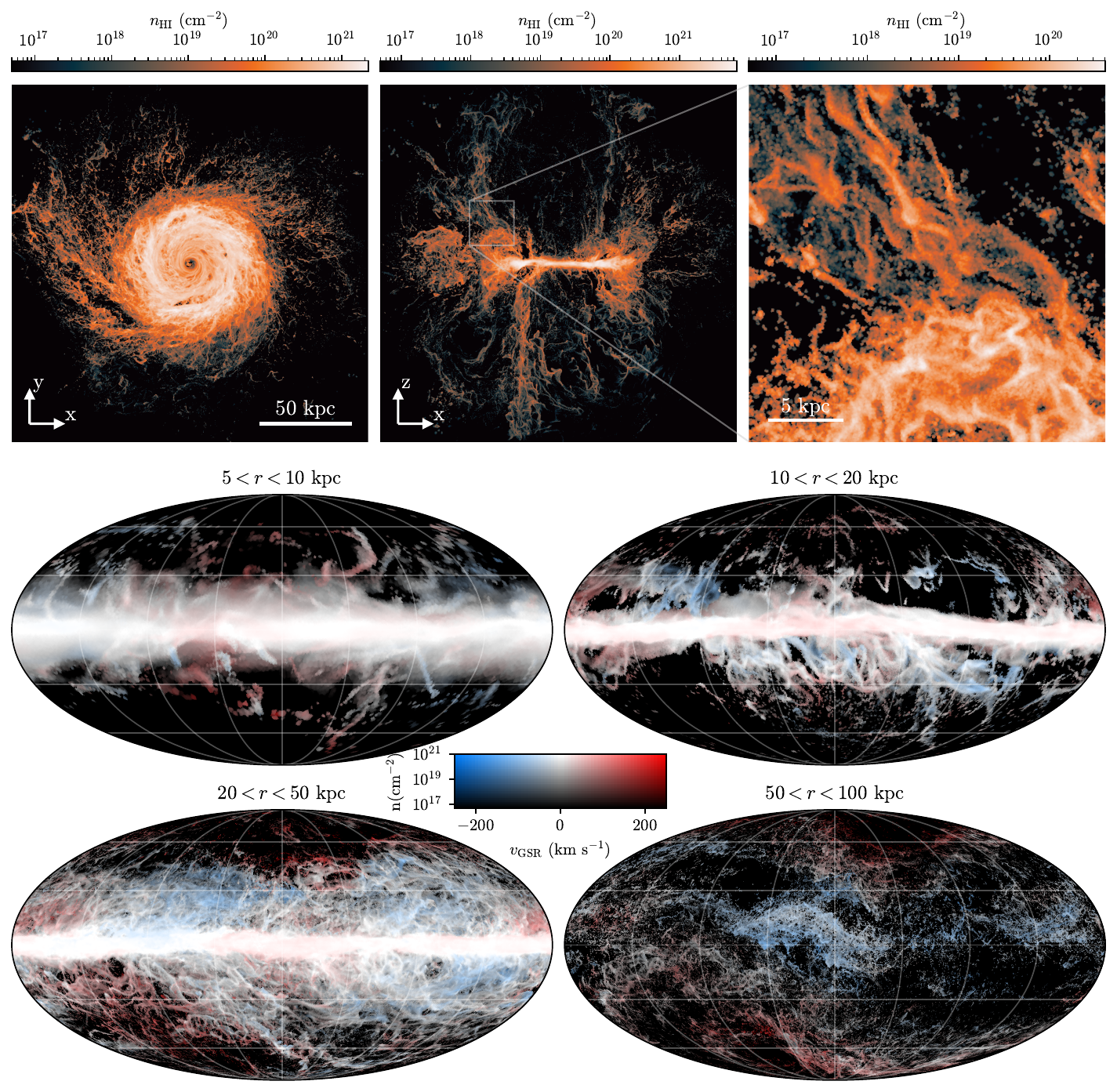}
    \caption{{Top:} \ionn{H}{i} column density projections in Cartesian coordinates \citep{Lucchini2026}. {Bottom:} On-sky images of the \ionn{H}{i} column density colored by radial velocity. Each of the four panels shows the gas at a different distance range, viewed from a solar-like position 8 kpc from the Galactic center.}
    \label{fig:projected_gas}
\end{figure}

The life cycle of gas in galaxies affects a wide range of important issues from the baryonic mass distribution of the universe to the formation of planets and the conditions for life. Successive generations of star formation and stellar feedback shape the structure and chemical content of galaxies. However, the origin of galaxies' gaseous star-forming fuel and how it is replenished or recycled over cosmological timescales is poorly understood.  The key to understanding these processes is the elusive reservoir of cold ($\lesssim 10^4$ K) gas in the circumgalactic medium \citep[CGM;][]{Tumlinson2017}, and its structure from sub-pc to 100 kpc scales.  Figure \ref{fig:projected_gas} shows this cold CGM gas in a high-resolution simulation of the Milky Way.

Cold CGM gas may arise from thermal instabilities in the hot halo, outflows from the disk via a ``galactic fountain,'' direct accretion from the IGM, and/or gas stripped from infalling satellite galaxies \citep[e.g.,][]{Maller04}. Cosmological simulations have probed these possible origins \citep{Keres:2009, Lucchini2024, Augustin2025}.  However, cold gas in cosmological simulations remains unconverged at currently achievable simulation resolutions \citep{Hummels2019, FaucherGiguere2023}, so detailed comparison with data is necessary to test theoretical predictions. The metal and dust content of this cold gas offers complementary ways of probing its origin.  Higher metal content potentially indicates an origin in the Milky Way disk \citep{Fox2015, Fox2023}.

The cold CGM's emission is generally too faint for current facilities, excepting 21 cm \ionn{H}{i} emission around our Galaxy. Detailed \ionn{H}{i} maps have revealed a population of discrete, high-velocity clouds (HVCs) at velocities inconsistent with galactic disk rotation \citep{Putman2012}.  A key challenge to interpreting \ionn{H}{i} observations is that distances to the gas are generally unknown or poorly constrained.

Gas absorption lines in spectra of hot stars have enabled upper and lower distance limits for many major cold cloud complexes \citep{Thom2006, Wakker2007, Thom2008, Wakker2008, Smoker2011}.  Hot stars have few intrinsic absorption lines, enabling straightforward identification of absorption from line-of-sight clouds.  However, using hot stars for 2D or 3D gas mapping is hampered by the low density of hot stars at high Galactic latitude.  Absorption in the spectra of background galaxies and quasars has been used to make 2D maps of cold gas throughout the Galaxy \citep{Murga2015}, but cannot be used for 3D tomographic mapping within the Galaxy.

Lower-mass main sequence stars and red giants are abundant and would make excellent backlights for tracing cold gas absorption, but have a forest of strong intrinsic absorption lines in their spectra.  The intrinsic absorption in these abundant stars precludes detailed 3D tomography of cold gas with low-dispersion spectrographs.  The two strongest tracers of cold gas in the optical are \ionn{Ca}{ii} and \ionn{Na}{i}, both of which also appear as strong, broad absorption in the spectra of cool stars.  At low spectral resolution ($R\sim3,000$), a relative velocity separation between the backlight star and cold gas of $\gtrsim100\kms$ is required to disentangle intrinsic absorption. At a resolution of $R\approx15,000$ the damping wings of the strong stellar photospheric lines are nearly fully resolved, and enable the separation of stellar and intervening absorber lines to much lower relative velocities.  Via will build the first tomographic map of the Milky Way's cold gas from millions of stellar backlights, a capability that no existing survey provides.

\subsection{Characterizing the Transient Universe}
\label{s:transients}

The Rubin alert stream started in February 2026, and, when fully operational, is expected to generate roughly 7 million alerts  {\it every night} for objects that have changed. Among these will be millions of supernovae, thousands of tidal disruption events, and an unknown number of more exotic and poorly understood explosive transients. LSST's 100-fold increase in transient discovery rate will enable the detection of rare phenomena with significantly higher cadence and the construction of statistically meaningful samples. Remarkably, there will be approximately one active supernova per sq. deg. at any given time with $r<24$.   In each case, classification and physical interpretation require spectroscopy LSST cannot provide and, for the faintest events, spectroscopy reaching LSST's own single-visit depth.

The drastic increase in discovery rate will enable---for the first time---population-level studies of extreme explosive engines and progenitor configurations, thereby revealing how massive stars interact, shed mass, and form compact remnants. Many of these events are extreme in luminosity, duration, color, or environment, and each offers insight into exotic central engines, pre-supernova mass loss, and/or the diversity of binary interaction channels.

LSST will discover thousands of rare and exotic transients, including hydrogen-poor superluminous supernovae (SLSNe), luminous red novae, fast blue optical transients (FBOTs), strongly interacting supernovae and likely never-before-seen events. These events probe the extremes of stellar evolution---testing models of stellar mass loss, binary interaction, and exotic central engines. The most luminous of these events ($M\lesssim-20$) will be detectable in LSST to cosmological ($z\gtrsim1.5$) distances \citep{villar2018superluminous}, enabling population studies across cosmic time. Many open questions exist about these rare, luminous populations, including the nature of their central engines \citep{lebaron2026most}, how progenitor properties and environments (such as metallicity) influence their formation and evolution \citep{nugent2025characterizing}, and the prevalence/timing of pre-explosion mass loss or eruptive activity \citep{gkini2025eruptive}. 

Rapid spectroscopic follow-up of gravitational-wave (GW) counterparts enables exciting frontier science.  The first joint detection of gravitational waves (GWs) and electromagnetic radiation from GW170817 opened up a new field of multimessenger astrophysics in 2017 \citep{abbott2017gw170817}. This event established neutron star mergers as a site of heavy \textit{r}-process nucleosynthesis \citep{cowperthwaite2017electromagnetic}, constrained the neutron star equation of state \citep{radice2018gw170817}, and delivered a standard-siren measurement of the Hubble constant \citep{palmese2024standard}.  In the years since GW170817, no additional kilonova has been definitively observed coincident with GWs, despite several GW detections of mergers involving at least one neutron star.  This situation is likely to change soon due to two developments: (1) LSST will include a target-of-opportunity mode designed to rapidly tile gravitational-wave localization regions, enabling the discovery of optical counterparts to compact object mergers \citep{Andreoni22}; (2) the fifth observing run of the LIGO, Virgo, and KAGRA detectors, expected to begin in 2028, is expected to detect dozens of neutron star mergers, with $\sim20$ detectable by LSST and LIGO over the course of the 2.5-year run \citep{andreoni2024rubin}.

LSST will discover thousands of tidal disruption events (TDEs), which occur when a star enters the tidal radius of a massive black hole (MBH) and is consumed \citep{phinney_tde, rees_tde, evans_tde}. These are a key probe of MBH demographics, accretion physics, nuclear stellar dynamics, and the circumnuclear medium of quiescent galaxies \citep{Gezari2021TidalEvents}. Uniquely, LSST will probe MBH populations both to high-redshift and the lowest masses. The rare ($10$s\,yr$^{-1}$ across the Rubin sky) overluminous TDEs \citep{Hammerstein2023TheSurvey}, which may be related to the subset of TDEs that launch highly energetic relativistic jets \citep{Andreoni2022AHole}, will probe MBH populations out to cosmological redshifts ($z\gtrsim 1$). LSST may be sensitive enough to discover a population of TDEs by the lowest mass MBHs, the elusive intermediate mass black holes (IMBHs), the demographics of which are a sensitive probe of supermassive black hole formation \citep{Greene2020Intermediate-MassHoles}. These probes will enable strong constraints on MBH formation, evolution, and growth.

\begin{figure*}[t]
  \centering
  \includegraphics[width=0.48\linewidth]{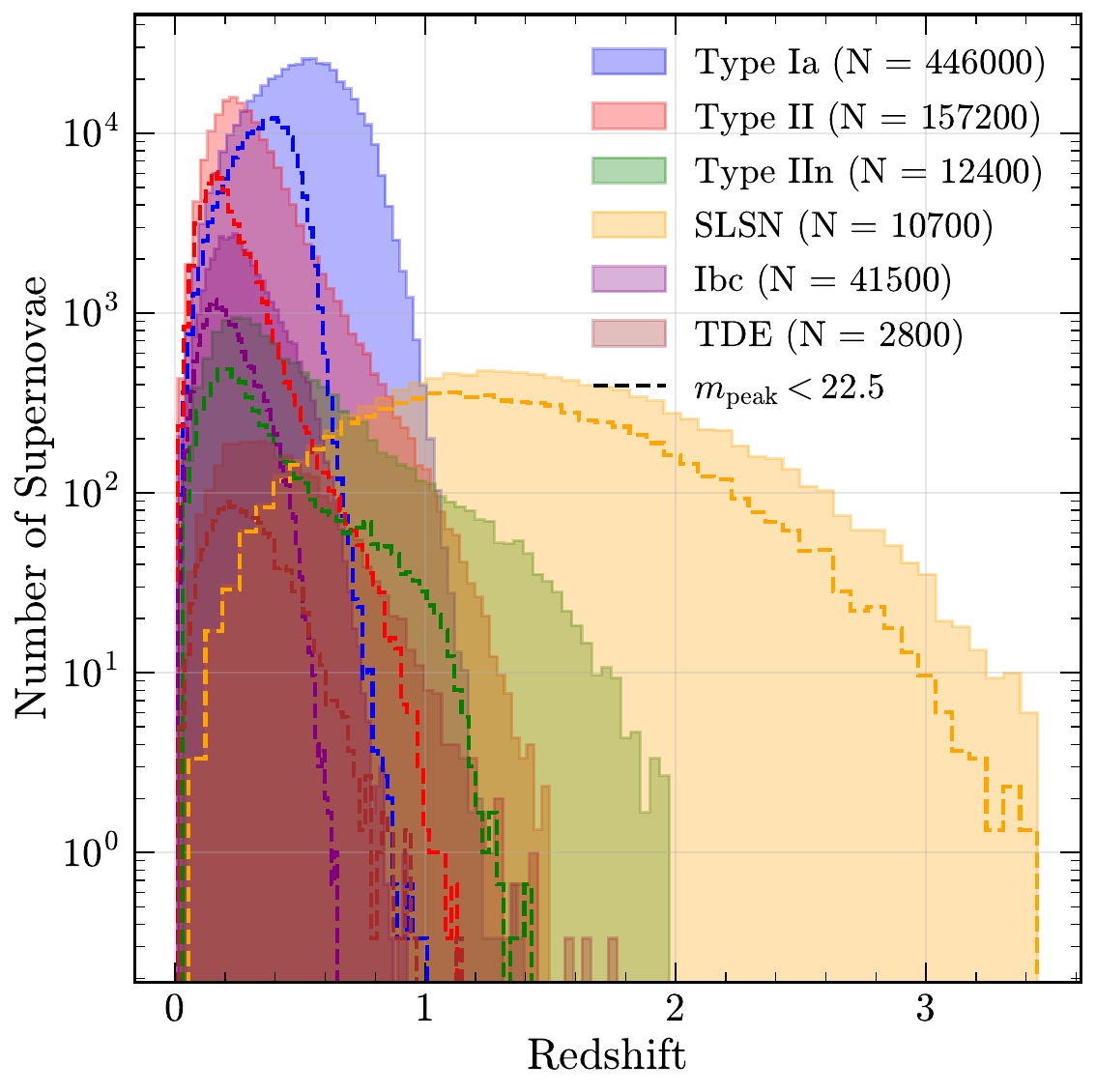}
  \hfill
  \includegraphics[width=0.48\linewidth]{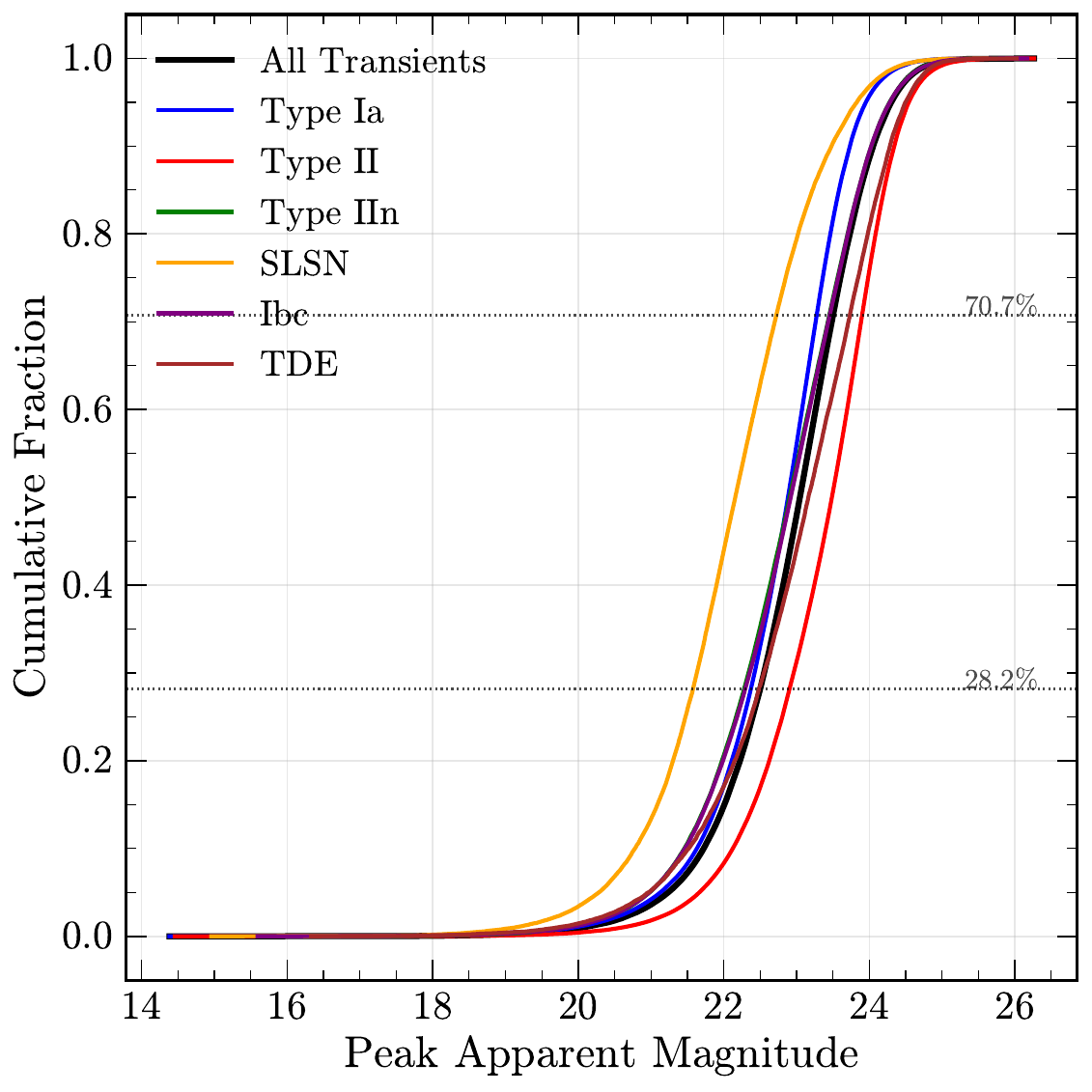}
  \caption{{Left:} Redshift distributions by transient class for a PLAsTiCC-simulated sample of realistic LSST transient light curves at a peak apparent-magnitude limit of $m_{\mathrm{peak}}=23.5$ \citep{kessler2019models}. Dashed curves show the reduced survey volume for a brighter limit of $m_{\mathrm{peak}}=22.5$, representative of 4MOST–TiDES. 
  {Right:} Cumulative distribution of peak apparent magnitudes for the same transient classes. Percentile markers indicate that 28.2\% and 70.7\% of all transients are brighter than $r\!\approx\!22.5$ and 23.5~mag, respectively.}
  \label{fig:bb-redshift-comparison}
\end{figure*}

LSST will be revolutionary, but photometry and light curves reveal only part of a complex story.  A deeper, physical understanding of the phenomena, including chemical composition and kinematics, requires contemporaneous spectroscopy.  There are several large-scale spectroscopic programs planned, but they are limited to relatively small aperture telescopes (e.g., 4MOST--TiDES; \citealt{frohmaier2025tides}).  An extensive program on larger telescopes with an efficient low-dispersion spectrograph capable of reaching the single-epoch depth of LSST ($r\approx24$) would be transformative for transient follow-up science by opening the full discovery space made available by LSST (see Figure~\ref{fig:bb-redshift-comparison}).

\subsection{The Via Project}
\label{s:overview}

Addressing the three topics described in \S\ref{s:streams}--\ref{s:coldgas} requires pushing observational limits in the same direction: moderately high-resolution spectroscopy ($R\gtrsim10,000$), with a stability of $\sim100\,\ms$, of faint stars across the Milky Way and beyond.
Extreme-precision radial velocity spectrographs---designed for detection of exoplanets orbiting low-mass stars---can easily achieve this precision, but are limited to the brightest stars and small samples.
On the other hand, massively multiplexed faint-object spectrographs---designed to measure redshifts for millions of stars and galaxies---do not have the stability required to reach $100\,\ms$ precision.
The regime of $100\,\ms$ radial velocity stability for millions of faint targets is uncharted territory.
Accomplishing both goals requires developing new instrumentation and executing a long-term observing program on large telescopes.

The Via Project emerged from these considerations, following conversations about future instrumentation at the MMT in the fall of 2022.  New instrumentation is clearly required, including a moderately high-resolution spectrograph with stability at the level of $100\,\ms$; high throughput to reach $G\approx21$ on 6.5\,m telescopes; flexible, fiber-based multiplexing of $>500$ over a 1\,deg FoV; and wavelength coverage spanning the velocity-sensitive \ionn{Mg}{i} triplet and the cold gas-sensitive \ionn{Na}{i} doublet (505--595 nm).  Observations spanning both hemispheres are essential to deliver the most robust constraints on dark matter models, to provide a complete census of faint satellite galaxies, and to deliver a map of cold gas across the sky.  The 6.5\,m Magellan/Clay telescope has a Cassegrain focus that is optically identical to the MMT, and is located in the opposite hemisphere.  This presented a unique opportunity to equip 6.5\,m telescopes in the north and south with identical instruments in order to conduct a dual-hemisphere five-year survey.

In \S\ref{s:transients}, we motivated fast followup of sources to the LSST single-visit depth.  A dual-hemisphere survey, spanning thousands of square degrees at $\sim1$\,hr depth on 6.5\,m telescopes is also well-suited to follow up transients detected by the LSST.  Transients will be so numerous that there will be $\sim1$ per Via FoV at any given time.   Being on-sky for a significant fraction of time will also enable rapid target-of-opportunity follow-up for the most interesting transients.    This science case motivated the addition of Boombox, an efficient low-resolution spectrograph covering a wide wavelength range to the Via instrument suite.

These instrument and survey characteristics will enable a wide range of exciting ancillary science that can be pursued in parallel to the core science goals, including measurement of the Ly$\alpha$ forest at $3<z<4$, characterization of polluted white dwarfs, chemistry and kinematics of metal-poor stars, a census of hot Jupiters and brown dwarfs at low metallicity, orbits of \gaia\ astrometric binaries, and integrated kinematics of unresolved low-mass galaxies to $z\approx0.1$.

By the summer of 2024 the Via partnership was defined and the basic parameters of the instrument and survey were set.  A driving goal of the project is to be on-sky in 2027 in order to fully harness the revolution that will be ushered in by LSST.  The rest of this document describes the instrumentation required to carry out this science program (\S \ref{instrument}), the data reduction and analysis plan (\S \ref{data}), the survey strategy (\S \ref{survey}), and a comparison to existing and planned instruments and surveys (\S \ref{comp}).  The key parameters for the Via Project, spectrographs, and survey, are given in the Table below.

\vspace{2cm}


\begin{tcolorbox}[
                colback=cornflowerblue!15,    
                  colframe=cornflowerblue!75!black, 
                  coltitle=white,    
                  title={\Large Via at a Glance}, 
                  center title, left = 1.4cm, right = 1.4cm]

    \centering

    \begin{tabular}{lr}
      Telescopes & 6.5m MMT \& Magellan/Clay \\
      Field of View & $1^\circ$ diameter \\
      Number of Robotically Positioned Fibers & 576 \\
      Number of Fiducial/Sky Fibers & 60 \\
      Fiber Core Diameter &  $\mathrm{200\,\mu m}$ ($1.2\arcsec$ on-sky)\\
      Primary Survey Size & $>500$ nights \\
      Number of Unique Sources & $\approx2,000,000$ \\
      Number of Repeat Observations & $10^5$ stars with $>5$ visits \\
      & $10^4$ stars with $>15$ visits \\
    \end{tabular}
  
    \vspace{0.15cm}

    \begin{tabular}{lrr}
      Spectrograph & Viaspec & Boombox \\
      \hline
         Wavelength Range & 505--595 nm & 360--1010 nm \\
         Resolution & $R \approx 15,000$ & $R \approx 1,000$\\
      Mean Instrument Throughput & 35\% &41\% \\
      Number of Fibers (incl. calibration) & 600 & 36\\
      Depth at SNR~$=3$ in 1 hr & $G=21$ & $G=23.5$\\
      Radial Velocity Stability & $<100\ms$ & $< 5\kms$ \\
          \end{tabular}

\end{tcolorbox}

\clearpage

\section{Instrument}
\label{instrument}

\subsection{Overview}

\begin{figure}[ht!]
    \centering
    \includegraphics[width=\textwidth]{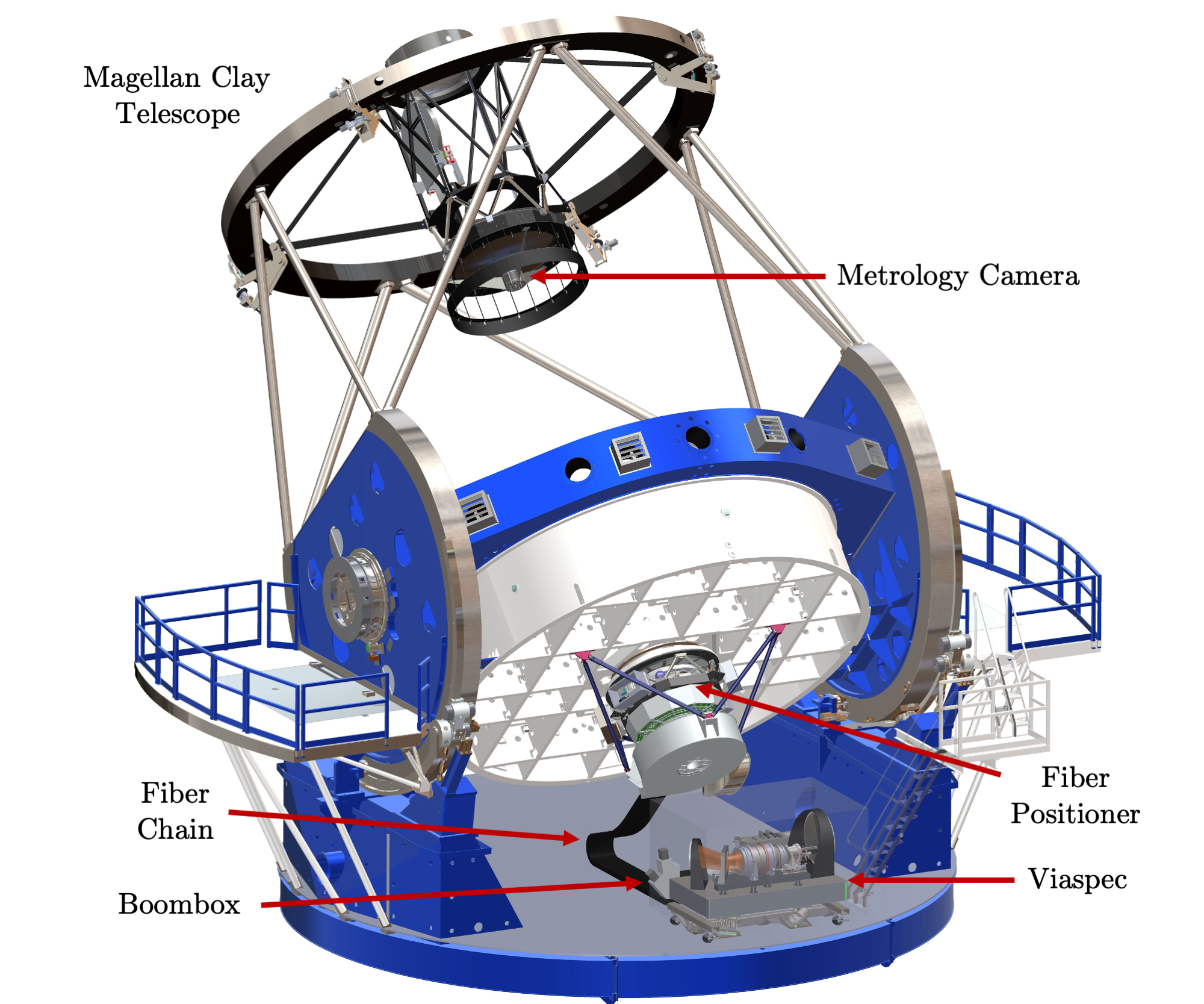}
    \caption{Overview of the Via instrument system, as shown deployed at the Magellan/Clay telescope.  The fiber-positioning system (FPS) is secured to the back of the primary mirror at the focus of the $f/5.3$ secondary.  The fiber run brings the optical fibers from the focal plane to the Viaspec and Boombox spectrographs, located on the azimuth disk.  The metrology system resides in front of the secondary mirror.  At the MMT the system is similar, except that the spectrographs are permanently located in a separate room and are fed by a much longer fiber chain.}
    \label{fig:instover}
\end{figure}

The science goals described in \S\ref{motivate} place three simultaneous demands 
on the instrument: radial velocity stability at the 100~m~s$^{-1}$ level, depth to 
$G \approx 21$ on 6.5m telescopes, and wide-field multiplexing over a $1\degree$ field 
of view. The transient follow-up program further requires an efficient low-resolution 
spectrograph capable of reaching the single-epoch depth of LSST ($r \approx 24$). 
Meeting these requirements drove the design of the 
Via instrument suite (Figure~\ref{fig:instover}), which consists of six key subsystems: (1) a focal plane hosting 576 robotically positioned science fibers (540 Viaspec, 36 Boombox) plus 60 fixed fiducial/sky fibers; (2) Viaspec, an $R\approx15,000$ stabilized bench spectrograph covering $505<\lambda<595$ nm with a 254 mm collimated beam; (3) Boombox, an $R\approx1000$ dual-channel spectrograph covering $360<\lambda<1010$ nm with a 50-mm collimated beam; (4) a fiber run connecting the spectrographs to the fiber-positioner system; (5) a camera-based metrology system to provide iterative fiber-positioning feedback; and (6) guiders and wavefront sensors.  This section describes these systems.

\subsection{Focal Plane \& Fiber Positioners}\label{sec:fps}

\begin{figure}[t!]
    \centering
    \includegraphics[height=0.41\textwidth]{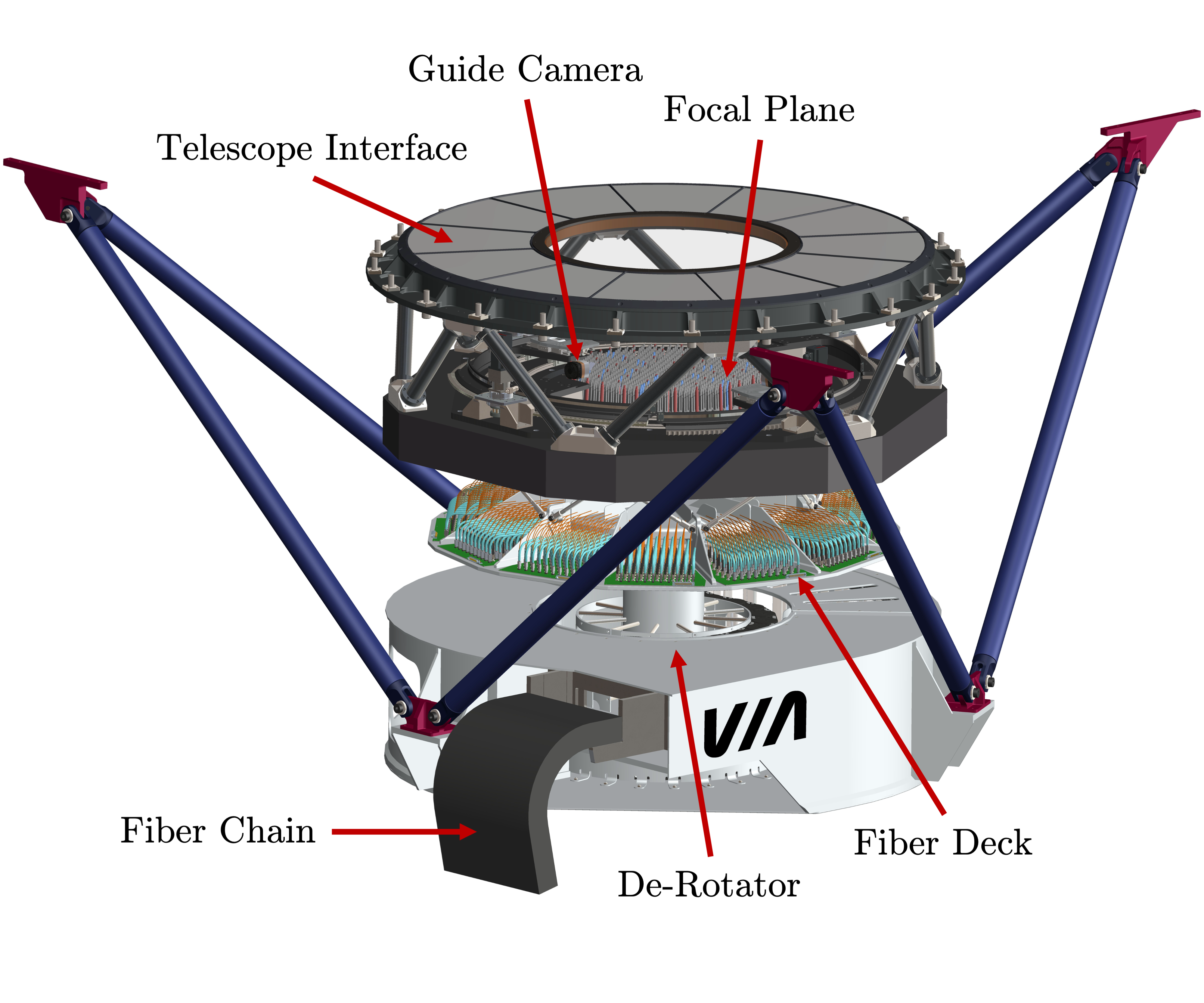}
    \includegraphics[height=0.41\textwidth]{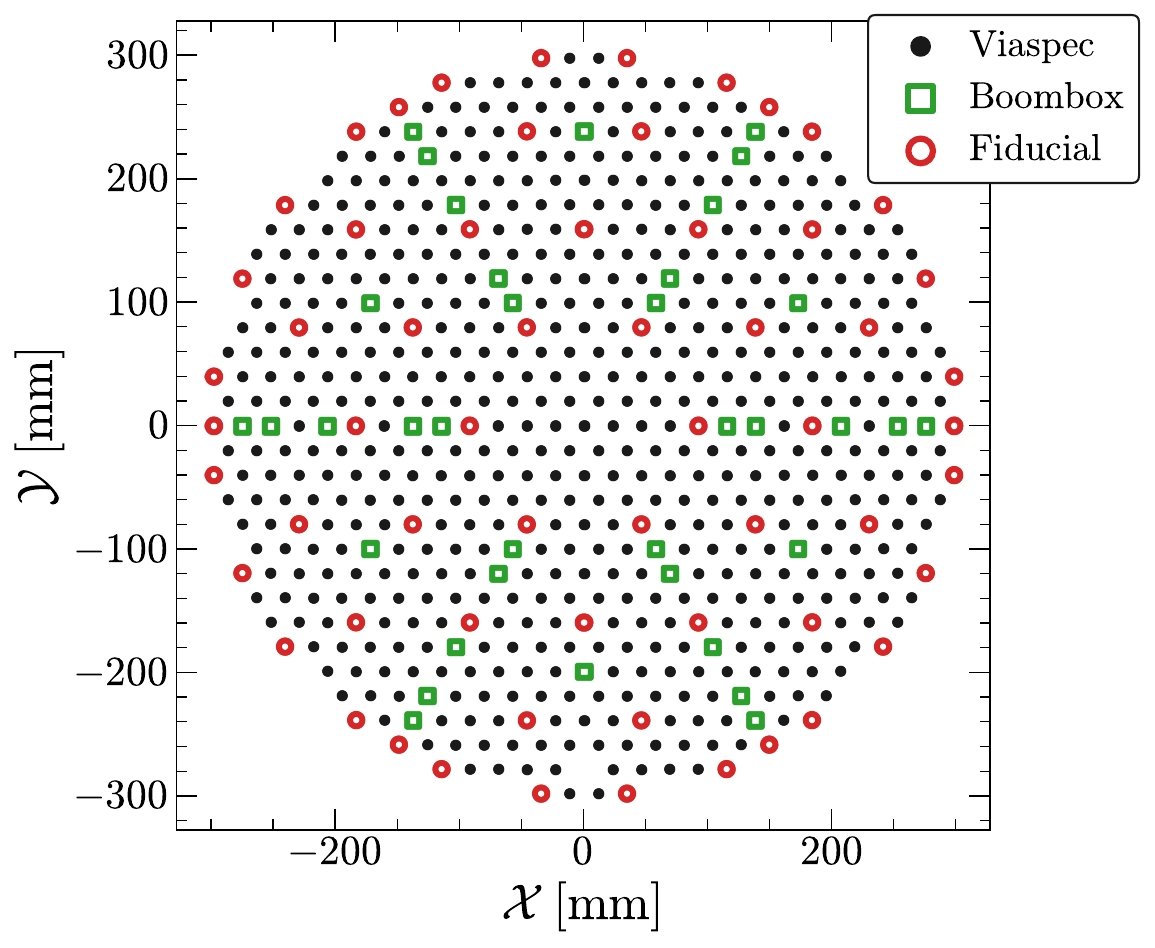}
    \caption{{Left:} CAD model of the Via fiber positioner in its full housing, with individual components labeled. 
    {Right:} Layout of fiber positioners and fixed fiducials at the focal plane.  
    The base positions of Viaspec ($N=540$) and Boombox ($N=36$) positioners are shown with circles and squares, respectively.  
    Fixed fiducials ($N=60$, open circles) will be used to set the physical coordinate system of the focal plane in the metrology camera and will also be used for fixed sky observations during normal science operations.}
    \label{fig:fps_cad}
    \label{fig:fps_layout}
\end{figure}

The Via focal plane (see Figure \ref{fig:fps_cad}) uses the $f/5.3$ Cassegrain foci of the MMT and Magellan telescopes, which deliver excellent image quality over a $1^\circ$ diameter FoV.  This corresponds to a focal plane with a physical diameter of $600$ mm.  Generating a regular hexagonal grid of fiber base coordinates within this circular diameter, a fiber-to-fiber separation (often referred to as {\em pitch}) of $23.5$ mm accommodates $636$ fiber positioners and fiducials in the focal plane.  

Pick-and-place robots that sequentially move multiple fibers or individual robotic actuators that move all of the fibers simultaneously are the most common technologies used for astronomy.  The Via fiber positioners use two-axis selective compliance robot arm (SCARA) positioners, often termed ``theta--phi'' positioners in astronomy.  Each actuator has an on-axis ``$\alpha$'' rotation with an offset secondary rotary ``$\beta$'' arm that carries the optical fiber.  With two rotational degrees of freedom, each actuator can place the fiber anywhere within its ``patrol region''.  By overlapping the patrol regions of individual actuators, gapless focal plane coverage can be achieved.  Theta--phi positioners have been successfully used in several recent multi-object spectrographs, including DESI and SDSS-V.

\begin{figure}
    \centering
    \includegraphics[width=0.9\linewidth]{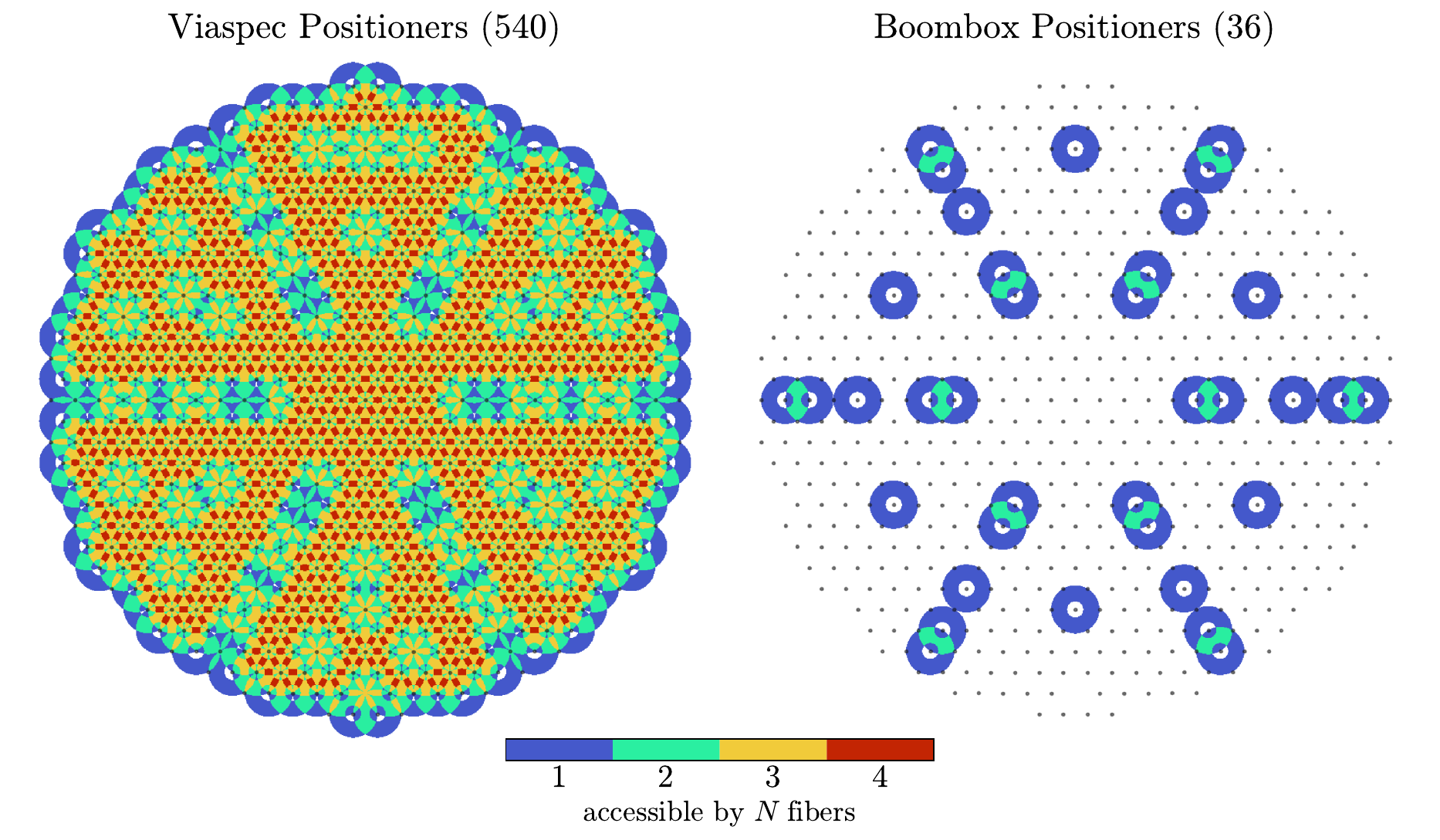}
    \caption{The Via focal plane, showing the patrol regions of Viaspec (left) and Boombox (right) fiber positioners, colored by the number of positioners that can access a given region.
    The Viaspec patrol regions are highly overlapping, with $\gtrsim 60\%$ of the focal plane accessible by $\geq 3$ Viaspec positioners. 
    Boombox positioners are arranged in a pattern of linear spokes, so that arbitrary arrangements of high-value transient targets can be observed by rotating and shifting the field (see Figure~\ref{fig:bb_fibers}). 
    }
    \label{fig:patrol}
\end{figure}

After a year-long R\&D program, we selected the ``Starling'' fiber positioners manufactured by Micro Precision Systems (MPS).  These positioners are similar to those used by the SDSS-V collaboration, with modifications including a new gearbox to prevent significant power-off motion and an EtherCat interface for much faster communication.  Each axis is driven by a brushless DC motor with Hall sensors.  The Hall sensors are calibrated to provide position encoding (accurate to $\approx 0.2^\circ$) in addition to motor commutation. The Hall sensors can detect motor stalls such as might be caused by $\beta$ arm collisions.  Given that the Starling $\alpha$ arm's reach is $7.64$\,mm, there are two possibilities for the $\beta$ arm length that ensure continuous coverage of the focal plane.  The $\beta$ arm can either be the same length as the $\alpha$ arm (allowing each positioner to reach its own center), or it can be long enough to enable each positioner to reach the center of its nearest neighbors.  The longer $\beta$ arm provides greater flexibility for placing multiple fibers in close proximity, and we adopt this configuration for Via (see Figure~\ref{fig:patrol}).  The separation between actuator centers at the focal plane is $22.92$ mm---different from the pitch at the mounting plate, due to the curvature of the mounting plate---and therefore the distance between the beta axis and fiber is set to $15.28$ mm.  Production actuators are scheduled to begin arriving in mid 2026.

The control theory for Via actuators is based on the \texttt{kaiju} optimization routine used by SDSS-V \citep{Sayres2021}.  Given a target list split into various priority classes, the controller allocates each actuator to its nearest and highest-priority star, and so on down the priority list until all fibers are allocated.  The controller then calculates the optimal path for each actuator to reach its final position, avoiding collisions between actuators and restricting the overall power draw of the FPS to within a certain tolerance (``annealing'').  We expect a re-configuration time of $\lesssim2$ minutes between consecutive fields.

Maximizing throughput places a stringent requirement on the positioning accuracy of the individual fiber robots. We estimate the magnitude of the mis-centering throughput loss by simulating the telescope's PSF on the focal plane.  The PSF is modeled from stellar images taken with Megacam at Magellan in $1.0\arcsec$ seeing.  Under the reasonable assumption that this PSF is dominated by atmospheric seeing, we fit this effective PSF with a von Karman turbulence model, and rescale this model to evaluate our mock seeing PSF at different seeing values.  Adopting a $1.2\arcsec$ fiber diameter at the focal plane, we numerically integrate the 2D PSF flux within the circular fiber aperture, over a range of offset positions.  To keep the throughput loss due to fiber mis-centering below $2\%$, the actuators must be capable of placing a fiber within $20\,\mu m$ of the commanded position.

In practice, this accuracy will be achieved with a metrology system that measures the absolute positions of fibers on the focal plane.  The metrology camera (MetroCam), mounted just in front of the secondary mirror, directly images the focal plane with a large CMOS detector and a custom lens.  Viaspec fibers can be back-illuminated at the slit with a movable back-illumination system. Each back-illuminated fiber image is sampled by approximately $2.5 \times 2.5$ MetroCam pixels.  We have forward modeled MetroCam images---including the full optical model of MetroCam in Zemax and background noise estimates in the image---and verified that the metrology system can centroid fibers to better than $\lesssim 5\,\mu m$ at the focal plane.

Metrology measurements will be placed on an absolute scale using the known locations of 60 fixed fiducial fibers on the focal plane (see Figure~\ref{fig:fps_layout}).  These fibers will be mounted in place of actuators on rigid structures whose absolute location on the focal plane will be measured with a coordinate measuring machine in the lab after the positioner is assembled.  The Via guide cameras also host fixed fibers that are imaged by MetroCam, which tie the focal plane coordinate frame to the celestial coordinate frame at any given pointing.  The fixed fiducial fibers serve a dual purpose as sky fibers, providing at least 60 sky spectra in every Via pointing.

The fiber positioners are extensively tested across a range of operating loads and temperatures.  During ``blind'' moves with no camera correction, the positioners are able to place fibers with a root-mean-square error of $\approx 50\,\mu m$ relative to the commanded positions.  After a single correction loop with a metrology camera, the error drops to below $< 5\,\mu m$ rms, easily meeting our positioning requirements.  This is only part of the total fiber error budget; other terms include the pointing/guiding accuracy of the telescope, the transformation of guider coordinates to focal plane coordinates, and the ability to accurately centroid fiber images in the presence of dome turbulence \citep[e.g.,][]{Schlafly2024turb}.  Ongoing work is forecasting and mitigating these errors, and the full fiber-positioning accuracy (from on-sky astrometry to fiber aperture losses) will be validated during commissioning \citep[following][]{Schlafly2024dither}.

To maintain accurate pointing and telescope focus, Via is equipped with two guiders and two Shack--Hartmann wavefront sensors that patrol a one arcminute-wide annulus centered 31.7~arcminutes off-axis.  
Each of the four units can move in an $80^\circ$ segment of the annulus, accessing 44 sq.~arcmin.~of sky.  
The Shack--Hartmann wavefront sensors sample the telescope pupil with $\approx65$ cm resolution at the primary, and are based on the Binospec wavefront sensors \citep{Fabricant2019}.  
The imagers for the guiders and wavefront sensors are Ximea MJ042MR-GP-P11-BSI cooled CMOS cameras with a 22 mm square format (larger than the required $\approx10$ mm), $11\,\mu m$ pixels, and 1.7e$^-$ read noise.

\begin{figure}[t!]
    \centering
    \includegraphics[width=\textwidth]{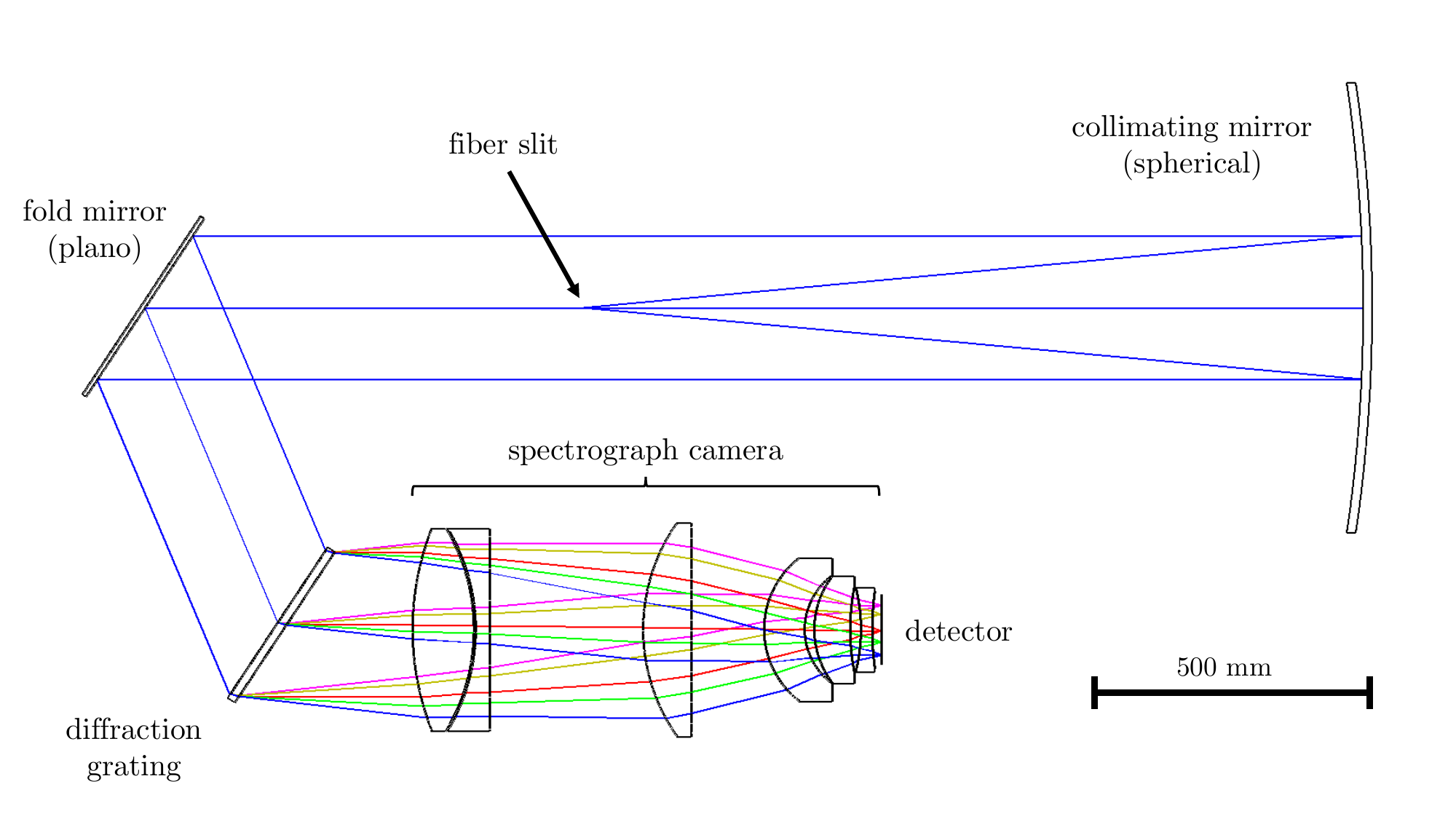}
    \caption{Zemax lens drawing of the Viaspec optics. This drawing shows a top-down view of the spectrograph optical prescription. The fiber slit is defined perpendicular to the page. Only a single on-axis field point is drawn and the ray colors differentiate between wavelengths of light.}
    \label{fig:optics}
\end{figure}

\subsection{The Viaspec Spectrograph}

\begin{figure}[t!]
    \centering
    \includegraphics[width=0.7\textwidth]{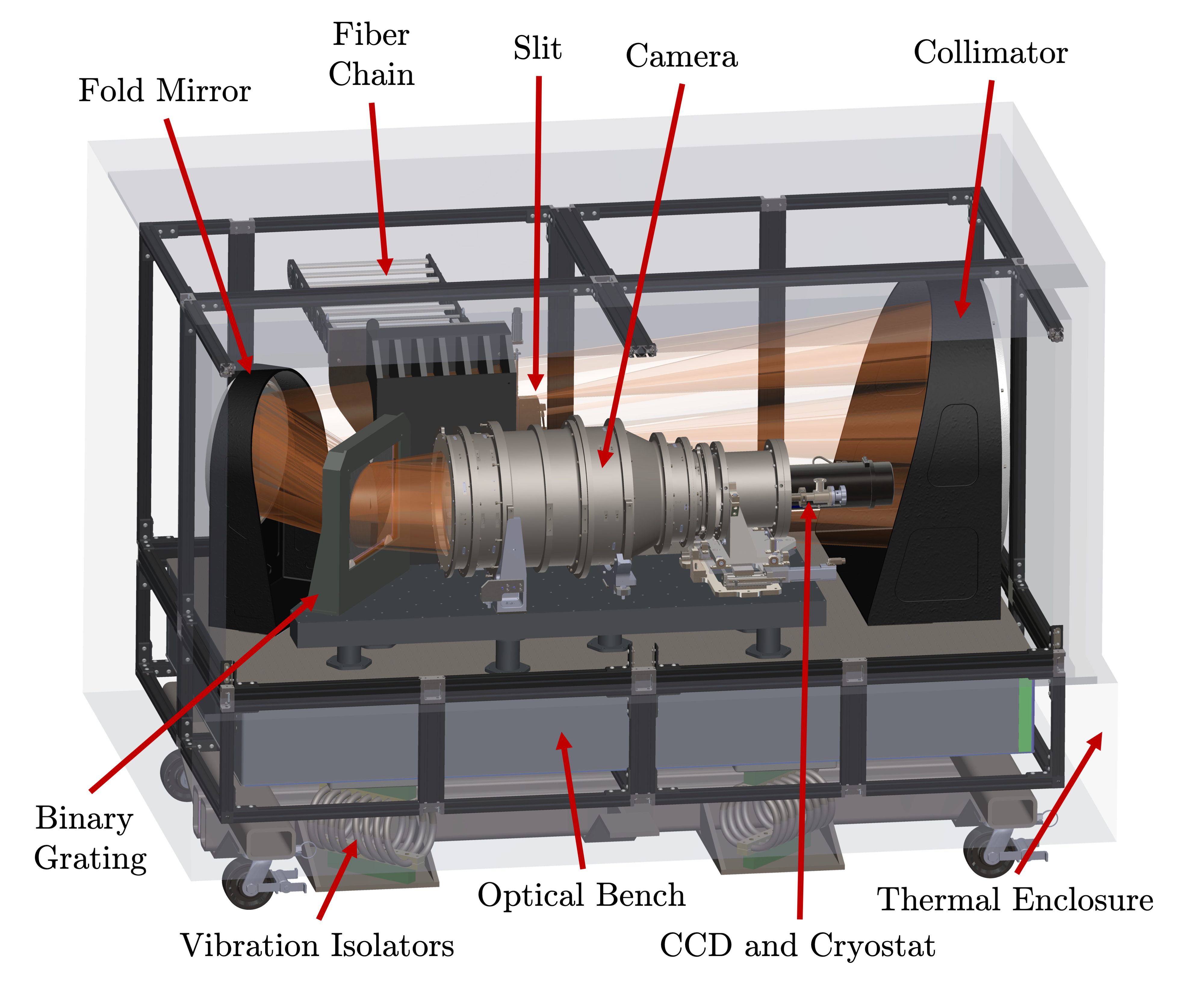}
    \caption{CAD model of the Viaspec spectrograph. The light path from the slit to the camera is shown in orange.}
    \label{fig:spec_cad}
\end{figure}

The fibers exit the FPS through a derotator so that the exiting fiber bundle only needs to accommodate changes in the telescope elevation if the spectrograph is mounted on the azimuth disk at Magellan or in the corotating building at the MMT. An energy chain protects the fibers in the derotator and in their path to the spectrograph.  The Via spectrograph is composed of a fiber slit, a collimator mirror, a fold mirror, a binary grating, and an imaging camera (see Figure~\ref{fig:optics} and Figure \ref{fig:spec_cad}).  At the spectrograph, the Via optical fibers are arranged into two parallel columns to form a pseudo-slit.  The fibers in each column are spaced by 800$\mu$m, and the two columns are offset by 400$\mu$m to provide a fiber spacing of 400$\mu$m.  

A rigid structure, called the {\em fiber shoe}, holds the fibers to form the spectrograph slit. The fiber shoe is removable from the bench spectrograph at Magellan, allowing the fiber-positioning system and the bench spectrograph to be moved and stored independently when not in operation. At the MMT, the fiber shoe need not be separated from the spectrograph when Viaspec is not in operation.  

\begin{figure}[t!]
    \centering
    \includegraphics[width=\textwidth]{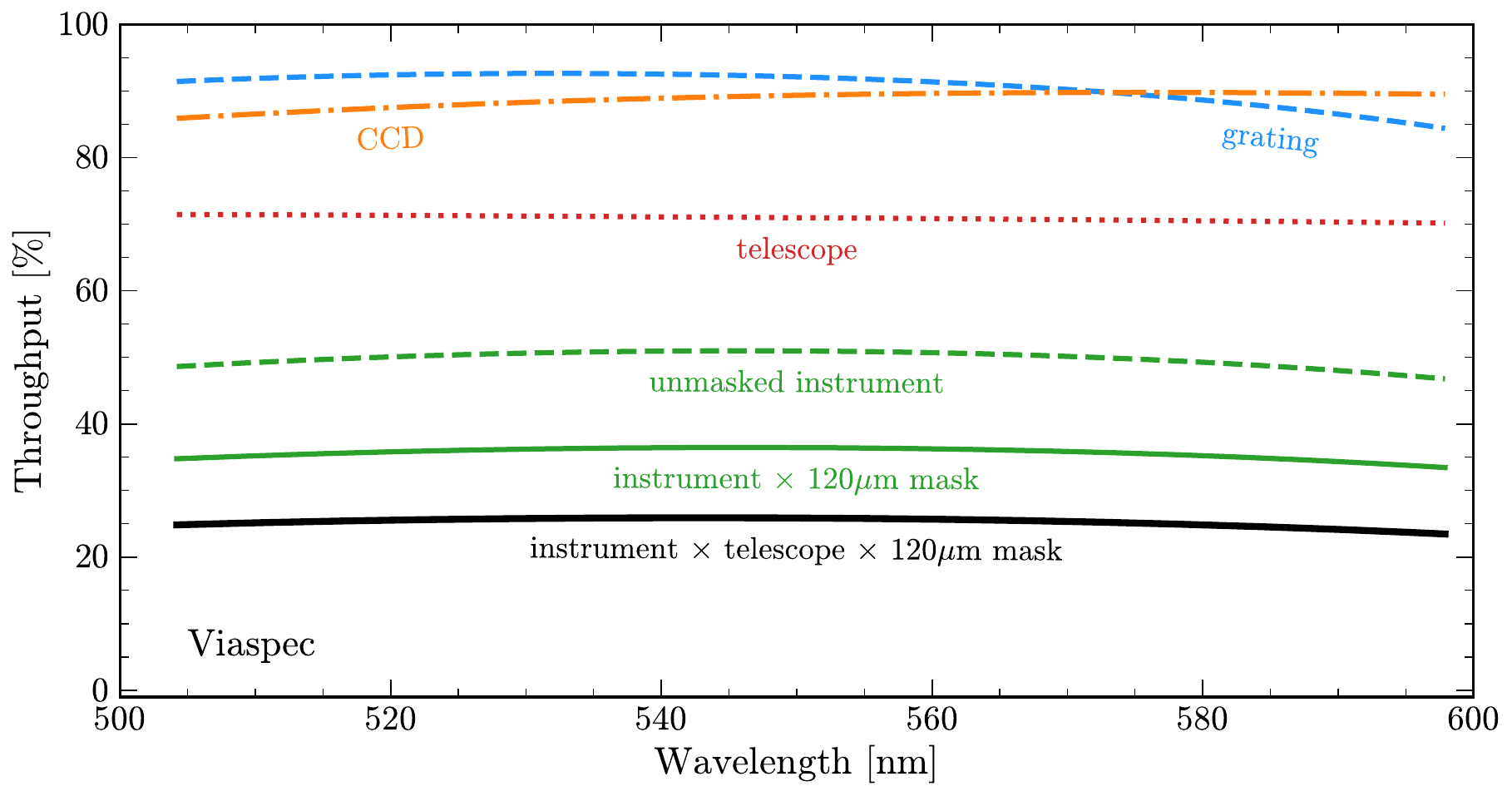}
    \caption{Throughput of different elements along the Viaspec optical path.
      The instrument throughput includes all components except the telescope, ADC (atmospheric dispersion compensator), and seeing-dependent aperture losses.  The total throughput includes telescope and ADC contributions.}
    \label{fig:tp_summary}
\end{figure}

An 800 mm diameter spherical mirror collimates the $\approx f/5.3$ output light from each fiber into a 254 mm diameter pupil on the face of the grating. A fold mirror allows a more compact spectrograph layout. The transmission grating is a binary grating manufactured by Plymouth Grating Laboratory in Carver, Massachusetts. The grating is etched into synthetic fused silica. Binary gratings are highly efficient when the grating pitch is approximately equal to the design wavelength. This condition is satisfied for the Viaspec gratings with a line density of 2000 l/mm and a design wavelength of 550 nm.  The total predicted throughput of the Viaspec spectrograph is shown in Figure \ref{fig:tp_summary}.

\begin{figure}[t!]
    \centering
    \includegraphics[width=\textwidth]{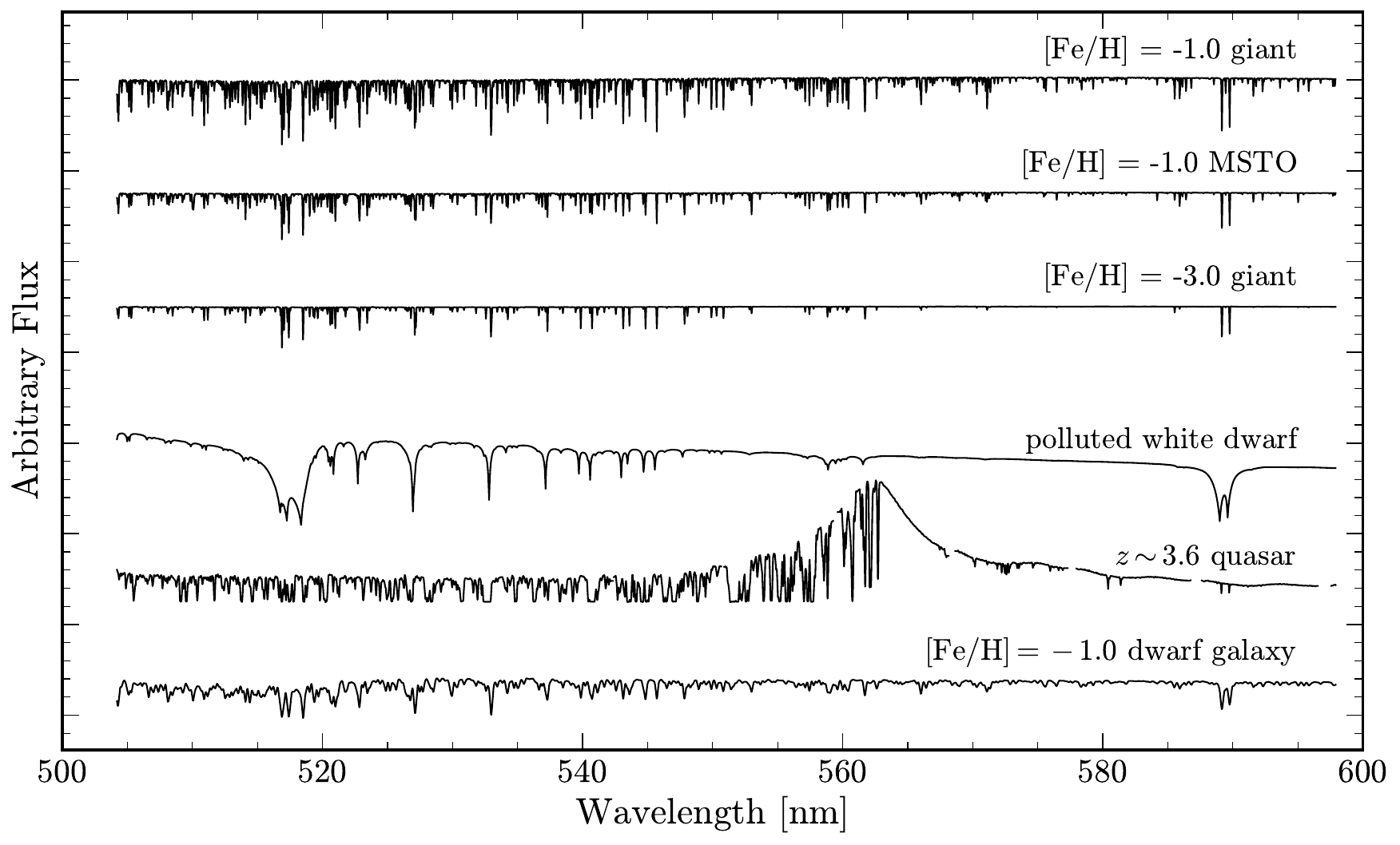}
    \caption{Example spectra demonstrating the Viaspec wavelength range and resolution, for a variety of core and ancillary science cases.
    The polluted white dwarf model is courtesy S. Blouin. 
    The quasar is a Keck HIRES spectrum of B1422+2309 courtesy A. Cowie and the Keck Observatory Archive.}
    \label{fig:spec_showcase}
\end{figure}

 The camera images the dispersed light from the grating onto an e2v CCD290-99 9K $\times$ 9K CCD. The six-element camera achieves well-corrected images. The narrow Via bandpass simplifies the color correction challenge. A biconcave field flattener is also used as the dewar window to reduce the number of surfaces.  Optimax is producing optics from optical glass supplied by Ohara.  The e2v CCDs were delivered in April, 2024.  
 Figure~\ref{fig:spec_showcase} showcases the Viaspec resolution and wavelength coverage with example spectra from various science cases.

\begin{figure}[t!]
    \centering
    \includegraphics[height=6.5cm]{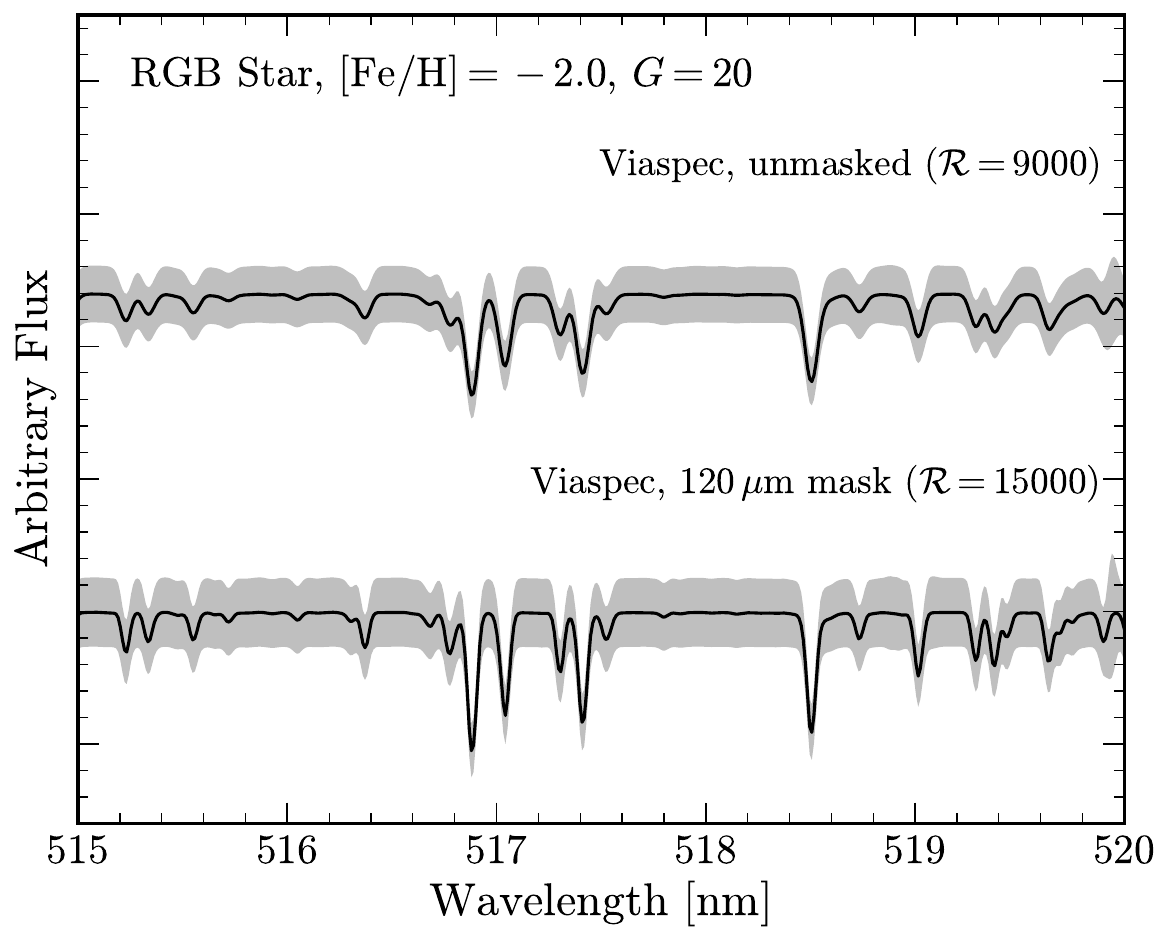}
    \includegraphics[height=6.6cm]{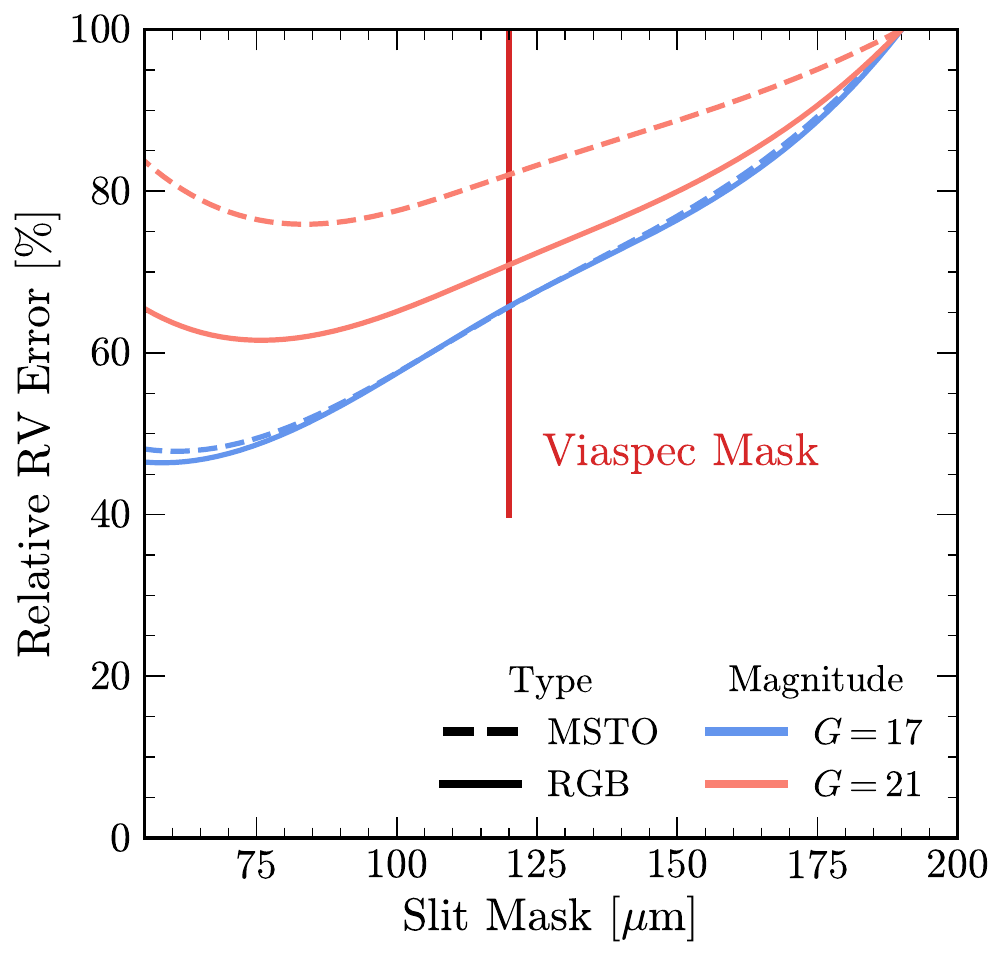}
      \caption{{Left:} Simulated spectra (with $1\sigma$ errors at $G = 20$ in dark time) of a metal-poor RGB star observed with unmasked Viaspec fibers (top), and with a $120\,\micron$ slit placed at the fiber output. {Right:} Improvement in the derived RV uncertainty as a function of the width of the Viaspec slit mask, relative to the unmasked case.}
        \label{fig:masktest}
\end{figure}

The spectral resolution---and consequently RV precision---of the Viaspec instrument can be increased by masking the output face of the fibers with a slit mask at the expense of light loss.  
Figure \ref{fig:masktest} shows the radial velocity precision gained by masking the Viaspec slit, as compared to the unmasked case. 
For this test we assume a metal-poor [Fe/H]$=-2.0$ red giant branch star observed for 1 hr.  
This test involves forward-modeling Viaspec spectra through the exposure-time calculator and fitting the data (see $\S$\ref{sec:etc} and $\S$\ref{sec:minesweeper}), and therefore self-consistently models the influence of masking on resolution and throughput.
At a mask width of $120\,\micron$ (for the $200\,\micron$ fiber diameter), the gain in RV precision for brighter RGB stars is $\approx 50\%$.
It is undesirable to use a mask width smaller than $\approx 100\,\micron$, since that would reduce the spectrograph's pixel sampling per resolution element below $3$ pixels.
We therefore use a $120\,\micron$ slit mask to maximize RV precision without sacrificing sampling, and remaining conservative about throughput loss. 
Our data simulations also demonstrate that stellar parameter estimates are not adversely affected by this choice of mask.
We emphasize that the mask is placed at the fiber \textit{output} in the spectrograph slit, and consequently delivers a constant throughput loss regardless of seeing. 

The Viaspec wavelength solution will be measured during science exposures with a continuous calibration unit mounted at the fiber positioner.  Argon lamp lines are relayed through 60 fibers evenly spaced along the Viaspec slit.  A variable neutral density filter adjusts the lamp line intensities to accommodate a range of science exposure times maintaining optimal count rates.

\subsection{The Boombox Spectrograph}

\begin{figure*}[tbp]
    \centering
    \includegraphics[height=0.44\textwidth]{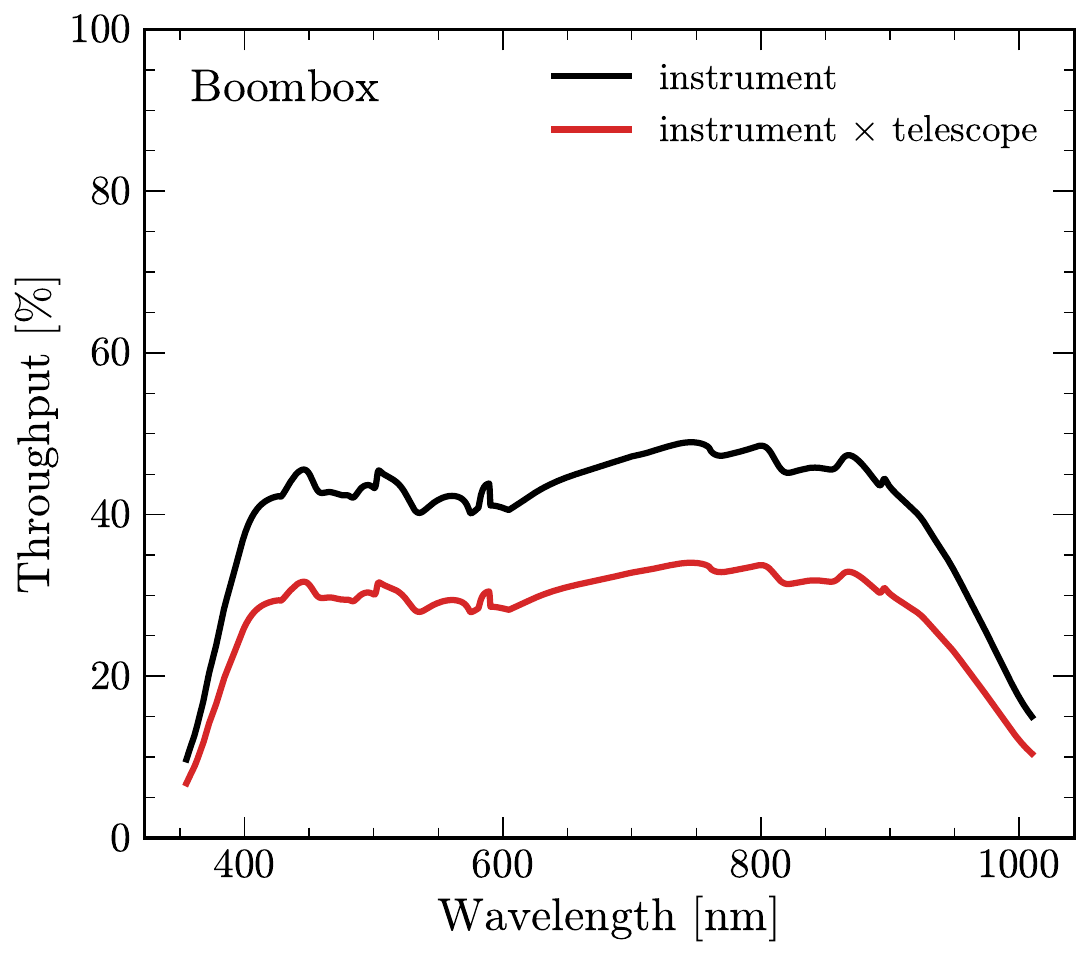}
    \includegraphics[height=0.44\textwidth]{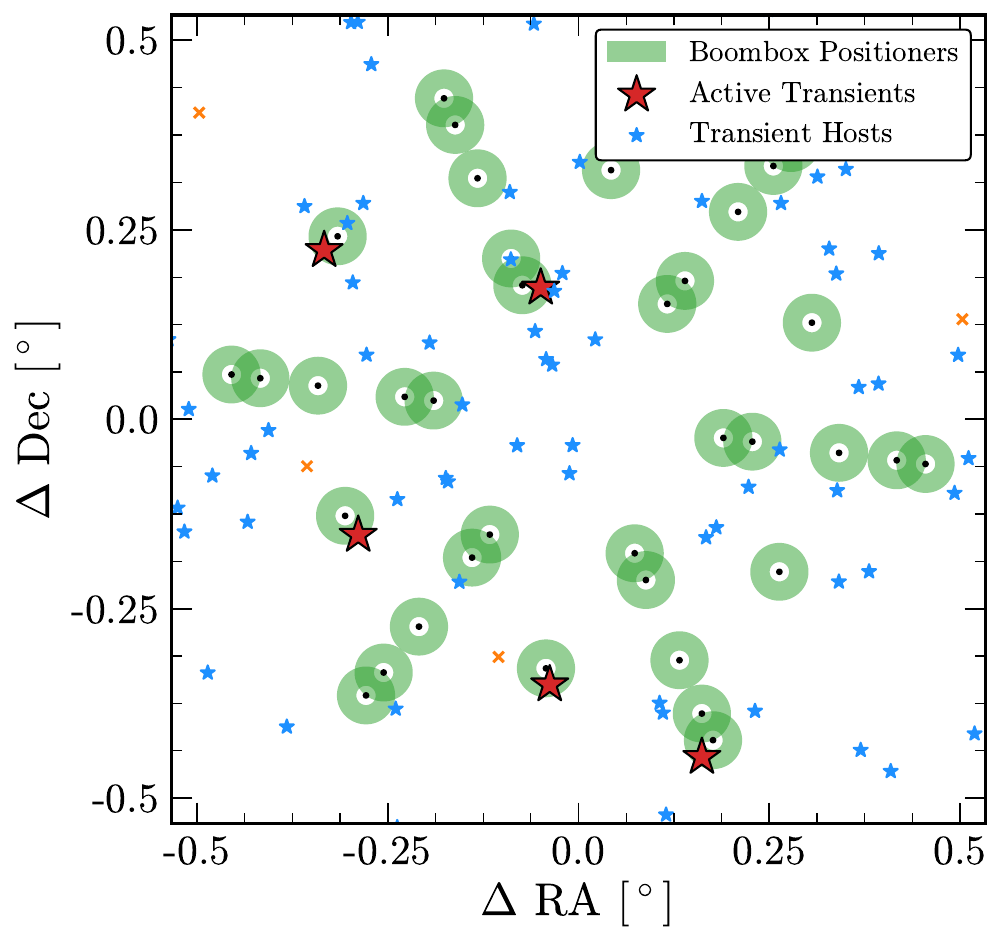}
    \caption{{Left:} Predicted throughput of the Boombox instrument, with and without the telescope contribution. The sharp features arise from the expected throughput of the binary gratings. {Right:} Boombox fiber layout in the Via focal plane. The 36 fibers are arranged in a spoke design, allowing for pointing flexibility. This sample field has 11 ``active'' transients (orange crosses) from a simulated LSST survey. With minor field re-centering and rotation, Boombox can observe 5 of these (red stars) simultaneously. Static transients (blue stars) are the host galaxies of previously active transients and will be used for redshift measurements and population studies---several are expected to fall into the patrol radii of Boombox positioners.  Ancillary science targets, such as metal-poor stars and extragalactic dwarf galaxies, will fill the remaining Boombox fibers.}
    \label{fig:bb_fibers}
    \label{fig:bb_snr_thru}
\end{figure*}

Boombox is a companion instrument to Viaspec for low-resolution spectroscopy, accepting 36 fibers from the focal plane. 
The 36 Boombox fibers are arranged throughout the Via focal plane in a spoke design  (Figure \ref{fig:bb_fibers}; right panel). 
This design has been optimized for targeting flexibility allowing modest field center and instrument rotation offsets. 
Along each spoke, pairs of targets separated by $\approx 20^{\prime\prime}$ to the field diameter of $1^\circ$ are accessed by field center translations. 
The spoke symmetry allows selection of any target with $\leq$$30^\circ$ field rotation. 
A subset of the Boombox fibers will be used as sky fibers, leaving the remainder for ancillary science targets.

\begin{figure}[tbp]
    \centering
    \begin{minipage}[l]{0.5\textwidth}
        \includegraphics[width=\textwidth]{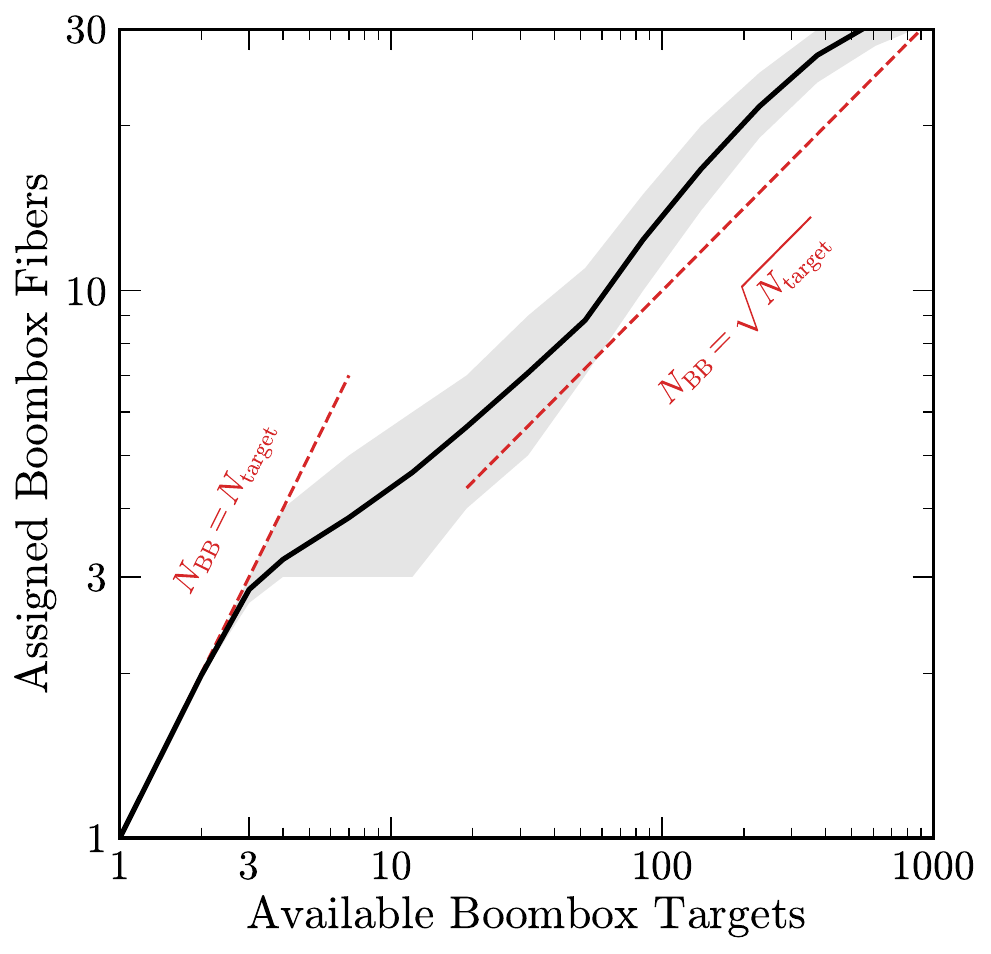}
    \end{minipage}
    \begin{minipage}[c]{0.45\textwidth}
        \caption{Number of objects that are assigned Boombox fibers, as a function of the total number of available Boombox targets. This is from a full simulation of the Boombox focal plane, including ``aiming'' the center and rotation of the field.
        The black line (shaded band) indicates the mean ($1\sigma$ distribution) from $1000$ random realizations. 
        The three highest-priority targets are guaranteed to be observed, and more than 10 fibers are assigned for target classes with more than 100 available targets.}
        \label{fig:bb_assign}
    \end{minipage}
\end{figure}

Figure~\ref{fig:bb_assign} shows a simulation of Boombox assignment for targets that are uniformly distributed across the FoV. 
The assignment algorithm is run on $1000$ random target distributions, with the field center and rotation being optimized to place the $10$ highest-priority targets as close to Boombox patrol regions as possible. 
Boombox is guaranteed to observe any $3$ high-priority targets by translating and rotating the field. 
For additional lower-priority targets, the importance of field aiming diminishes, and assignment relies on a target fortuitously landing within the patrol region of a Boombox fiber positioner. 
Any target class with $\gtrsim 100$ candidates per Via FoV can expect to receive $\gtrsim 10$ Boombox fibers.

\begin{figure}[tbp]
    \centering
    \includegraphics[width=0.9\textwidth]{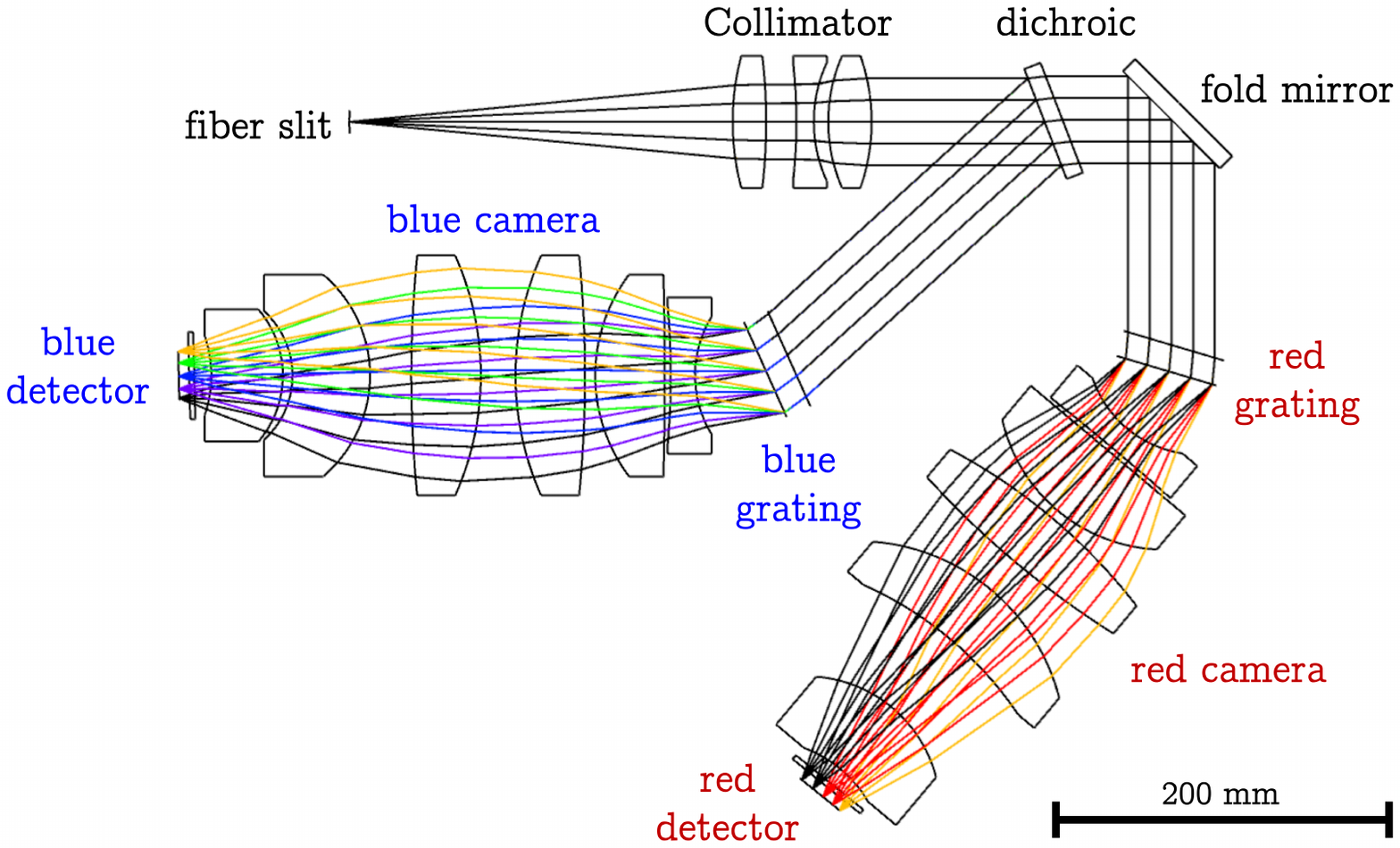}
    \caption{Zemax lens drawing of the Boombox optics. This drawing shows a top-down view of the spectrograph optical prescription. The fiber slit is defined perpendicular to the page. Only a single on-axis field point is drawn and the ray colors differentiate between wavelengths of light.}
    \label{fig:bb_optics}
\end{figure}

\begin{figure}[tbp]
    \centering
    \includegraphics[width=0.8\textwidth]{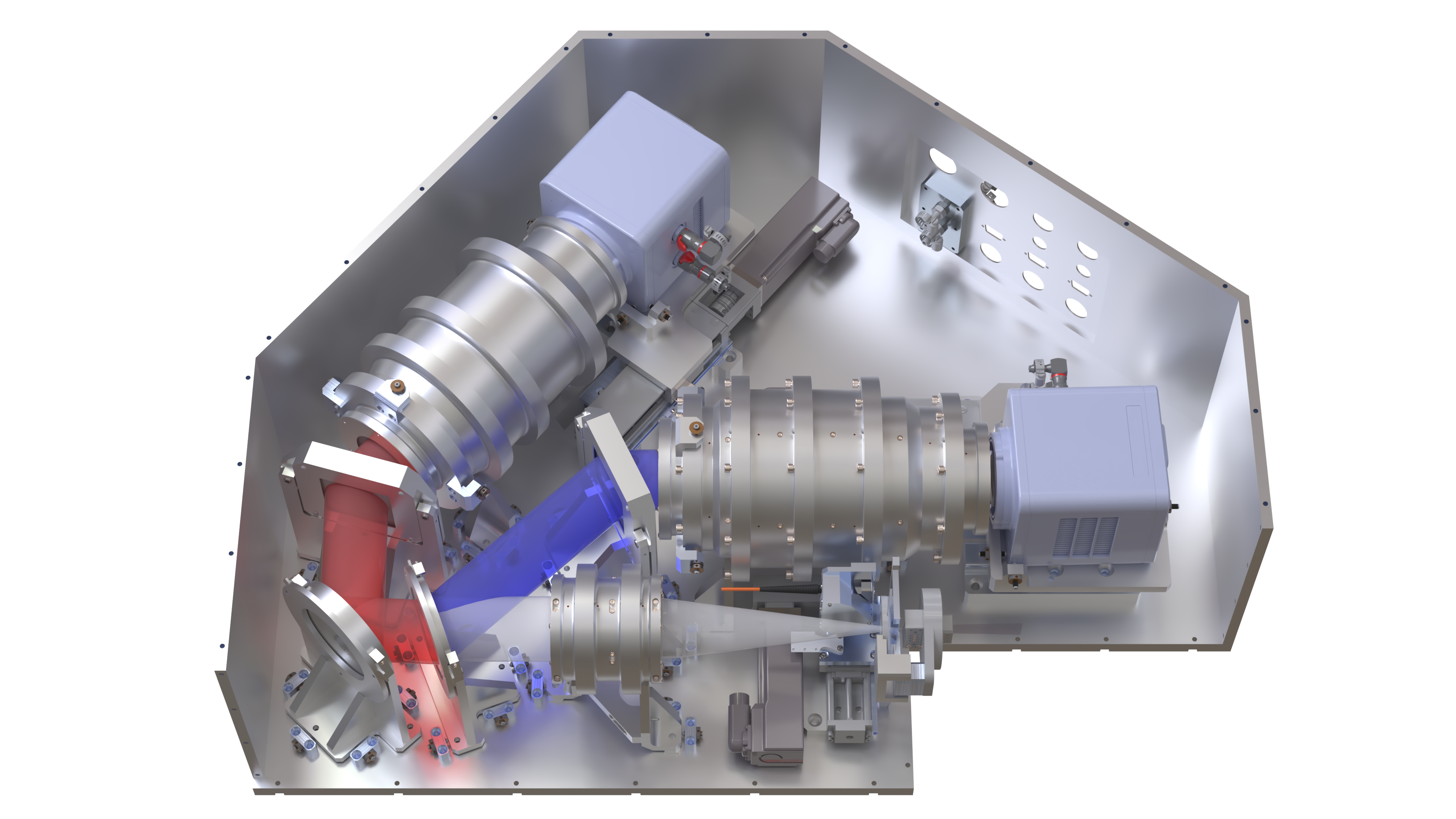}
    \caption{The Boombox mechanical design. Light enters through the fiber slit (bottom right) and is collimated (bottom middle) before being split into red and blue channels by a dichroic. In both channels the light is dispersed by a grating and then re-imaged through a custom camera onto the detectors (light gray). The camera and collimator assemblies are made up of a stack of precision lens bezels for fine alignment of the lenses. The first lens of each camera is adjustable during laboratory alignment. Other details shown include focus stages for each camera and a backlighting mechanism that can be deployed in front of the fiber slit, providing a shutter, flat field lamp, and back-illumination to the fibers.  }
    \label{fig:bb_cad}
\end{figure}

Boombox fibers are routed from the focal plane through the same fiber chain as the Viaspec fibers, branching off at the end of the fiber chain and routing to the Boombox spectrograph. 
The Boombox spectrograph consists of a two-channel optical system: a blue channel (3600 to 5900 \AA) and a red channel (5700 to 10100 \AA), as shown in Figure \ref{fig:bb_optics}. Figure~\ref{fig:bb_cad} shows the mechanical design, including the lens-barrel assemblies, focus stages, detectors, and calibration/back-illumination mechanism. The light from the 36 fibers is collimated by a refractive triplet and split into the two wavelength channels with a dichroic mirror. Light is dispersed with red and blue optimized  binary transmission gratings manufactured by Plymouth Grating Laboratory. The dispersed light is imaged onto the CCD sensors with custom $f/1.5$ refractive cameras. The camera lenses have all spherical surfaces to reduce cost.  

Both Boombox channels use the Teledyne LANSIS 261 CCD camera. The LANSIS 261 is a fully integrated camera package with readout electronics and cooling, built around the e2v CCD261-04 back-illuminated deep-depleted CCD. The red detector quantum efficiency is higher than for typical back-illuminated CCDs. The detector is composed of an array of $2048 \times 256$ pixels with 15 $\mu$m pitch for an active area of $30.72 \times 3.96$~mm. For the simulations below (see \S \ref{sec:etc}), we use an rms read noise of 5~e$^-$ and a conservative dark current of 18~e$^-$ hr$^{-1}$ pixel$^{-1}$.

\begin{figure}[tbp]
    \centering
    \includegraphics[width=\textwidth]{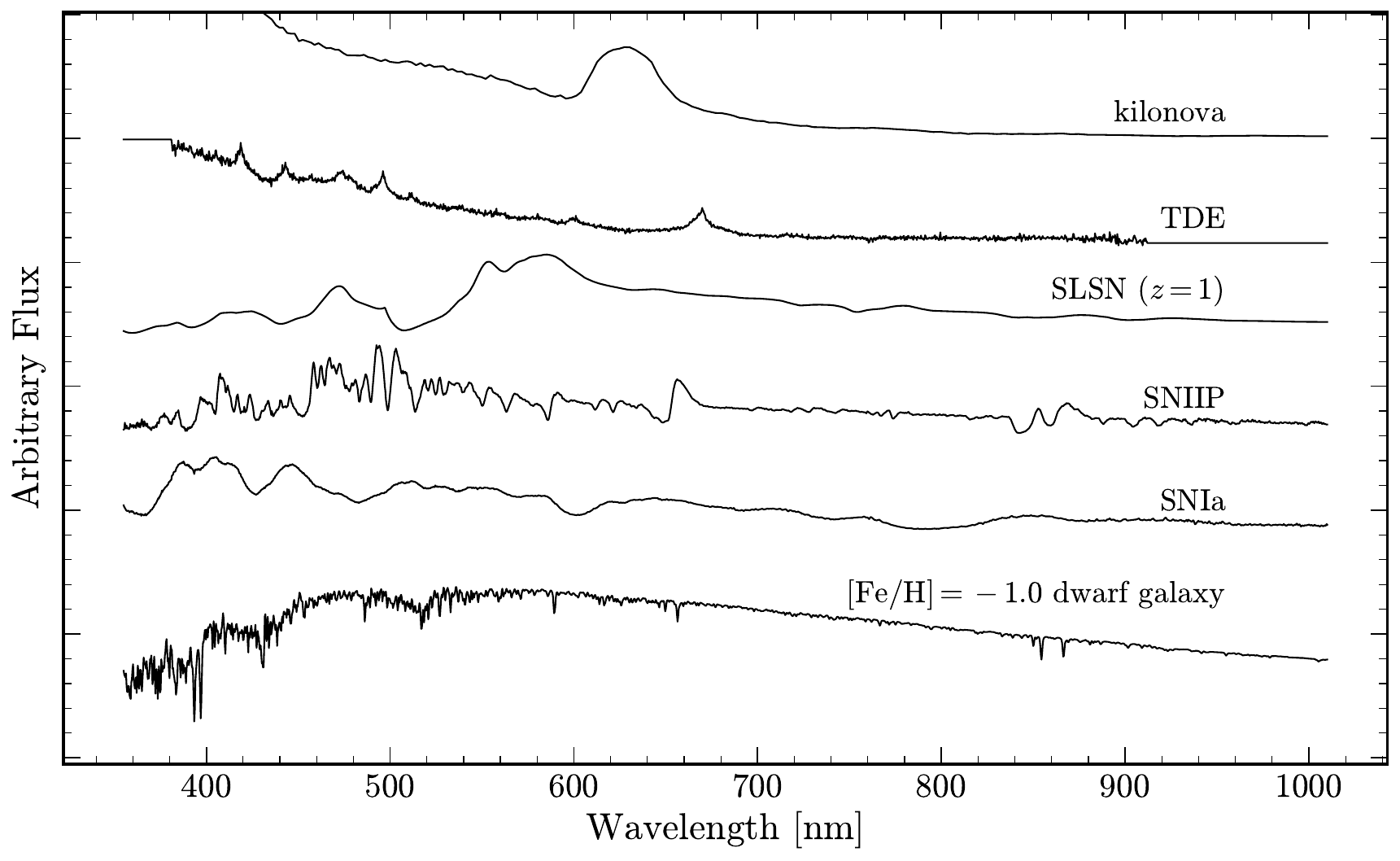}
    \caption{Example spectra demonstrating the Boombox wavelength range and science cases, including kilonovae, tidal disruption events, supernovae, and dwarf galaxies. The supernovae are examples from the PLAsTiCC database \citep{kessler2019models}.  }
    \label{fig:spec_showcase_bb}
\end{figure}

The total system throughput, with and without the telescope, is shown in the left panel of Figure \ref{fig:bb_snr_thru}.  The estimated total instrument throughput peaks around 40$\%$ in the blue channel and 43$\%$ in the red channel. In dark time at the Magellan telescope, this leads to an estimated limiting magnitude in one hour of G=23.5 for SNR $=5$ per resolution element (assuming 1\arcsec{} seeing in dark time). Figure \ref{fig:spec_showcase_bb} shows example simulated spectra for the Boombox wavelength range and resolution for a variety of target classes. Section~\ref{sec:etc} simulates the Boombox end-to-end spectroscopic performance. Figure~\ref{fig:bb_noisy} shows simulated Boombox spectra for a representative Type~IIP supernova, while Figure~\ref{fig:etc_snr} summarizes the corresponding signal-to-noise performance as a function of target brightness and observing conditions.

\vspace{1cm}

\subsection{The Fiber Run}\label{sec:fiber}

\begin{figure}[tbp]
    \centering
    \includegraphics[width=\textwidth]{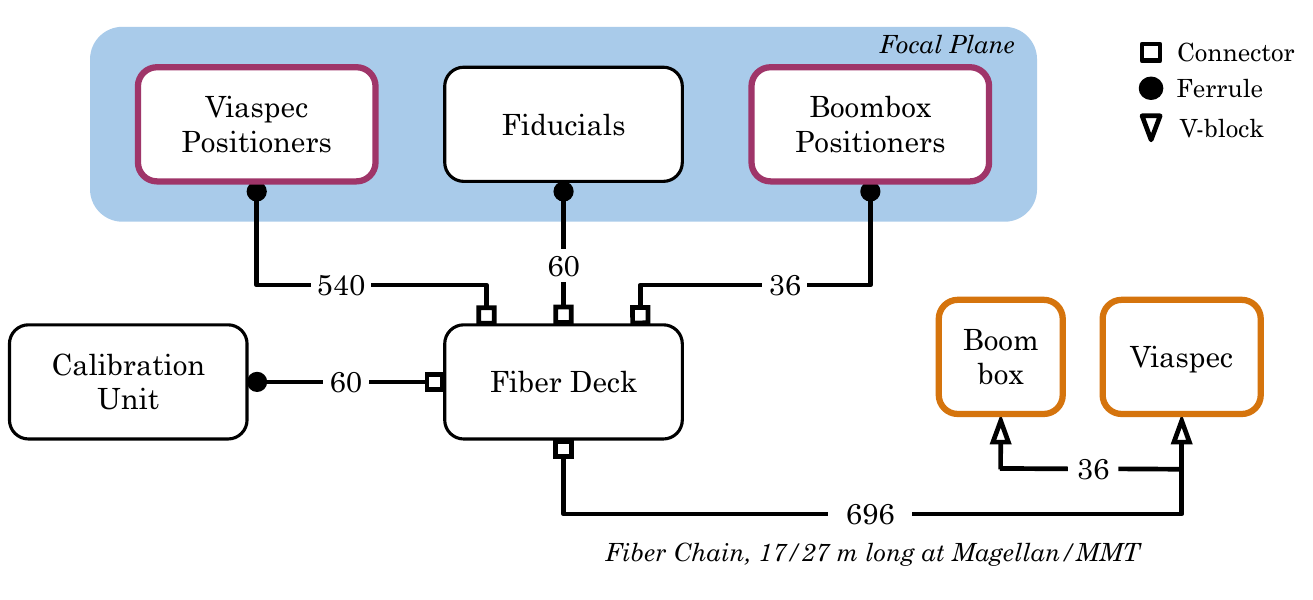}
    \caption{Schematic layout of fiber optic cables connecting the focal plane system to Viaspec and Boombox.  The fiducials are fixed position fibers used to measure the absolute positions of all fibers, and are fed to Viaspec, where they are also used as sky fibers during science observations.  The calibration unit contains Pen-Ray lamps that are also fed to Viaspec.  The number of fibers between each connection is shown along each line. The different fiber terminations utilized are indicated in the legend.}
    \label{fig:fibers}
\end{figure}

Via uses fiber optic cables to relay light from the focal plane to the spectrographs. Figure \ref{fig:fibers} is a schematic layout of the fiber-based connections between various components of the instrument.  A fiber consists of a silica core, a cladding, and a polyimide buffer. Via uses a Polymicro FBP series 200 $\mu$m core, multi-mode, step-index optical fiber with low attenuation over a broad optical spectrum.  Incident light propagates through the core by total internal reflection.  Fibers azimuthally scramble light efficiently, and less efficiently radially across the fiber core.  Due to bending, pinching, stresses, or imperfect end finishes of the fiber, light may exit the fiber at a faster focal ratio. This effect, known as focal ratio degradation (FRD), will reduce the throughput of Viaspec, since light faster than $f/5.28$ is obstructed at the grating. 

The fiber run has two parts: a short actuator ``pigtail'' at the fiber positioner, and a long fiber run to the spectrograph.
These parts are connected with mated ferrule connectors (FCs). 
The 0.95m fiber pigtail connects robotic actuators and fiducials at the focal plane to the fiber deck (see Figure \ref{fig:fps_cad}). 
The fiber pigtail is encased in 23 AWG Teflon tubing, protecting the fiber from the motion of the actuator. 
The pigtail terminates in a 15 mm long LC ferrule at the actuator end (the tip of the actuator's second arm), and a standard FC connector at the fiber deck end. 

The main fiber run is $17$ meters long at Magellan and $27$ meters at the MMT.  The MMT run is longer because the spectrographs are permanently located in an insulated room, whereas at Magellan the spectrographs are mounted on the azimuth disk during operation. 
The fibers in the main run are protected with 20 AWG Teflon tubing.  Multiple thermal breaks allow for the differential expansion of the Teflon tube and the silica fibers.  The fibers in the main run are further protected by woven Nylon tubing in groups of 11.  The main run fibers are protected in an energy chain that controls the bend radius of the fibers and constrains the bend to one dimension. This design is adapted from the successful Hectospec fiber run at the MMT.
At the spectrograph end, the fibers are mounted in V-grooves that form the spectrograph slit. 

Survey instruments like DESI adopt fusion splicing to connect their fiber positioners to their spectrographs, which minimizes connection loss at the cost of a complicated and lengthy installation process \citep{Poppett2024}. 
Due to their repeatable nature, fiber connectors are much more efficient from an installation perspective, but their throughput losses have historically been $> 10\%$ \citep{Farr2022}.
The Via instrument uses unionized FC connectors to simplify the installation procedure, and to facilitate the quick replacement of faulty/damaged actuators during instrument servicing runs. 
To minimize connector throughput loss, we use tightly toleranced commercial connectors and place strict requirements on the polishing quality of the fiber ends. 
Our lab testing indicates power losses of $< 5\%$ in the union between the pigtail and long fiber run.  
The total throughput of the fiber system---including FRD, transmission loss, and connector loss---is budgeted to be $\approx 80\%$. 

\subsection{Data Simulator}\label{sec:etc}

\begin{figure*}[t!]
    \centering    \includegraphics[width=1.\textwidth]{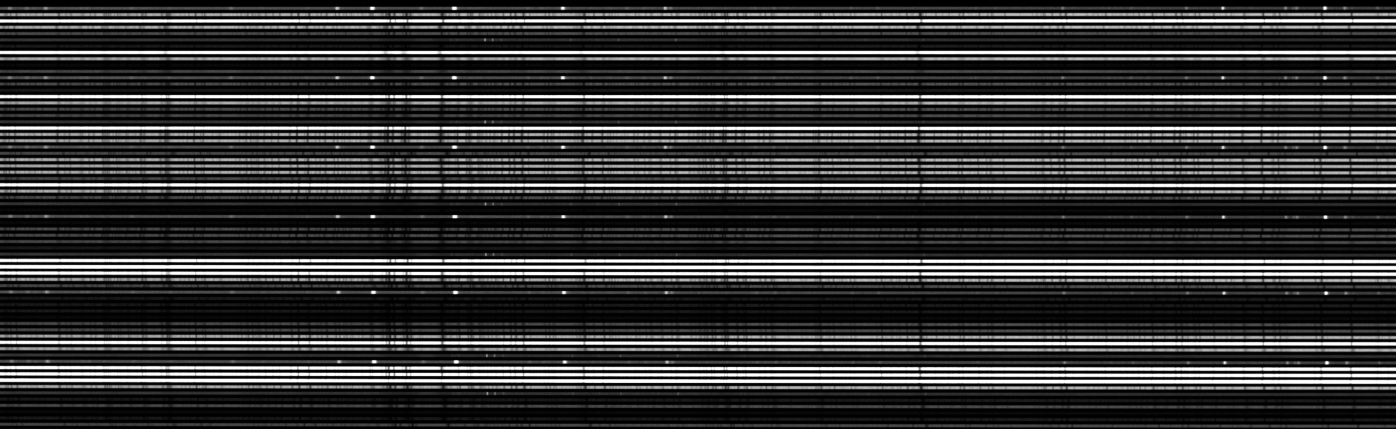} 
    \caption{A small portion of a simulated raw 2D Viaspec image from the data simulator.  The image is zoomed in on 66 fibers near the \ionn{Mg}{i} triplet (located near the center of the image).  Six calibration fibers are clearly visible as emission line spectra.}
    \label{fig:2d}
\end{figure*}

We have developed a full simulation of the telescope, Viaspec, and Boombox to realistically predict raw 2D detector images (Figure \ref{fig:2d}) and extracted 1D spectra and SNR for varying source brightness and observing conditions.  Versions will be available for both Viaspec and Boombox.  The 1D simulator will be used in a dynamic exposure-time calculator (ETC) at the telescope to increase survey efficiency. 

As input stellar models, we use high-resolution theoretical spectra from the C3K spectral grid \citep{Cargile2020}. 
For galaxy spectra we adopt high-resolution spectra constructed from the same stellar spectral grid and the FSPS stellar population synthesis program \citep{Conroy09a}.  
To simulate Boombox data, we use template SN spectra from the Weizmann Interactive Supernova Data Repository (WISeREP; \citealt{Yaron2012}).

We adopt wavelength-dependent transmissions for each instrument component (see Figures \ref{fig:tp_summary} and \ref{fig:bb_snr_thru}).  Telescope mirror reflectivities are taken as average measured values for the MMT and Magellan---in practice, they vary by up to $\approx 10\%$ over the cleaning and aluminization cycle.  Atmospheric extinction is dependent on both wavelength and airmass; we adopt extinction measurements from the CTIO observatory and scale them to the requested airmass. 
Perhaps the most significant---and variable---throughput term is the aperture loss at the fiber face due to atmospheric seeing. 
In the Via data simulator, we model the combined 2D PSF of the telescope and atmosphere, and also allow for offsets between the fiber aperture and PSF center. 
Although these losses will be operationally minimized, an accurate offset model can (for example) enable observations of stars much brighter than the spectrograph's dynamic range would normally accommodate.

\begin{figure*}[t!]
    \centering \hspace*{-0.43cm}\includegraphics[width=0.905\textwidth]{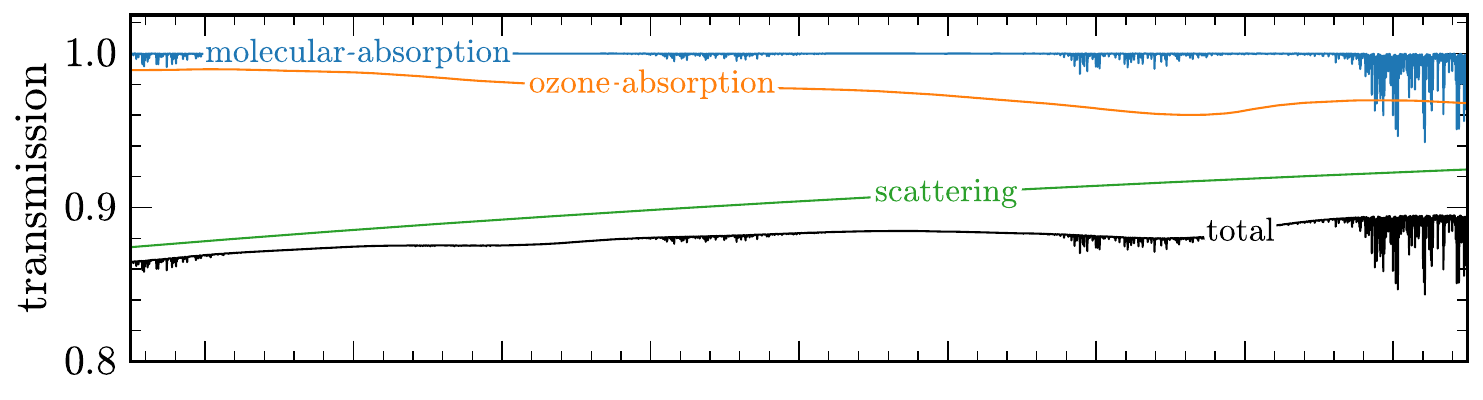}
    \includegraphics[width=\textwidth]{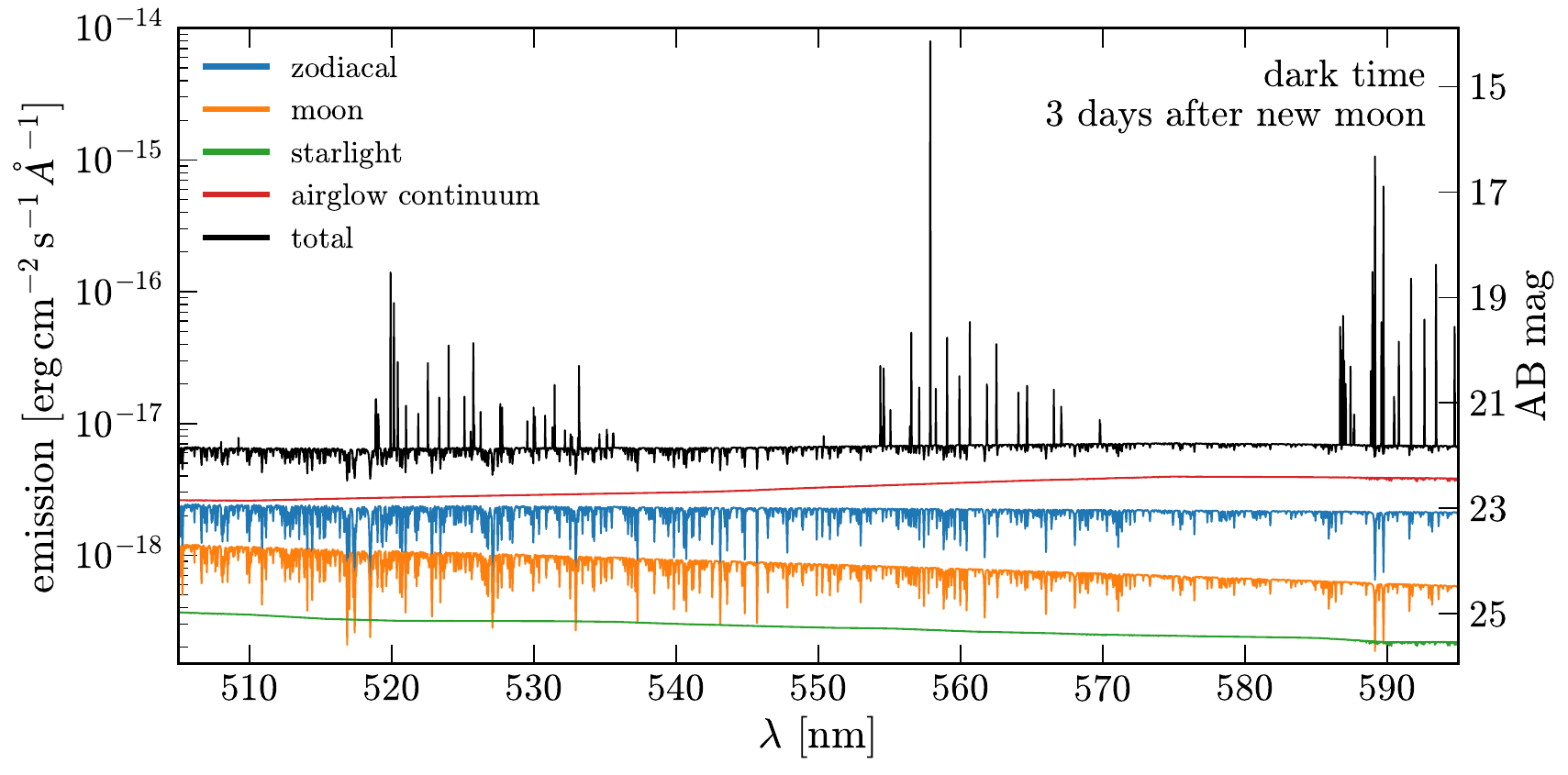}
    \caption{Sky model from the ESO SkyCalc program, in dark time (3 days after new moon), for the Viaspec wavelength range and resolution.   The top panel shows the sky transmission from molecular absorption lines (telluric features).  Including all absorption and scattering components, the total sky transmission in the Viaspec wavelength range is $\approx 88\%$ at zenith.  The bottom panel shows the various sky emission components in flux units.  The right axis indicates the equivalent AB magnitude.  Ecliptic coordinates of $(135 ^\circ, 45 ^\circ)$ are assumed when calculating the zodiacal emission.}
    \label{fig:sky}
  \end{figure*}

\begin{figure*}[t!]
    \includegraphics[width=\textwidth]{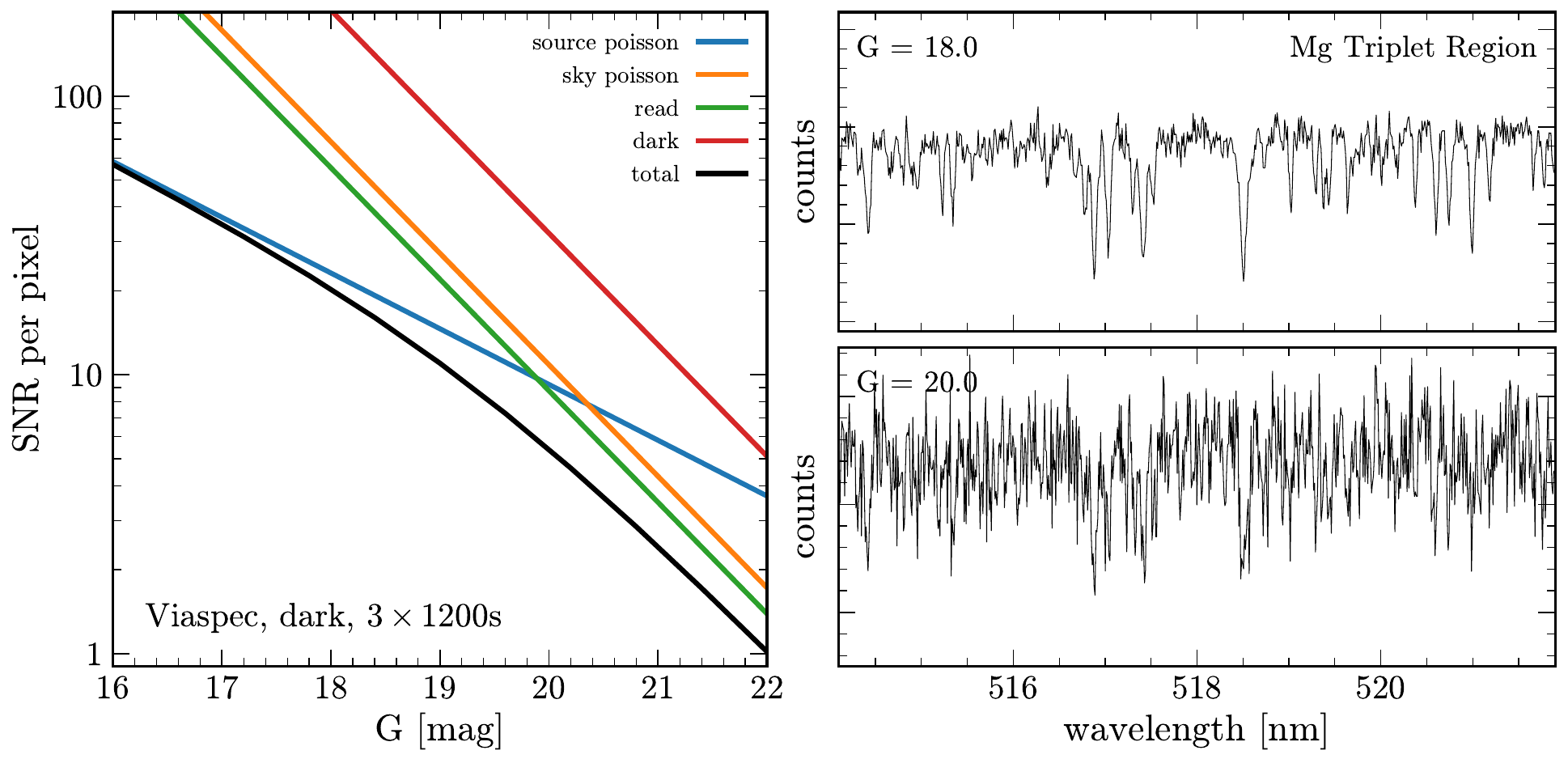}
    \caption{{Left:} Fractional noise contribution to Viaspec spectra from different sources, assuming dark time for the fiducial 1~hr survey configuration. 
    Since the instrument is read-limited in dark time for 20-minute exposures, dark survey programs may use longer individual exposures, improving SNR at the faint end. 
    {Right}: Simulated Viaspec spectra for a metal-poor K giant ($T_{\mathrm{eff}} = 4500$K, $\log{g} = 2.0$, $\text{[Fe/H]} = -1.0$) during conditions of $1.0\arcsec$ seeing for a 1-hr exposure.
    }
    \label{fig:etc}
    \label{fig:etc_noise}
\end{figure*}

\begin{figure*}[t!]
\includegraphics[width=\textwidth]{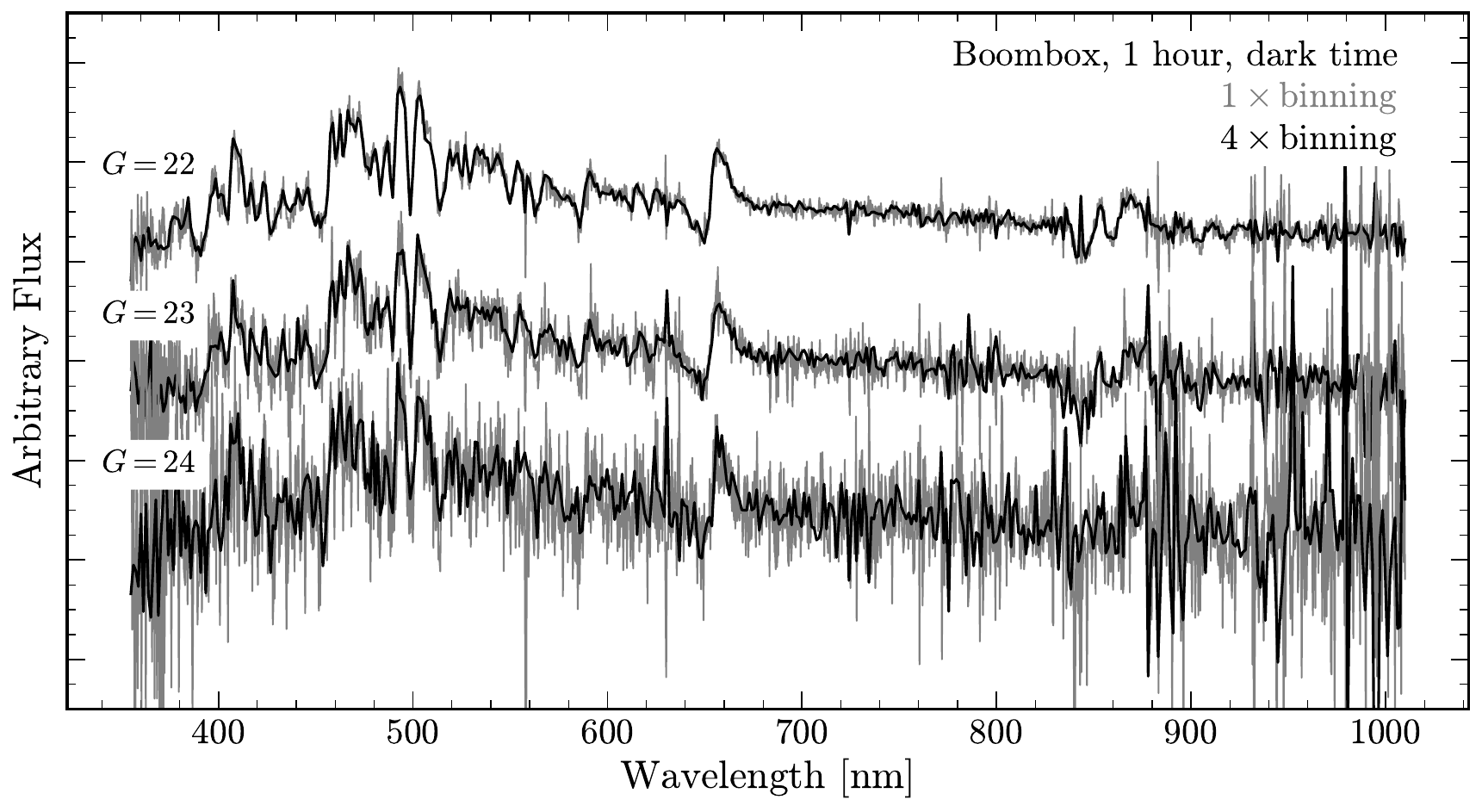}  
    \caption{Simulated Boombox spectra for a Type IIP supernova observed for 1 hr in nominal dark conditions. The $G$ magnitude of each spectrum is indicated, and spectra are shown with $1\times$ and $4\times$ spectral binning.}
    \label{fig:bb_noisy}
\end{figure*}

The optical model for each spectrograph is used to construct PSFs for fibers along the slit, allowing realistic simulation of the 2D fiber images on the detector.  The slit can be masked by any amount to reduce the effective fiber output below $1.2\arcsec$---the data simulator self-consistently calculates the resulting increase in resolution and reduction in throughput. As discussed in the preceding section, we adopt a $120\,\micron$ mask width (for a $200\,\micron$ fiber diameter) in Viaspec.  The Boombox fibers are unmasked. 

The sky flux is modeled with the ESO SkyCalc utility, which includes contributions from sky emission, zodiacal light, and lunation \citep{Noll12}.  
Figure~\ref{fig:sky} illustrates the various contributions to the sky model. 
In this document, {\em dark time} refers to 15\% fractional lunar illumination (FLI; $\approx 3$ nights after the new moon), and {\em bright time} refers to 100\% FLI, unless otherwise noted.  
When calculating the variance of the sky model, we include the Poisson noise of the sky flux and the read and dark noise contributions to the measured sky in each sky fiber.  
The lunar component of the sky flux is scaled to any lunar phase using the \cite{Krisciunas1991} model.
For Viaspec, the read and dark noise components assume an e2v MB2 CCD operating at $\approx -100^\circ$\,C.  We use an RMS read noise of $3$~e$^-$ modified as appropriate for spectral and spatial binning, and a dark current of $1$~e$^-$\,hr$^{-1}$ per pixel.  
For Boombox we use a read noise of $5$~e$^-$ and a dark current is $18$~e$^-$\,hr$^{-1}$ per pixel. Boombox's much lower resolution means that these are not dominant terms in the error budget.

\begin{figure*}[t!]
    \centering
    \includegraphics[height=0.47\textwidth]{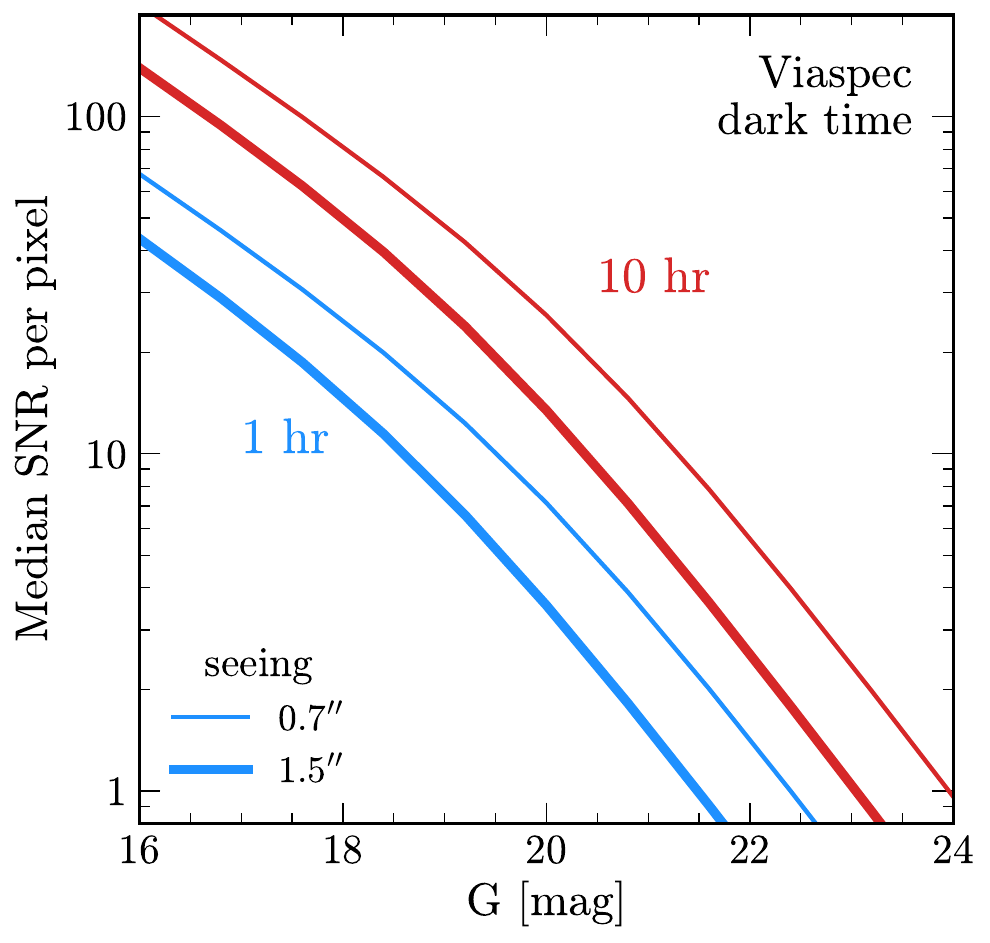}
    \includegraphics[height=0.47\textwidth]{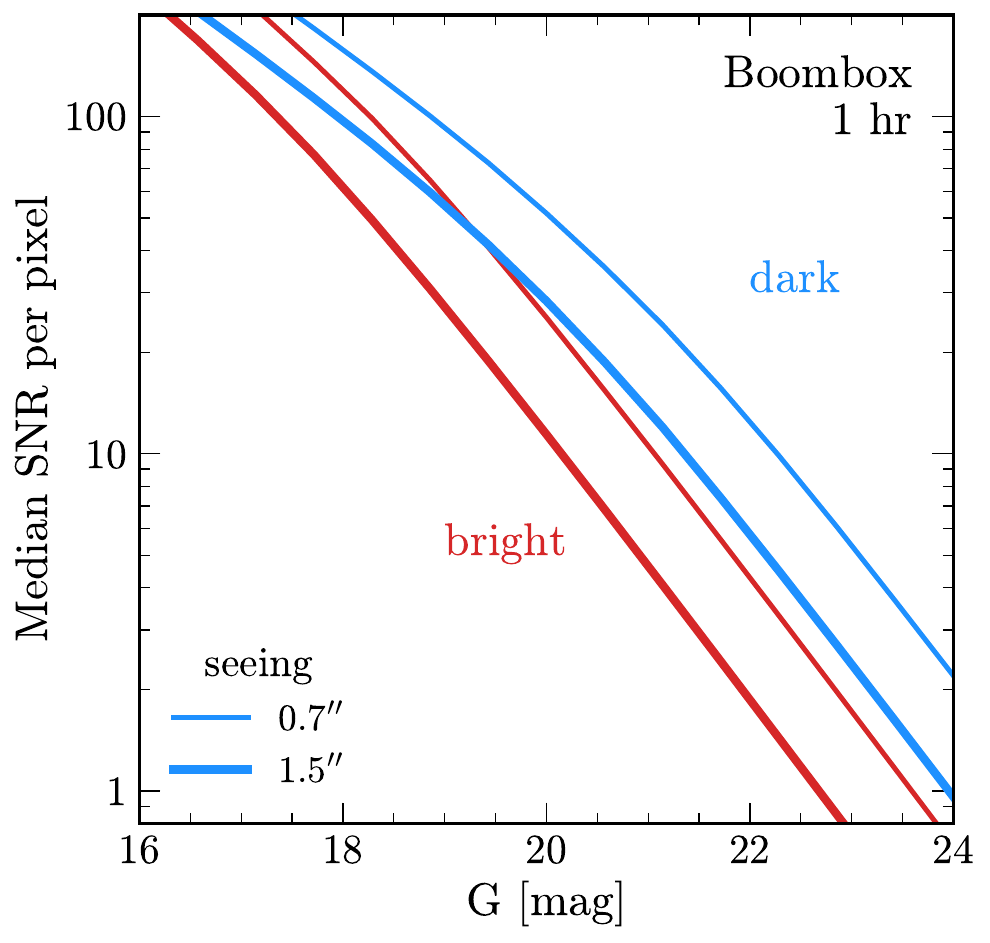}
    \caption{{Left:} Median Viaspec SNR versus $G$ magnitude for a variety of seeing conditions in dark time (0\% lunar illumination), at the survey depth of 1 hr and the deep-drilling depth of 10 hr.
   {Right:} Analogous SNR curves for Boombox with 1-hr exposures, shown for dark and bright time.   }
    \label{fig:etc_snr}
\end{figure*}

The left panel of Figure~\ref{fig:etc_noise} illustrates the contributions to the Viaspec noise model in dark time. 
With individual exposures of 20~min in dark time, CCD read noise limits the SNR for stars fainter than $G \approx 20$. 
This motivates longer individual exposures, potentially at the cost of cosmic ray rejection.
In bright time, sky brightness dominates the SNR.
Predicted 1\,hr Viaspec spectra for a typical metal-poor K giant ($T_{\mathrm{eff}} = 4500$ K, $\log{g} = 2.0$, $\text{[Fe/H]} = -1.0$) in $1.0\arcsec$ seeing are shown in the right panel of Figure \ref{fig:etc}, zoomed in to the \ion{Mg}{1} triplet region.

Figure~\ref{fig:bb_noisy} shows corresponding 1\,hr Boombox spectra of a type IIP SN template from \cite{Yaron2012}, at three different apparent magnitudes.
For faint supernova classification, Boombox can be binned at the readout level to boost the SNR at the cost of resolution, as shown in Figure~\ref{fig:bb_noisy}. 
Figure \ref{fig:etc_snr} shows trends of median signal-to-noise ratio as a function of $G$-band magnitude for Viaspec and Boombox. 
Note that these are SNR values per-pixel, and the SNR per resolution element is $\approx \sqrt3$ better than the values shown here. 

\clearpage

\section{Data Reduction and Analysis}
\label{data}

\subsection{Data Reduction Strategy}

Data will be processed with a custom pipeline that supports real-time diagnostics, reproducible calibrations, and seamless integration with the stellar parameter pipeline (see $\S$\ref{sec:minesweeper}).
Raw CCD frames will be calibrated to remove detector systematics to prepare for spectral extraction. Calibration first corrects electronic crosstalk between readout amplifiers and then removes amplifier-specific bias drifts using the overscan regions. Cosmic ray removal uses standard algorithms \citep[e.g., L.A. Cosmic,][]{vandokkum:2001}. A master bias frame, constructed from nightly stacks of zero-exposure images is then subtracted to remove fixed-pattern noise. Flat fields correct for pixel-to-pixel sensitivity variations and illumination inhomogeneities.  A light source with diffuser on the Via back illumination system will be used for flat-field illumination.  The correction is applied to both science and calibration exposures.

Viaspec’s line-spread function (LSF) and relative fiber throughput are central to the calibration pipeline. This will be achieved through two-dimensional modeling of twilight sky exposures, which will provide high signal-to-noise illumination measurements across the full focal plane. The LSF will be parameterized as a continuous function of both wavelength and spatial position, enabling accurate forward modeling during spectral extraction. Simultaneously, a smooth throughput model will be fit to the normalized twilight fluxes, capturing fiber-to-fiber transmission differences.  The LSF model will also provide an accurate wavelength solution for the science data.  Absolute throughput can be recovered by observing \gaia-calibrated stars.
The final stage of the pipeline will involve extracting one-dimensional spectra from the calibrated two-dimensional CCD images. Two complementary extraction strategies will be implemented, each optimized for different scientific and operational goals: flat-relative optimal extraction and spectro-perfectionism.

For quicklook (at the mountain), and routine extractions, a flat-relative optimal extraction algorithm will be used \citep{zechmeister:2014}. This method will leverage spatial profiles derived from normalized flat fields to perform a variance-weighted summation along the cross-dispersion direction.  This approach maximizes signal-to-noise while preserving photometric accuracy. Extraction weights will incorporate both the spatial illumination profile of each fiber and the estimated pixel variance, yielding statistically optimal results under Gaussian noise assumptions.

For applications requiring the highest possible fidelity in both wavelength and flux, a spectro-perfectionism algorithm will be employed \citep{bolton:2010}. This method will forward-model the two-dimensional science data by convolving a trial one-dimensional spectrum with the empirically determined LSF from the twilight data, and fitting this model directly to the pixel values. This approach will achieve near-lossless extraction by accounting for LSF asymmetries, pixel-level correlations, and overlapping fiber profiles. Though computationally intensive, spectro-perfectionism provides unmatched accuracy for precision spectroscopic analysis.

Via fibers are well-separated spatially at the detector allowing us to extract each spectrum independently of its neighbors, simpler and more robust than extraction in instruments with closely packed fibers.  Care will be taken during fiber assignment to ensure that very bright and very faint targets are not assigned to adjacent fibers at the slit, further limiting the influence of cross-talk on faint spectra.  Although this document has assumed $1\times$ spatial binning to be conservative, in practice it is likely that we will be able to bin two pixels in the spatial direction, leading to a $\approx \sqrt{2}$ reduction in the read noise, the dominant noise source in dark time (see Figure~\ref{fig:etc_noise}).

\subsection{Wavelength Calibration}
\label{calib}

The Viaspec velocity resolution at the detector is $\sim 5.5\kms$ pix$^{-1}$ (this is the pixel sampling, not the width of the LSF).  To achieve $100\ms$ radial velocity accuracy the wavelength solution must be stable and accurate at the 2$\times$$10^{-2}$ pix level (RMS).  For comparison, state-of-the-art spectrographs deployed for exoplanet detections require wavelength solution stability at the level of $10^{-3}$--$10^{-4}$ pix.  Achieving this high level of stability typically requires vacuum operation, very high thermal stability, and very stable spectrograph benches.  We do not require these techniques for Viaspec's more modest stability target.

Via will use twilight spectra to define the absolute wavelength scale.  Solar twilight spectra are frequently used for flat-fielding, but less commonly for measurement of the wavelength scale.  The advantage of using twilight data is that the incoming light path is identical to our science sources.  We have demonstrated with MMT/Hectochelle observations of radial velocity standard stars that this procedure delivers an absolutely calibrated wavelength accuracy to better than $30\ms$, corresponding to 0.015 pix (rms) for Hectochelle. 

To monitor drifts in the wavelength calibration during the night, we have developed a secondary calibration approach.  Sixty fibers will feed light during each exposure from a continuous calibration unit located in the fiber positioner (see Figure \ref{fig:fibers}).  The calibration unit contains Argon Pen-Ray lamps feeding an integrating sphere through a variable neutral-density filter. 
The filter modulates the intensity of the calibration light to avoid saturating the Viaspec detector, based on the desired exposure time. 
Lamp light from the integrating sphere is imaged onto the sixty calibration fibers via a two-lens telecentric $f/5$ relay, mimicking the illumination from the telescope. 

This calibration procedure will be blind to fiber-to-fiber variation in the PSF throughout the night, which can induce subtle changes to the LSF.  There are various features imprinted on the data that will be used to check LSF variations throughout the night, including the bright \ionn{O}{i} sky line at 557.7 nm, the telluric absorption lines at $>590$ nm, and the weak but numerous absorption lines in the zodiacal light spectrum (see Figure~\ref{fig:sky}). 

Th-Ar arc lamps are suspended from the secondary mirror support structure and can be used as backup wavelength calibration sources if necessary.

\subsection{Sky Subtraction}

Figure \ref{fig:sky} shows the sources of sky emission over the Viaspec wavelength range at $R=15,000$ in dark time at $45^\circ$ above the ecliptic plane.  The zodiacal light is brighter by nearly one magnitude in the ecliptic, resulting in an overall brighter sky continuum by $<0.5$ mag.  Sixty fiducial fibers and unassigned actuator fibers allow us to measure the sky level during exposures.  We will use the MADGICS pipeline  \citep{Saydjari2023} to build an empirical time and spatially dependent sky model that is superior to both standard non-local sky subtraction methods and PCA-based methods.  

Objects to the  nominal survey depth of $G=21$ are comfortably brighter than the night sky continuum in dark time, and there are only a handful of bright sky lines in the $500$--$600$ nm wavelength range.  The deepest proposed observations both for Viaspec and Boombox, reaching $G\approx23-24$, will be $\approx10$\% of the sky continuum.  Continuum spectroscopy with fibers at $\approx1$\% of the sky has been demonstrated in various contexts \citep[e.g.,][]{Bundy22}, and so we have confidence that our stringent calibration program will enable continuum spectroscopy at the faint end of the survey.   Careful measurements of fiber-to-fiber throughput variations are key to avoid systematic biases in the inferred sky level.

\subsection{Flux Calibration}

Owing to its wide wavelength coverage, flux calibration of the Boombox spectra would be scientifically advantageous. \gaia\ data has made flux calibration relatively straightforward; as of DR3, the \gaia\ BP/RP low-resolution spectroscopic data provide flux-calibrated spectra across the entire sky to $G\approx17$.  The BP/RP spectra are accurate to $\approx2$\% (0.02 mag) across the Boombox wavelength range of $360-1010$ nm \citep{Montegriffo23}.  By observing several $15<G<17$ stars in each field we will be able to flux calibrate the Boombox spectrograph.  Small differences in throughput between the reference and target fibers will be eliminated by normalizing the spectra to the observed magnitudes of the targets.  Although less scientifically important, the same procedure will be adopted to flux calibrate the Viaspec spectra.

\subsection{Stellar Parameter Pipeline}\label{sec:minesweeper}

Stellar parameters, including radial velocities, spectrophotometric distances, and abundances ([Fe/H], [$\alpha$/Fe], and [Na/Fe]) will be measured using the \MS\, program \citep{Cargile2020}.  \MS\, combines spectral libraries and stellar isochrones to simultaneously fit for stellar parameters along with distance and reddening.  Custom spectral libraries are computed using the \texttt{ATLAS} and \texttt{SYNTHE} programs.  We use the \texttt{MIST} v2 stellar evolution database for isochrones \citep{Choi16, Dotter16}.  \MS\, adopts a Bayesian framework to fit the continuum-normalized spectrum and the broadband photometry (including Pan-STARRS, \gaia, 2MASS, {\it WISE}, SDSS, and eventually {\it Roman} and Rubin/LSST data, where available).  \gaia\ parallaxes are used as a prior.  \MS\, has been used previously to derive stellar parameters for the H3 and SDSS-V spectroscopic surveys.

\begin{figure*}[t!]
    \centering
    \includegraphics[width=\textwidth]{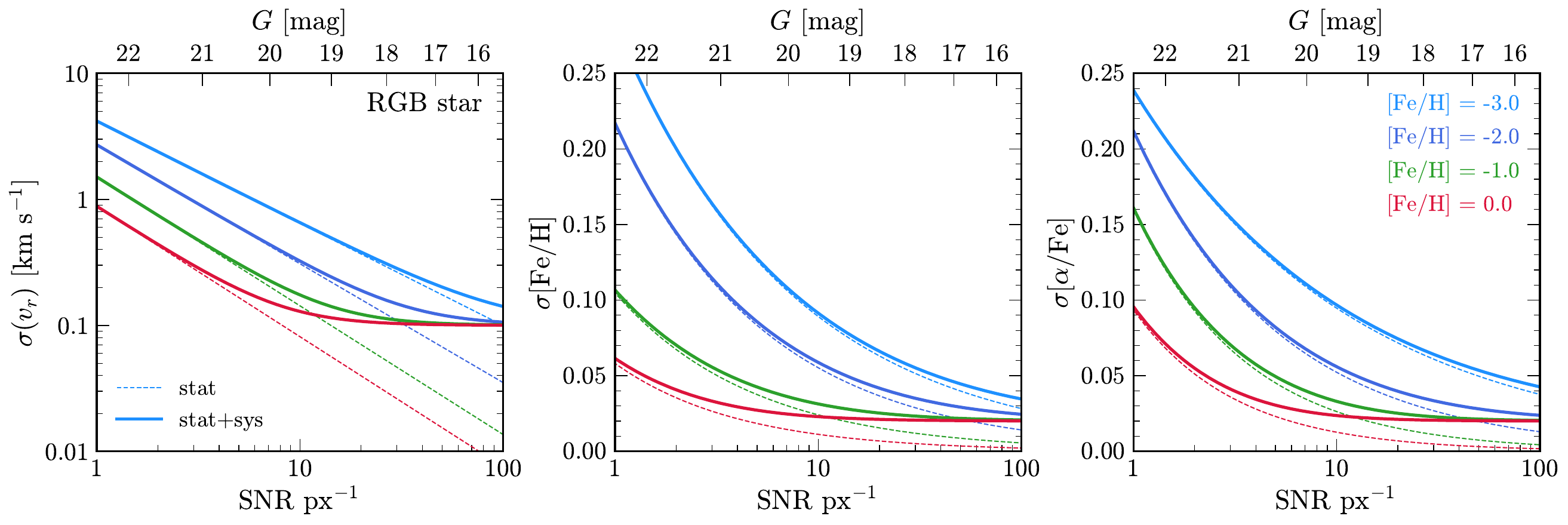}
    \includegraphics[width=\textwidth]{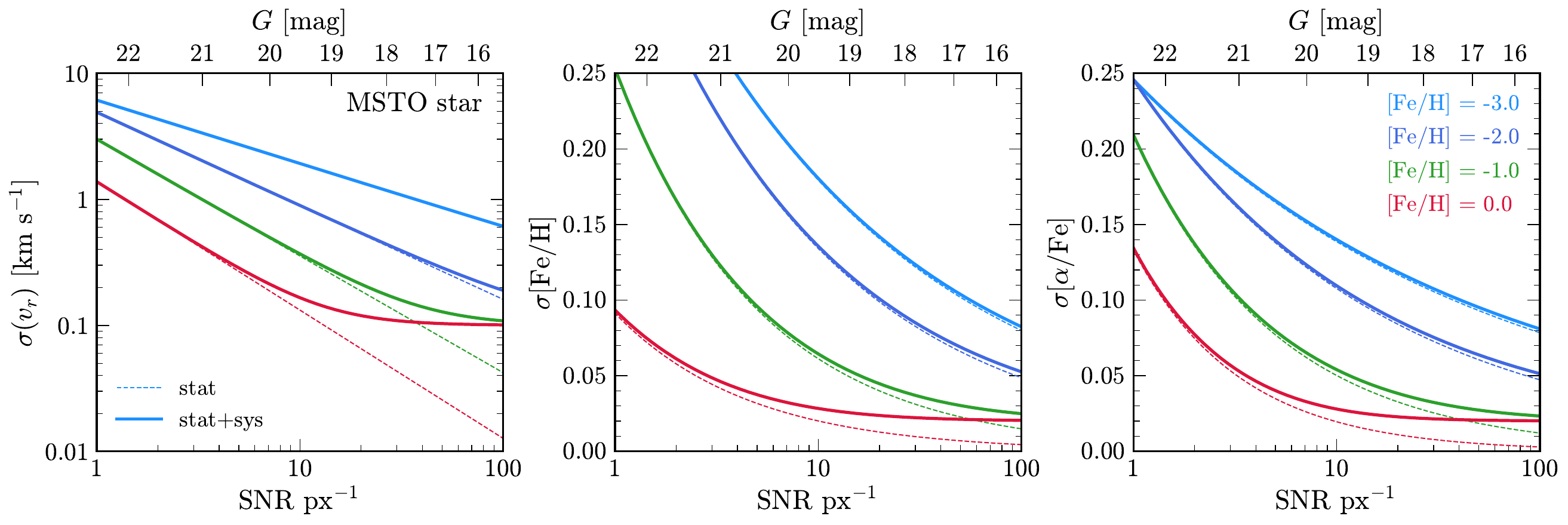}
    \caption{Predicted radial velocity (left), [Fe/H] (center), and [$\alpha$/Fe] (right) precision for mock red giant branch (RGB) and main-sequence turn-off (MSTO) stars, as a function of the SNR per pixel. 
    The top axis shows corresponding \gaia\ $G$ magnitudes for 1 effective hour of observations.
    The metallicity used when generating each mock is indicated by the line colors provided in the legend.  
    Errors are shown with and without expected systematic uncertainties added in quadrature: $100 \ms$
    for RVs (based on our instrument requirement), and 0.02 dex for abundances (based on data from the H3 Survey).}
    \label{fig:figMS}
\end{figure*}

Modeling the sodium abundance is a new extension to \MS; we will explore several options for measuring this parameter during the first year of the survey.  An accurate model for this feature is important in order to measure reliable column densities of cold gas from \ionn{Na}{i} absorption that is unassociated with the stellar photosphere.  Purely theoretical models are unlikely to produce reliable results, as the \ionn{Na}{i} doublet at 589.0/589.6nm is known to suffer from substantial NLTE effects \citep{Lind11}.  We plan to develop empirical relations between \ionn{Na}{i}-doublet equivalent width and temperature, gravity, and metallicity in order to model the line strength of this important feature.

In order to estimate stellar parameter precision vs. exposure time and magnitude, we fit mock spectra using the \MS\, program.  Mock Viaspec data are generated using the spectrum simulator described in $\S$\ref{sec:etc}.  We include photometry in our modeling from large area optical surveys with data available at the expected brightness of Viaspec stars; namely, SDSS, Pan-STARRS, \gaia, {\it WISE}, LSST, and other ancillary photometry where available.  We also include the \gaia\ parallax, which will be low SNR for many of our targets, but is still very useful to distinguish dwarf and giant stars.  We show in Figure \ref{fig:figMS} the radial velocity, [Fe/H], and [$\alpha$/Fe] precision estimated from modeling the grid of mock spectra, at a variety of input metallicities.  The top panel shows a moderate-luminosity red giant branch star and the bottom panel shows a main sequence turnoff star.  The precise temperatures and surface gravities are self-consistently calculated for the equivalent evolutionary phase at each metallicity, using \texttt{MIST} isochrones \citep{Choi16}.

\subsection{Measuring True Radial Velocities}

Deriving accurate ``true motion'' radial velocities at the level of $\sim100 \ms$ is a key science requirement for both the streams and dwarf galaxy science cases.  There are a multitude of effects that make it challenging to derive velocities at this level from the Doppler shift of absorption lines arising from the photospheres of stars \citep[e.g., see discussion in][]{Lindegren03}.  These include an accurate and stable wavelength solution, mitigation of sources of astrophysical radial velocity ``noise'' (convection, p-mode oscillations, and binarity), and offsets due to the gravitational redshift and convective blueshift. In this section we discuss our approach to mitigating these astrophysical effects.  In some applications (e.g., astrometric binaries and exoplanet host vetting), radial velocity stability---not absolute velocity accuracy---is the key and generally less demanding requirement.

\begin{figure}[t!]
\centering
\includegraphics[width=\textwidth]{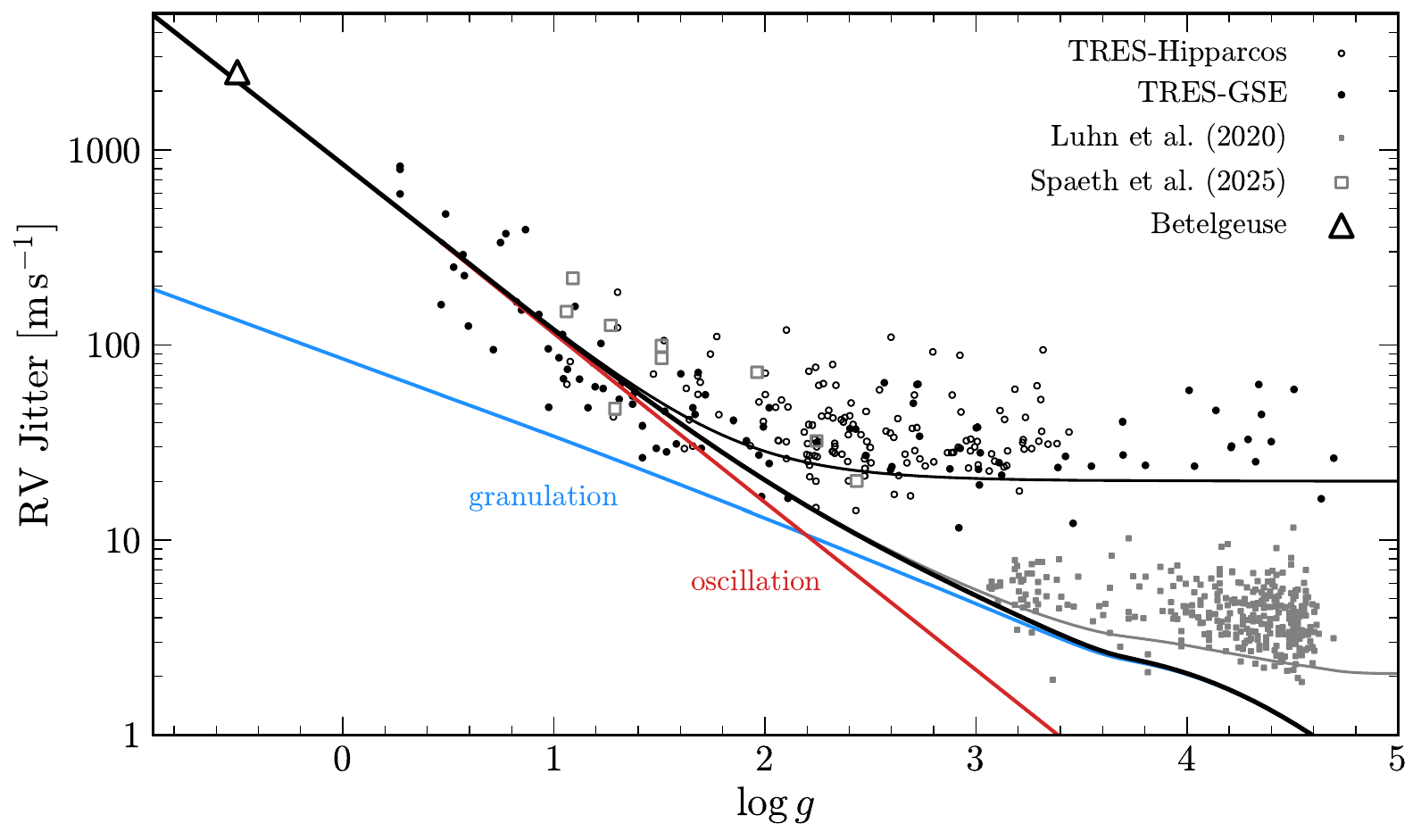}
\caption{Radial velocity jitter as a function of surface gravity for giants observed with the TRES spectrograph (D. Latham, priv. comm), giants from \cite{Spaeth2025}, and dwarfs observed with Keck/HIRES \citep{Luhn20, Luhn23}.  The TRES sample includes bright giants from the Hipparcos survey, and metal-poor giants associated with the Gaia--Sausage--Enceladus (GSE) merger. Binary orbital solutions have been fit to the TRES data and subtracted off where applicable to isolate the photospheric jitter. The supergiant Betelgeuse is included following \cite{MacLeod25}.  Models for granulation (dotted line) and oscillation-based (dashed line) jitter, as well as their quadrature sum (solid line) are shown.  Grey lines show models with an error floor of $2\ms$ and $10\ms$, appropriate for the HIRES and TRES data, respectively.  The rms RV variation remains at or below the $100\ms$ level for all spectral types except the most luminous giants with $\log{g} < 1$.}
\label{fig:jitter}
\end{figure}

\subsubsection{Granulation and Oscillation-induced RV variation}
\label{jitter}

Surface convection and p-mode oscillations at the surface cause the line-forming layers of a star to have a non-zero velocity relative to the true space motion of the star.  This velocity shift varies on short timescales, and manifests as a systematic error or ``jitter'' in single-epoch radial velocities.  In supergiants, this jitter can be in excess of $ 1\kms$ \citep{Gray08, Pugh13, MacLeod25}.  The strength of this effect correlates strongly with luminosity \citep{Cote96, Carney03, Luhn20}.

To explore the impact of surface-layer RV jitter, we have analyzed RV data of Hipparcos giants over a 10-year baseline (from the TRES spectrograph; D. Latham, priv. comm.) that have internal velocity accuracy of $\sim10\ms$.  We fit these data with binary models and measure the RMS RV with respect to the best-fit model.  The modeling framework has a prior on the RV semi-amplitude that strongly favors no binary, so that if the data do not show strong evidence for binarity then the model returns a solution for a single star.

Figure \ref{fig:jitter} shows the resulting RMS RV as a function of surface gravity (determined by converting absolute magnitudes to log $g$ using \texttt{MIST} stellar models).  Additional data are shown from \citet{Luhn20}, based on Keck/HIRES spectra with internal uncertainties of $\sim2\ms$, and from \cite{Spaeth2025} for well-studied giants.  We compare against models for the granulation and oscillation signals, and their quadrature sum \citep{Luhn20, Yu18}.  

The intrinsic RV jitter is less than $100\ms$ for all but the very most luminous giants.  Fortunately, such stars are rare and will not comprise a large fraction of our final sample (if necessary, we can excise such stars from our most sensitive analyses).  Furthermore, this effect is random, so we can also average multiple stars or multiple epochs of a single star to further reduce this source of noise.

\subsubsection{Unresolved Binary Stars}\label{sec:binaries}

Another important source of velocity scatter in stars is the presence of unresolved binary stars.  We estimate the magnitude of this effect by simulating a population of binary stars and ``observing'' them with the Viaspec observing cadence and RV uncertainties, to quantify our binary detection efficiency.  Specifically, we generate an initial population of binaries using the \texttt{COSMIC} code \citep{Breivik2020, Breivik2021}, using the multi-dimensional binary distribution from \cite{Moe2017}.  This includes eccentricities and a mass-dependent binary fraction, with an overall binary fraction (number of binaries divided by number of singles and binaries) of $\approx 30\%$.  We generate inclinations following a $\cos(i)$ isotropic distribution, and produce radial velocities for each binary with Kepler's equations.

This population of binaries is ``observed'' with a simulated Via survey.  Each system is observed for three epochs: the second epoch is 30--90 days after the first, and the third is 3--4 years after the second.  This emulates a plausible survey strategy in which fields are revisited after a couple of months to catch obvious high-amplitude binaries, and then get an additional epoch near the end of the survey to catch longer-period binaries.  For a given uncertainty in the individual RV measurements, we can ``detect'' that a source is a binary---and remove it from any kinematic analysis---if the observed RV distribution is significantly broader than that implied by the measurement errors themselves.  In practice, this involves a simple chi-square test on the observed RVs, with a binary being considered detected if the p-value of the chi-square test is $< 0.1$ (prioritizing sensitivity of the binary removal over specificity).

\begin{figure*}[t!]
    \centering
    \includegraphics[height=0.45\textwidth]{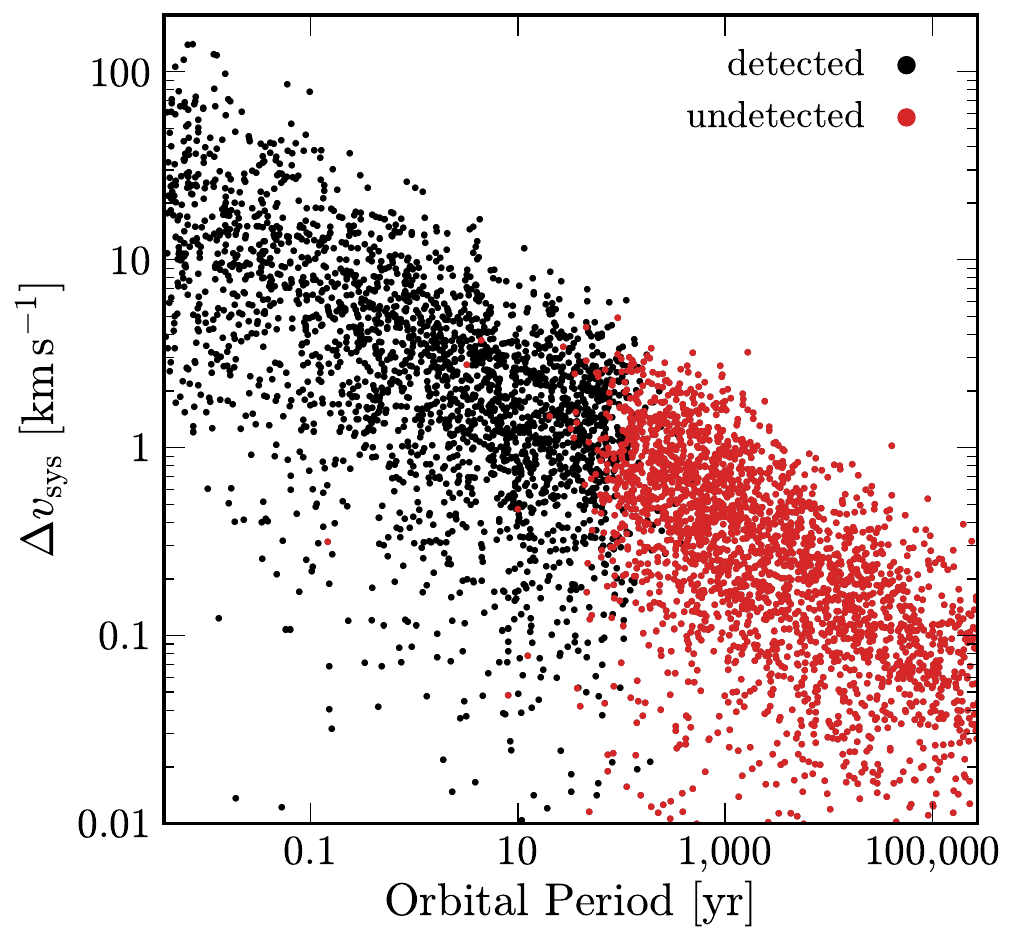}
    \includegraphics[height=0.45\textwidth]{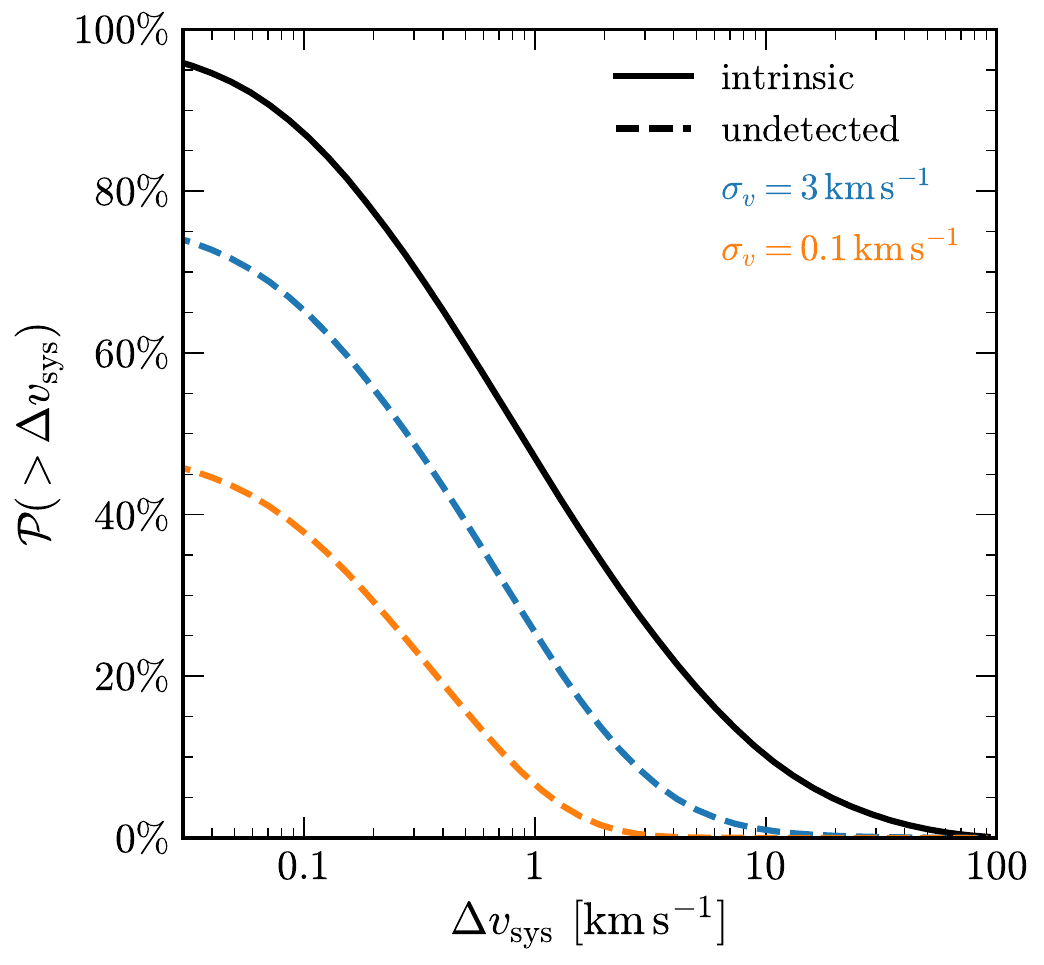}
    \caption{Left: Period distribution of binaries in our numerical simulation.  
    $\Delta v_\mathrm{sys}$ is the offset between the mean RV measured by the survey (if we assumed all epoch measurements come from a single star), and the intrinsic center-of-mass velocity of the binary.  
    Note that this is different from the binary semi-amplitude $K$, which is the worst-case value of $\Delta v_\mathrm{sys}$ for a given system. 
    Right: The probability that a binary system in our simulation has $\Delta v_\mathrm{sys}$ larger than a given value.  The black curve shows the intrinsic value in the simulation, and the colored curves show the contamination probabilities after three observed epochs, for two example per-epoch RV uncertainties $\sigma_v$.}
    \label{fig:bindetect}
  \end{figure*}

Figure \ref{fig:bindetect} illustrates the results of this experiment.  The left panel shows our input binary distribution.  The quantity $\Delta v_\mathrm{sys}$ is the offset between the mean RV measured by the survey, and the intrinsic systemic velocity of the binary.  This quantifies the jitter added to any kinematic analysis if an undetected binary is assumed to be a single star.  The right panel shows the probability that a binary has $\Delta v_\mathrm{sys}$ larger than a given value.  To turn this into the probability that an observed {\it{system}} is a binary, these quantities can be multiplied by the overall binary fraction.  Colored curves indicate this probability after binaries have been removed following our revisit scheme.  For example, if the overall binary fraction is 30\%, then $30\% \times 50\% = 15\%$ of the stellar systems in our simulation are binaries with $\Delta v_\mathrm{sys} > 1~\kms{}$.  Under this assumed binary fraction, by the end of the survey after three epochs have been observed, the contamination from undetected $\Delta v_\mathrm{sys} > 1~\kms{}$ binaries falls to 8\% (2\%) if the per-epoch RV uncertainty is $3 \kms$ ($0.1 \kms$).

Dynamical processing of binaries in dense stellar clusters will tend to result in the disruption of wide binaries.  In direct N-body simulations of globular cluster streams, this effect reduces the contamination from wide binaries by up to an additional 60\% \citep{Phillips2026}. Using a similar survey strategy and binary rejection scheme as described above, these simulations predict that undetected binaries could contribute $\sim 0.1~\kms{}$ of velocity dispersion to the observed streams. 
Modeling and predicting this effect in detail will require more comprehensive observations of the intrinsic binary population in cluster environments. 

\subsubsection{Radial Velocity Zeropoint of the Photosphere}

There are several physical mechanisms that shift the radial velocity of stellar absorption lines relative to the true stellar velocity.  Without correcting for these effects, the measured radial velocity will not be a true measure of the space motion of the star.

The first effect is the gravitational redshift induced as photons escape from the stellar potential well.  This effect scales as $GM/(Rc)$, or $633\, (M/\msun) (R/\rsun)^{-1} \ms$.  So long as the spectroscopic parameters can be measured with reasonable accuracy, the gravitational redshift can be readily removed. 

The second important effect is caused by convective motions at the photosphere.  The hotter material rising upward is brighter than the cooler falling material.  This results in a net blueshift (and asymmetry) in spectral lines.  This effect is in addition to the overall RV jitter induced by convective motions discussed in \S\ref{jitter}.  While the strength of this effect depends on the formation depth of the spectral line, there is an overall average effect that traces the convective velocity (which, in mixing length theory, scales as $v_c\propto \teff^{32/9} g^{-2/9}$).  We can therefore expect this blueshift to depend on temperature, metallicity, and surface gravity \citep{AllendePrieto13}.  This effect also depends on the spectral resolution, as the shift manifests as a line asymmetry.  At $R=20,000$, the effect is reduced by half when compared to $R>100,000$ \citep{AllendePrieto13}.

\begin{figure}[t!]
\centering
\includegraphics[width=\textwidth]{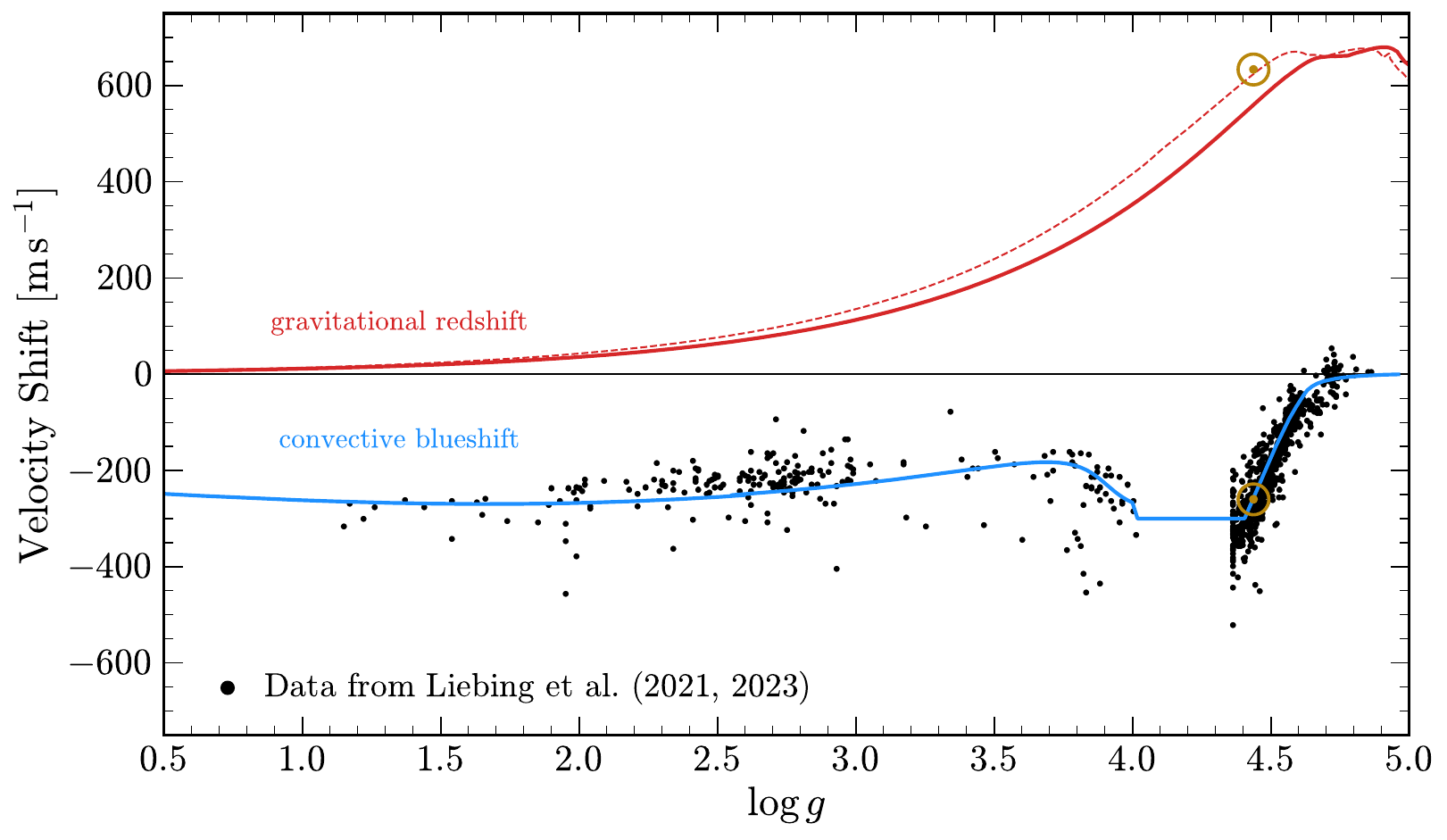}
\caption{Velocity shift relative to the restframe of the photosphere as a function of stellar surface gravity.  The gravitational redshift effect is shown along an isochrone for [Fe/H]$=0$ at 5 Gyr (dashed red line) and [Fe/H]$=-2$ at 10 Gyr (solid red line). Convective blueshifts measured by \citet{Liebing21, Liebing23} are shown as open symbols.  A simple model for convective blueshifts is shown as a blue line, in which the shift scales with the convective velocity predicted from mixing length theory \citep{Dalal23}.  Measured values of both quantities for the Sun are indicated by $\odot$.}
\label{fig:rvshift}
\end{figure}

Figure \ref{fig:rvshift} shows measured convective blueshifts of giant and dwarf stars from \citet{Liebing21, Liebing23}, as well as the convective blueshift of the Sun (yellow star).  A simple model for convective blueshifts is also shown (blue line), in which the shift is proportional to the convective velocity predicted by mixing length theory \citep{Dalal23}.  For giants, the shift is relatively constant at $-200\ms$; for dwarfs the convective blueshift varies strongly (and predictably) with surface gravity.

For giant stars (log $g<3$), the gravitational redshift is small and the convective blueshift is nearly constant.  For dwarf stars both of these effects very strongly correlate with surface gravity.  However, so long as the surface gravity is reasonably well-measured, we anticipate being able to correct for both of these effects at the level of $\lesssim 100\ms$.

\subsubsection{Radial Velocity Validation}

Our radial velocity accuracy and stability will be verified by monitoring bright radial velocity standard stars. We have compiled a full-sky grid of standard stars for this purpose from the Keck \citep{nidever:2002, chubak:2012}, CfA \citep{latham:2002}, Geneva-Copenhagen \citep{casagrande:2011}, and \gaia\ RV calibration \citep{soubiran:2018} archives.  We will also observe solar system asteroids, which have absolute velocities known at the $\approx 1\ms$ level \citep{Zwitter07}. In addition, we will observe a significant sample of stars with published radial velocity measurements from large spectroscopic surveys such as APOGEE.  This sample will allow us to further assess both the radial velocity zeropoint and radial velocity errors \citep[e.g., Figure 11;][]{geha2026b}.

\clearpage

\section{The Via Survey}
\label{survey}

The Via Survey aims to observe at least $500$ nights, combining time at the 6.5m MMT and Magellan/Clay telescopes over five years. Accounting for weather losses, seeing variations, and other effects, this amounts to more than $\sim 3300$ \emph{effective} hours (ehr) of science time for the survey.  One ehr delivers the same SNR at $G=20$ as 3$\times$20 min exposures under nominal conditions (defined as $1.0\arcsec$ seeing, an airmass of 1.2, during dark time).  Observations will be divided into four key projects that address the main science goals described in \S\,\ref{s:streams}-- \S\,\ref{s:transients}.  These key projects will determine the pointing centers for Via Survey observations.  In most cases the number of targets associated with each project is far less than the number of fibers available.  The remaining fibers will be allocated to a suite of ancillary science cases. In this section we describe the key projects, ancillary science, our approach to the survey selection function, and simulations of the five-year survey.

\begin{deluxetable*}{lccccc}[h!]
    \centering
    \label{tab:surveys}
    \tablecaption{Simulated Via Survey Properties}
    \tablehead{\colhead{Survey} & \colhead{Unique Tiles} & \colhead{Total Visits} & \colhead{Effective Hours} & \colhead{Area [sq. deg]} & \colhead{Primary targets per pointing}}
    \startdata
SPS & 1000 & 2250 & 1500 & 750 & $\approx 5$--$100$\\
CGS & 800 & 850 & 400 & 800 & $>300$\\
DGS & 150 & 880 & 1000 & 150 & $\approx 5$--$40$\\
TFS & 800 & 800 & 400 & 800 & $\approx 1$--$5$\\
\hline
Total & 2750 & 4780 & 3300 & 2500 & \\
    \enddata
\tablecomments{SPS: Stream Perturbation Survey, CGS:  Cold Gas Survey, DGS: Dwarf Galaxy Survey, TFS: Transient Follow-up Survey}
\end{deluxetable*}

\subsection{Key Projects}
\label{s:key_projects}

Four key projects define the primary Via Survey.  Targets associated with the key projects will be assigned priority 1 or 2 in the fiber-assignment algorithm, with priority class 2 reserved for fainter and/or less probable members of the target class. The expected time allocation for these key projects is given in Table \ref{tab:surveys}.

\subsubsection{The Stream Perturbation Survey (SPS)}
\label{s:sps}

\begin{figure*}[t!]
\centering
\includegraphics[width=\textwidth]{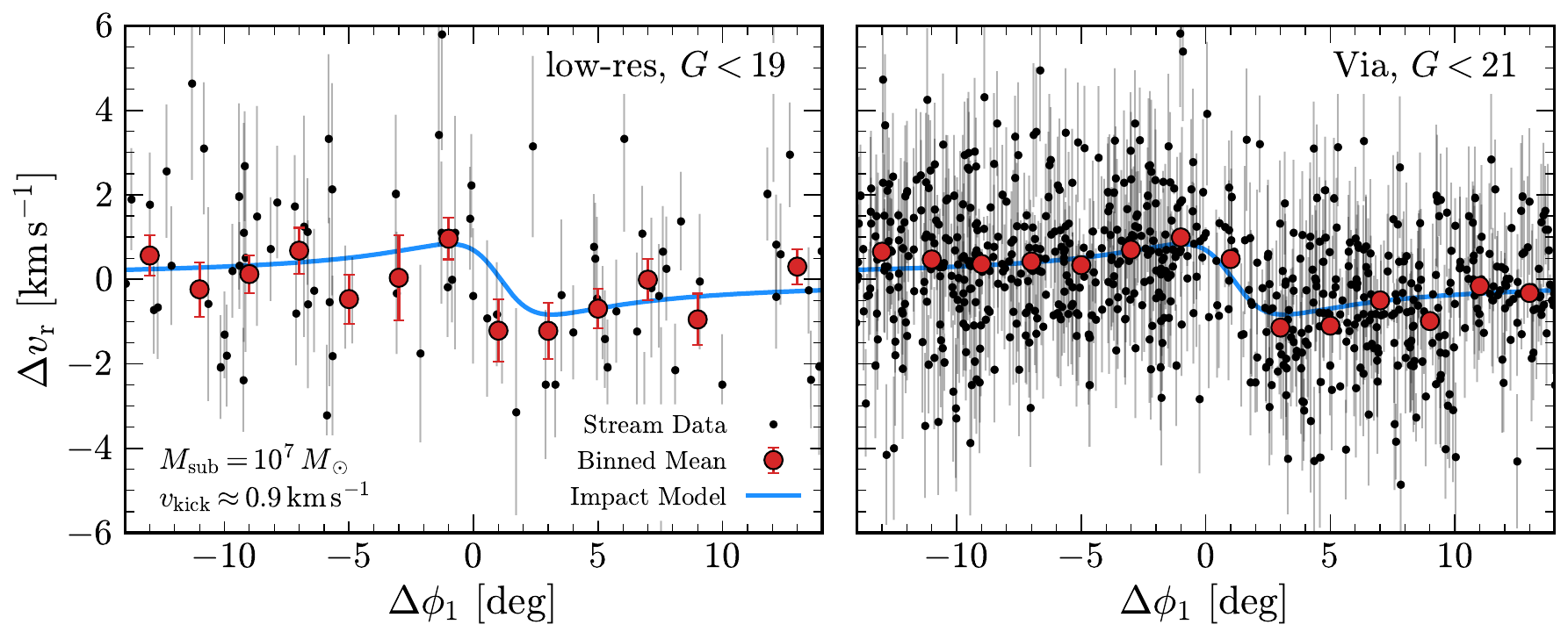}
\caption{Idealized simulation of a stream after directly encountering a $10^7\,M_\odot$ subhalo, showing the radial velocity perturbation as a function of longitude along the stream. 
The stream is sampled as a $10^4 M_\odot$ stellar population at a distance of 20~kpc (analogous to the Pal-5 stream), and magnitude-dependent observational uncertainties are added for two survey configurations.
The red line fits the characteristic $1/\phi_1$ velocity trend to the mock data. 
{Left:} Representing current surveys, the stream is observed down to $G < 19$, with RV errors similar to the DESI and SDSS-V surveys.  {Right:} Representing the Via Survey, the stream is observed down to $G < 21$, with RV errors from our instrument simulations (see $\S$\ref{sec:minesweeper}).  The improved depth and sampling ($\approx 10 \times$ more stars) and increased RV stability ($0.1\kms$ floor instead of $1\kms$) make Via ideally suited to detect subhalo--stream interactions.  }
\label{fig:stream_rvs}
\end{figure*}

The Stream Perturbation Survey (SPS) will target cold stellar streams in the Milky Way with the goal of measuring the abundance, mass, and size of dark-matter subhalos below the threshold for galaxy formation through the unique velocity signatures their impacts leave on stellar streams.
In this section, we: (1) demonstrate that Via will have the velocity precision and sensitivity to detect $\approx10^6\,\msun$ subhalos, and thus reach the regime of completely dark subhalos, and (2) estimate the number of stellar streams needed to account for the stochasticity of subhalo impacts and other confounding factors, and thus robustly measure the properties of low-mass subhalos.
We discuss our initial strategy for selecting stream targets, and possible modifications following Rubin/LSST observations.

Every subhalo flyby imparts a characteristic $1/\phi_1$ signature in the velocity profile of a stellar stream \citep[e.g.,][]{erkal:2015, lu:2025, nibauer:2025}.
However, the expected amplitudes of these signals are small, and for the low-mass subhalos, beyond the reach of the existing facilities.
Figure \ref{fig:stream_rvs} illustrates the velocity structure imprinted by a $10^7\,M_\odot$ subhalo on an idealized stellar stream at 20 kpc, simulated using the techniques described in \cite{nibauer:2025}.
The left panel shows the measurements possible with existing large spectroscopic surveys like DESI (\citealt{DESICollaboration2016, Koposov2026}), observing  to $G = 19$ with a $1\kms$ velocity precision floor.
The right panel shows mock observations of the same simulated stream down to $G = 21$, with a $100\,\mathrm{m\,s^{-1}}$ velocity precision, matching the planned depth and accuracy of Via observations.
The gain from Via is twofold: high velocity precision, and deeper observations leading to a larger sample of stars.  
Figure \ref{fig:stream_rvs} shows the additional subhalo--stream interactions that Via can reveal. 
The Via velocity stability is also crucial for detecting binary motion that can confound any kinematic analysis (see $\S$\ref{sec:binaries}). 

Beyond individual impacts, the nature of dark matter is also encoded in the cumulative effect of all past perturbations on a stellar stream. 
Dark matter subhalos are distinct from other stream perturbers at a population level, with different models predicting different overall abundances, mass functions, spatial distributions, and kinematics.  
As a result, summary statistics that trace the cumulative effects of all perturbations---such as the total number of stream gaps, or a power-spectrum of stream density variations---can constrain the abundance of subhalos as an excess over baryonic perturbations \citep{carlberg:2012,banik:2019,banik:2021}.  
On average, in a single stream we expect $\mathcal{O}(1)$ subhalo impacts to produce a prominent gap, $\mathcal{O}(10)$ to leave an individually detectable signature (e.g., the velocity signal shown in Figure \ref{fig:stream_rvs}), and $\mathcal{O}(100)$ to contribute to cumulative observable effects (e.g., dynamical heating; \citealt{Nibauer:2025heating}).  
The total number of dark matter perturbations in a given stream depends on a number of factors, and in general increases with the stream length, its Galactocentric distance, and orbital alignment with a recent major merger, such as the LMC \citep{erkal:2016,barry:2023,arora:2024}.  
Perturbations from dark matter are easier to interpret in retrograde streams, as those are less sensitive to perturbations from baryonic structures (which are typically prograde, e.g., GMCs, spiral arms, bar).  
To account for these variations in the subhalo impact rates and their stochastic nature, robust conclusions regarding the nature of dark matter will require a joint analysis of multiple stellar streams.

\begin{figure*}[t!]
\centering
\includegraphics[height=0.45\textwidth]{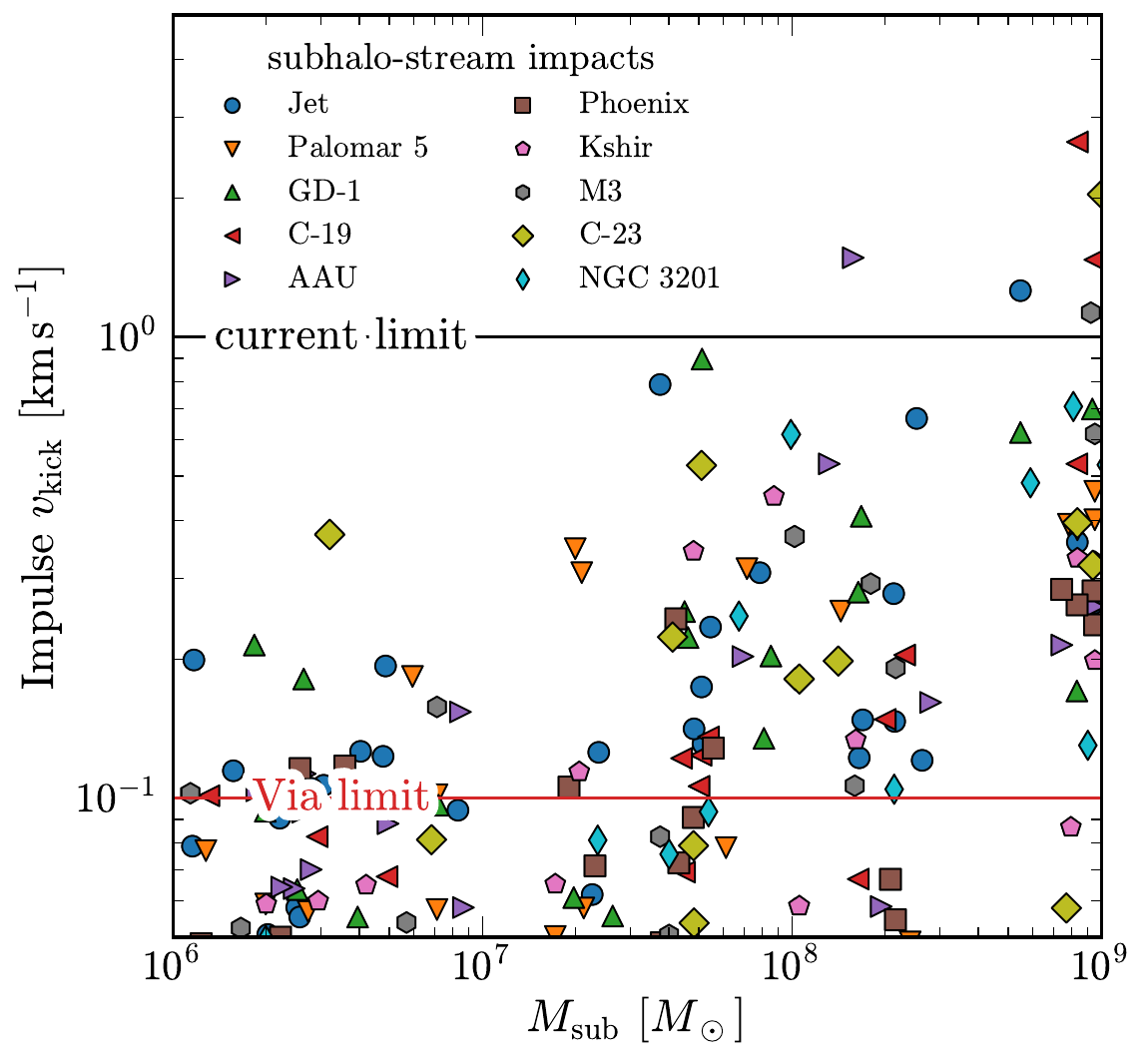}
\includegraphics[height=0.45\textwidth]{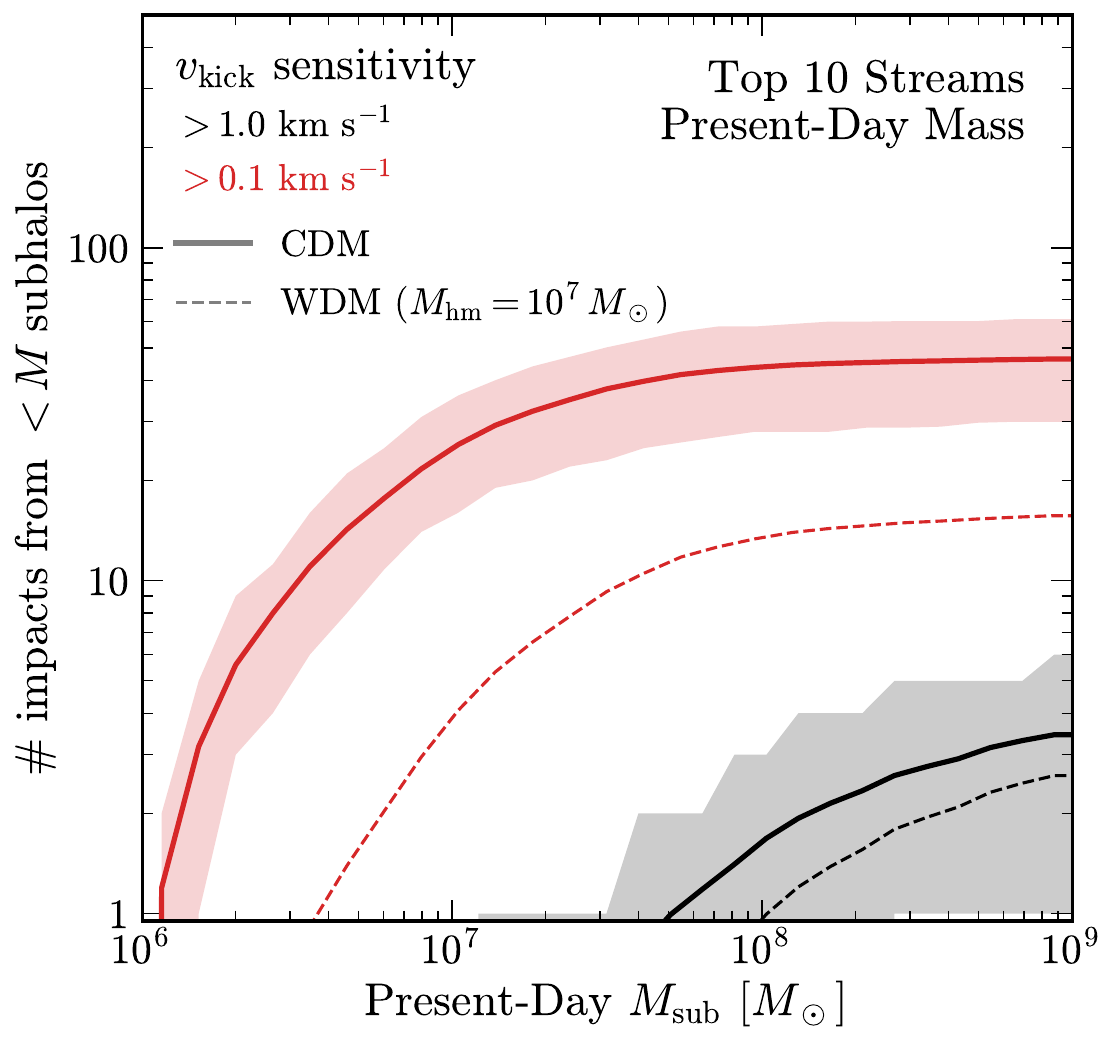}
\caption{{Left:} Simulation of subhalo--stream impacts showing the impulsive kick velocity $v_\mathrm{kick}$ as a function of subhalo mass at the time of the impact.
The subhalos are drawn from a semi-analytic simulation of a Milky Way-like galaxy assuming cold dark matter (CDM), and the streams are simulated to match real streams in the Milky Way (Chandra et al, in prep).
Radial velocity stability limits for current surveys ($1\kms$) and the Via Survey ($0.1~\kms$) are indicated.
{Right:} The cumulative mass function of detectable subhalos for two different limiting $v_\mathrm{kick}$ velocities (colors), and two assumed dark matter models (dashed vs dotted). 
A sensitivity to $v_\mathrm{kick} \gtrsim 0.1\kms$ impacts increases the number of detectable perturbations from $< 10^9\,M_\odot$ subhalos by an order of magnitude, and reveals currently undetectable impacts from $10^6$--$10^8\,M_\odot$ subhalos.}
\label{fig:vkick_msub}
\end{figure*}

Figure \ref{fig:vkick_msub} shows predictions of the subhalo population around a Milky Way-mass galaxy using the semi-analytic \textsc{galacticus} code \citep{Benson2012}.
Simulated streams matching the present-day properties of known streams (length, stellar mass, dynamical age) are evolved in this simulation, generating predictions of impulsive impacts for each stream (Chandra et al., in prep; see also \citealt{barry:2023, Adams2025, lu:2025}). 
The left panel of Figure \ref{fig:vkick_msub} shows the impulsive velocity kick imparted by each impact as a function of the subhalo mass $M_\mathrm{sub}$, for the top 10 Milky Way streams in terms of impact rate.
The right panel of Figure \ref{fig:vkick_msub} shows the cumulative distribution of subhalo masses that are detectable based on a sensitivity to two $v_\mathrm{kick}$ thresholds corresponding to current (black) and Via (red) observations.  
Reaching $\sim 100\,\mathrm{m\,s^{-1}}$ velocity precision reveals an order-of-magnitude more impacts than current limits, including those from tens of subhalos in the $M_\mathrm{sub} \lesssim 10^7\,M_\odot$ regime that are expected to be entirely devoid of stars.
We also show dashed curves that assume a warm dark matter suppression (with a half-mode mass of $10^7\,M_\odot$) of the subhalo mass function, illustrating how a sensitivity to low-mass subhalos can differentiate these dark matter models. 
 
The SPS will focus on detailed observations of approximately $10$--$20$ particularly constraining streams.
An initial reconnaissance campaign will obtain complete coverage of these streams to $G=21$, requiring $\approx1000$ unique pointings (1000 ehr).
Stream member candidates will be selected based on parallax, proper motion, and color--magnitude diagram filtering \citep[e.g.,][]{Starkman23, tavangar:2025}.
Based on published membership catalogs, we expect $5$--$30$ members at $G<21$ per Via FoV \citep{bpw:2025}.
Stream widths vary between $\sim0.1$--$1$\,deg, so we will tile most streams with two fields in the transverse direction.
Transverse fields will partially overlap to ensure continuous coverage and allow additional exposures for stars along the centerline of the stream in the overlapping regions.
Followup observations will be used to obtain greater depth or area in constraining regions of the stream (e.g., showing density variations in the LSST data), while simultaneously improving binary star identification via multi-epoch radial velocities.
Another $\approx500$ pointings (500 ehr) will be dedicated to this followup effort.  

Within the first year of the survey, Via will observe multiple stellar streams and deliver measurements of small-scale dark matter structure in the Milky Way. 
If perturbations from low-mass, truly dark subhalos are unambiguously detected, this would constitute a major validation of the cold dark matter paradigm.
Alternatively, a finding of kinematically unperturbed (or weakly perturbed) streams would place strong limits on the existence of low-mass dark matter substructure, and favor alternative dark matter models.

\subsubsection{The Dwarf Galaxy Survey (DGS)}
\label{s:dgs}

The Via Dwarf Galaxy Survey (DGS) will target  Milky Way satellite galaxies down to the critical mass threshold of galaxy formation, providing a key spectroscopic complement to the scores of photometric discoveries anticipated from LSST, Euclid, and Roman.

The most important measurements that DGS will make for satellite galaxies are \textit{dynamical masses} derived from the velocity dispersions of their resolved member stars. These dynamical masses serve as proxies for present-day halo masses, directly separating dark-matter-dominated ultra-faint dwarf galaxies from the morphologically similar population of baryon-dominated star clusters.  These masses also trace the matter power spectrum on small scales and enable a clean comparison of observed satellite galaxies to their counterparts in galaxy formation simulations and/or semi-analytical models \citep{Simon2019b}. These kinematic measurements also inform efforts to detect dark matter from its annihilation into standard model particles by enabling estimates of dwarf galaxy {\em J-factors}, which quantify the astrophysical component of the expected gamma-ray flux from annihilation \citep[e.g.,][]{Evans2016, PaceStrigari19}.
\par The key observational requirements for determining precise and accurate dynamical masses are sizable samples of radial velocity measurements with precisions smaller than the magnitude of each galaxy's intrinsic velocity dispersion, obtained across multiple epochs to mitigate potential biases from spectroscopic binaries. For the fiducial limiting case of a $M_* = 200 \ M_\odot$, $r_{1/2} = 10$~pc ultra-faint dwarf with a mass-to-light ratio of $M/L = 50~ M_\odot L_\odot^{-1}$, the predicted stellar velocity dispersion is just $\sigma_v = 1.0 \kms$; thus, per-star velocity uncertainties down to $\sigma_v \approx 0.5 \kms$ or smaller are a strict requirement. At this precision, Via should easily resolve the velocity dispersions of known ultra-faint dwarfs (and dwarf galaxy candidates) for which only upper limits are currently available \citep[e.g.,][]{Kirby2013,Buttry2022,Simon2024, Geha2026,Cerny2026}.

\begin{figure*}[t!]
    \centering
    \includegraphics[height=0.55\textwidth]{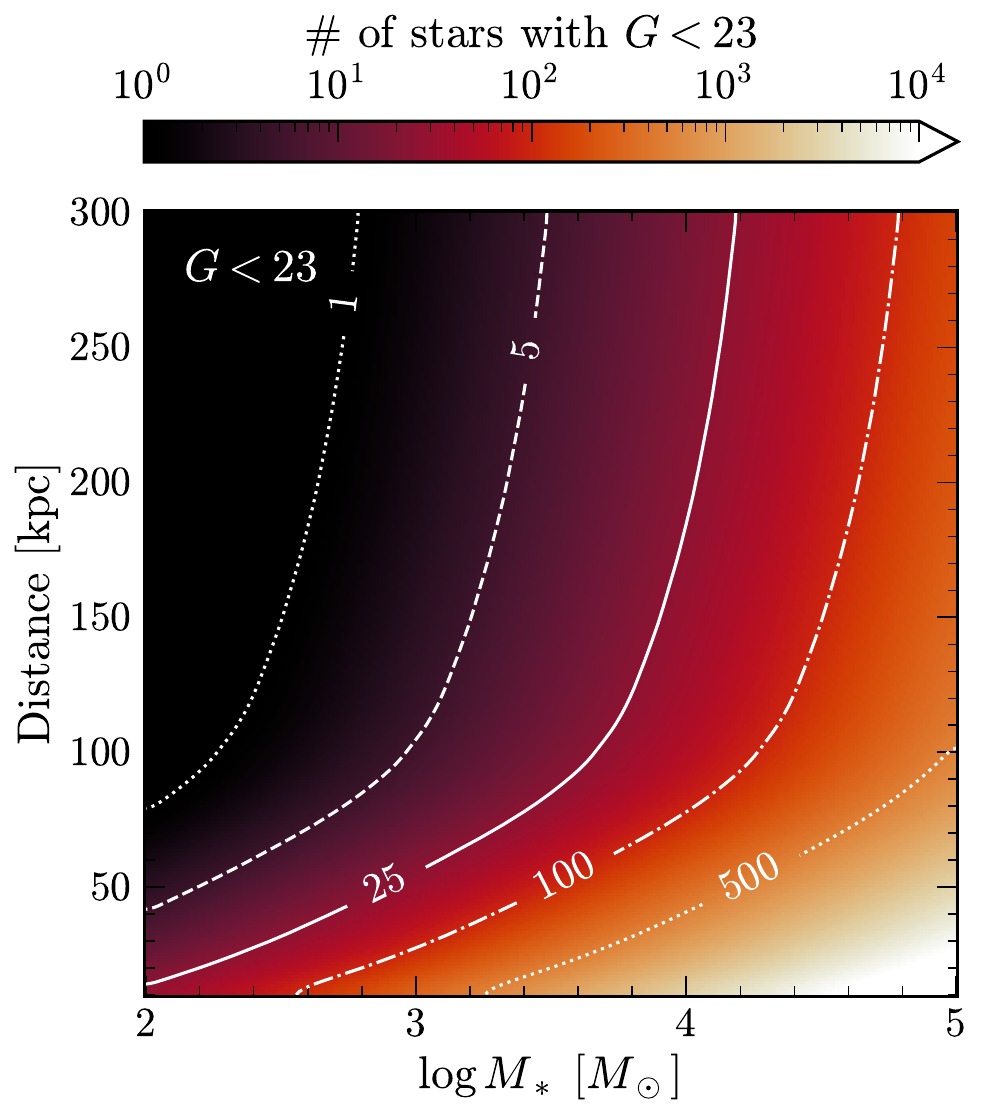}
    \includegraphics[height=0.55\textwidth]{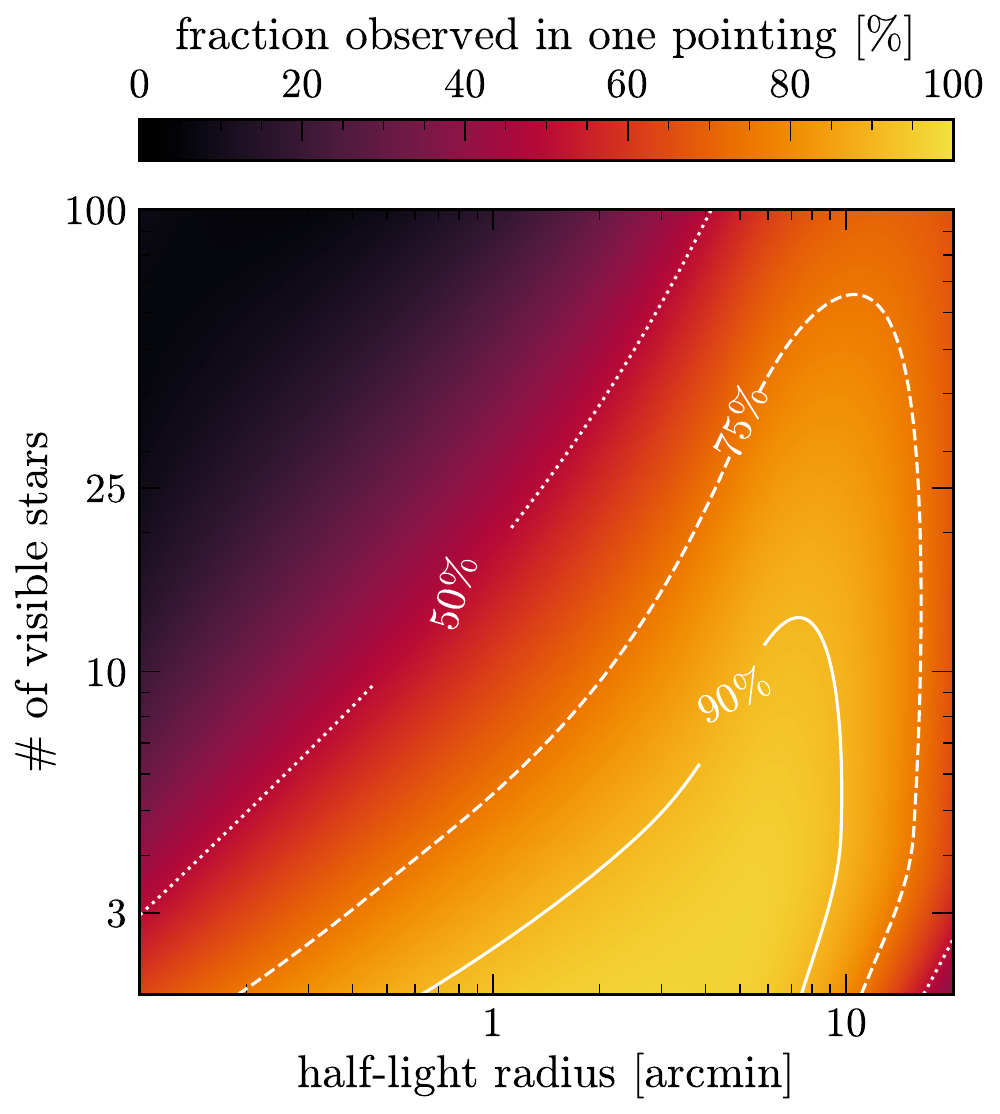}
    \caption{Left: Number of targets in dwarf galaxies of varying mass and distance, for a $G < 23$ limiting magnitude corresponding to 10\,hr Viaspec depth. We assume a [Fe/H]~$=-2.0$, 12 Gyr stellar population and a Kroupa IMF for this test.   Right: Fraction of dwarf galaxy stars observable by the Via fiber positioner in a single pointing, as a function of the number of visible stars and the apparent half-light radius. These are realistic simulations of the Via FPS, including fiber collision avoidance.  Dwarf galaxy stars are assumed to follow a Plummer distribution. }
    \label{fig:dwarf_ntarg}
  \end{figure*}

In tandem with precise velocity measurements, Via will obtain iron abundances ([Fe/H]), $\alpha$-element abundances (especially Mg), and likely a small sample of additional elements (notably C and Ba) for the majority of dwarf galaxy stars it targets.  These abundances directly aid the dynamical mass measurements by helping isolate interloping metal-rich Milky Way field stars with velocities overlapping those of dwarf galaxy members.  Moreover, for the faintest systems for which resolving velocity dispersions will be the most challenging, the presence of significant internal metallicity dispersions can be used to independently separate dwarf galaxies from (nearly) mono-metallicity globular clusters \citep{WillmanStrader2012}.  The metallicity and abundance distributions of individual systems and the global mass--metallicity relationship for dwarf galaxies also offer a window into early galactic chemical evolution and the effects of feedback, tidal and ram pressure stripping, and reionization on the faintest galaxies \citep[e.g.,][]{Kirby2013, Agertz2020, CollinsRead22, Sandford2024}. Large metallicity samples amassed by Via will expand the search for ultra-metal-poor member stars that preserve the chemical signatures of Population III nucleosynthesis \citep[e.g.,][]{Jeon2017, Hartwig2018, Rossi2024, Chiti2025}.

\begin{figure}[t!]
    \centering
    \includegraphics[height=0.45\textwidth]{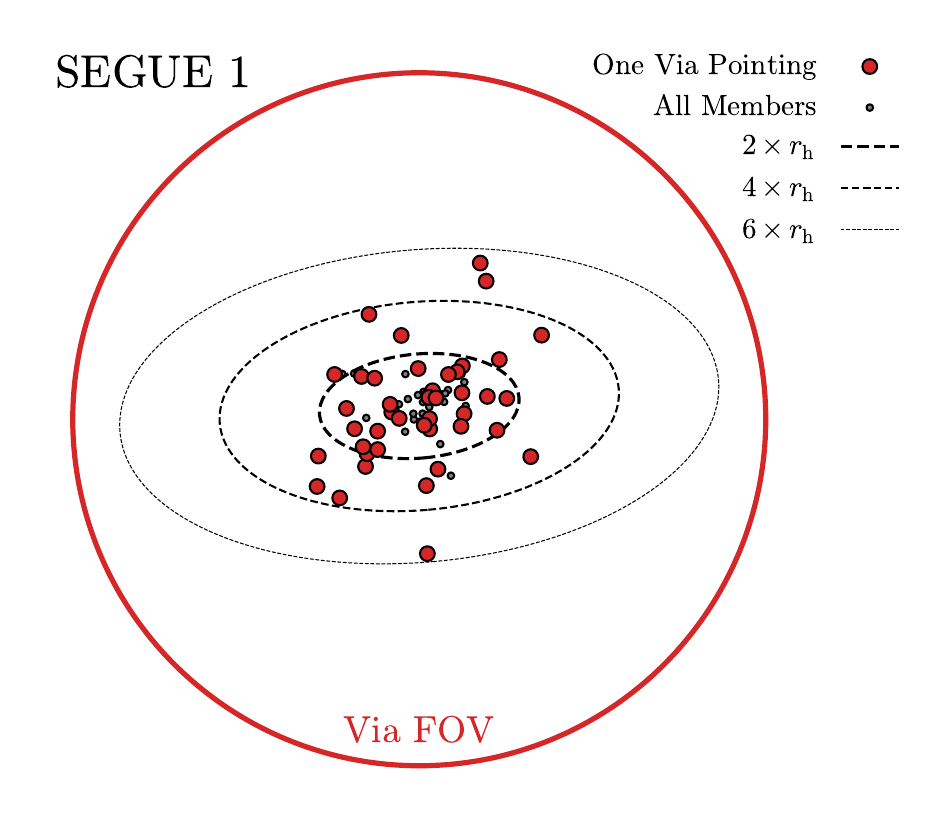}
    \includegraphics[height=0.45\textwidth]{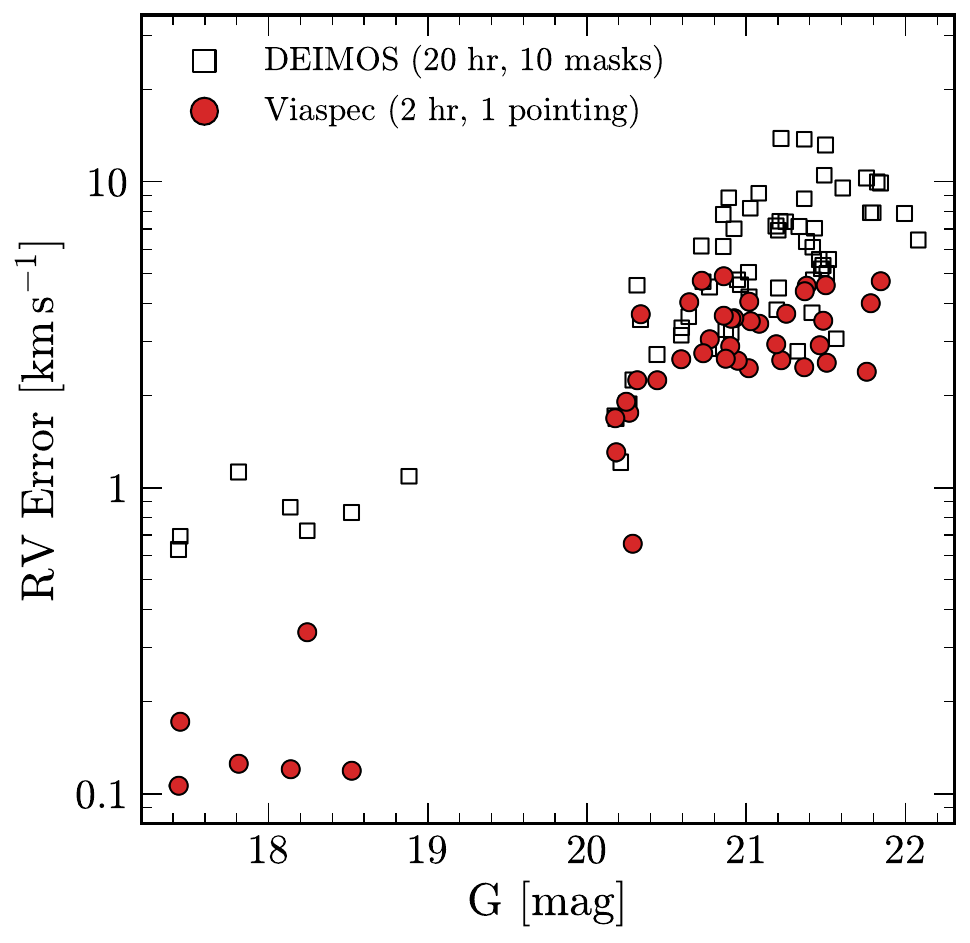}
    \caption{Simulating Via observations of the most dark matter dominated galaxy known, Segue 1. 
    {Left:} Spatial distribution of Segue~1 members, observed with a realistic simulation of the Via fiber positioner. 
    The 1-degree FoV of the fiber positioner is shown in red, and ellipses denote 2, 4, and 6 half-light radii of Segue~1. 
    A single pointing can observe 43 members (red points), and 5 pointings (fiber configurations) are sufficient to observe all 68 members (gray points). 
    {Right:} Predicted radial velocity accuracy vs magnitude for a single 2\,hr pointing, observing the stars shown in the left panel. 
    This simulation incorporates the magnitude, spectral type, and metallicity information of each star, observed and fitted with the full Via spectrum simulator (see $\S$\ref{sec:etc}). 
    Via can deliver up to $\sim$5--9$\times$ better RV precision for red giant branch stars and more than $2\times$ better precision for subgiants and upper main-sequence stars (in half the time, assuming nominal observing conditions) compared to state-of-the-art Keck/DEIMOS observations of Seg 1 (open black squares). 
  }
    \label{fig:seg1}
\end{figure}

\begin{figure}[t!]
    \centering   \includegraphics[width=\textwidth]{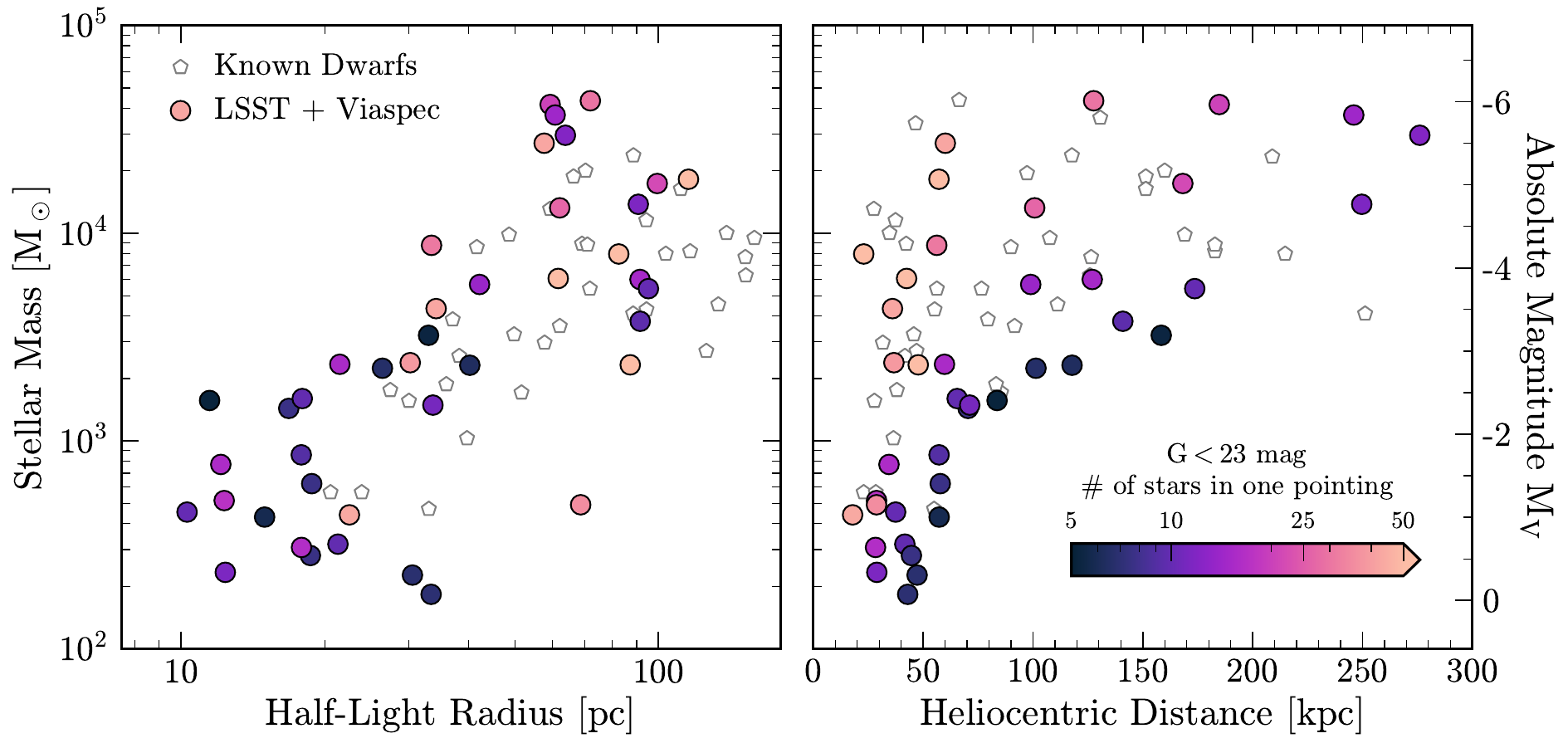}
    \caption{Forecast for the Via survey of dwarf galaxies discovered by LSST, showing stellar mass as a function of half-light radius (left) and heliocentric distance (right). Open markers show currently known dwarfs with measured velocity dispersions \citep{Pace2024}, and colored markers show predicted dwarfs that LSST will discover, and for which Via will deliver $> 5$ stellar spectra in a single pointing.  LSST predictions are drawn from the semi-analytical model of \citet{Manwadkar2022}.  }
    \label{fig:via_lsst}
\end{figure}

To assess the landscape of Via-accessible stars in the Milky Way's satellite galaxies, we simulated stellar populations across a range of stellar mass and distance  assuming \texttt{MIST} isochrones \citep{Choi16} and a \cite{Kroupa2001} IMF. 
Figure \ref{fig:dwarf_ntarg} shows the number of dwarf galaxy targets brighter than Via's 10-hour depth of $G=23$ in the plane of stellar mass and heliocentric distance, setting the parent sample of targets in each satellite.  However, not all stars are observable due to fiber-positioning constraints.
Realistic fiber-positioning experiments that consider the Via fiber-positioning system geometry are used to refine our predictions for Via dwarf galaxy observations (see $\S$\ref{sec:fps}). 
The right panel of Figure \ref{fig:dwarf_ntarg} quantifies the predicted fraction of stars that Via can target as a function of dwarf galaxy angular half-light radius. 

Figure~\ref{fig:seg1} provides an example of Via observations using the distribution of member stars in Segue 1---one of the faintest and smallest confirmed dwarf galaxies. 
The right panel of Figure~\ref{fig:seg1} shows the predicted velocity uncertainty from a single Via pointing, based on our instrument model (see $\S$\ref{sec:etc}). 
This figure demonstrates the major improvement in velocity precision and overall observing efficiency expected relative to the current state-of-the-art spectroscopic dataset from Keck/DEIMOS \citep{geha2026b}. 
In Figure \ref{fig:via_lsst}, we apply the same framework to an entire simulated population of dwarfs resembling those detectable from LSST to forecast the expected member star yield across the complete size--luminosity plane (assuming a 10~hr depth of $G = 23$). 

Although these calculations have assumed standard Plummer radial profile models, hydrodynamical simulations predict that low-mass dwarf galaxies should commonly host extended stellar populations in the form of tidal features  and stellar halos \citep[e.g.,][]{Tarumi2021, Deason2022, Go24, Goater24, Querci25} that might represent substantial contributors to a dwarfs' total stellar mass budget \citep{Andersson2025}. Observationally, the sample of ultra-faint dwarfs with unambiguous evidence of tidal features and/or halos remains small ($<20\%$ of known satellites; \citealt{Simon2019,jensen_2024}; see e.g., \citealt{Li2018,Chiti2021,Sestito2023} for illustrative detections).  However, in many cases the outskirts of these systems have been poorly sampled by existing spectroscopic observations due to field-of-view or depth constraints. With its large field of view and multiplexing, Via will probe the existence of these features in every satellite that it targets, providing a statistical sample that can be used to test simulation predictions. We expect that these observables can be folded into abundance matching analyses, though further theoretical work is needed to explore this possibility.

A key strength of Via's dwarf galaxy science case is homogeneity: all observations will be carried out with the same instrument, velocity and abundance pipelines, and a self-consistent stellar targeting strategy. The result will be a complete, homogeneous sample of dynamical and chemical measurements for a large sample of currently known and newly discovered dwarf galaxies from LSST, Euclid, Roman, and other future surveys.

The Dwarf Galaxy Survey aims to characterize the dynamics and chemistry of the Milky Way satellite galaxies down to the very threshold of galaxy formation. This dual-hemisphere, multi-epoch survey is split into three overlapping components totaling $1000$ ehr:

(1) The {\it DGS-Complete} component will measure stellar velocity dispersions, dynamical masses, and metallicity distributions for every Milky Way satellite galaxy ($D_\odot < 300$ kpc) down to an absolute magnitude of $M_V = -3.5$, including both currently known systems and LSST, Euclid, and Roman discoveries.  This absolute magnitude limit is chosen to ensure a minimum of five or more observable member stars in a satellite at any distance in the Milky Way halo.  These observations will robustly confirm new discoveries as bona fide dwarf galaxies and will yield a homogeneous dynamical mass function that can be compared with theoretical predictions. Each satellite will be observed for at least two epochs to allow rejection of large-amplitude binaries, with per-epoch effective exposure times of 0.5--4 hours depending on each dwarf's luminosity and distance. 

(2) The {\it DGS-Extreme} component will target a large number of the faintest Milky Way satellites ($M_V > -3.5$) with a goal of identifying and confirming the least massive galaxies in the universe. Unlike DGS-Complete, DGS-Extreme will have no explicit requirement on population completeness and will prioritize nearby systems ($D \lesssim 100$ kpc) where $\gg 5$~member stars can be detected even for the faintest satellites. For newer discoveries, DGS-Extreme will combine shorter reconnaissance observations focused on separating out obvious interloping ultra-faint star clusters (e.g., based on their mean metallicities) followed by deeper observations of the most promising extremely low-mass galaxy candidates. For existing candidates, DGS-Extreme will prioritize resolving the velocity dispersions of systems for which prior instruments lacked the precision. 

(3) The {\it DGS-Deep} component will return to $\sim 20$--$40$ of the most interesting confirmed dwarf galaxies from DGS-Complete and DGS-Extreme for up to $10$ additional hours per dwarf to expand dwarf galaxy stellar member samples to enable a wider range of science cases. These include more sophisticated mass modeling, deeper and/or wider searches for tidal features and stellar halos, and measurements of metallicity distribution functions and $\alpha$-abundance tracks. By splitting this additional exposure time across multiple epochs, these observations will also provide better identification of spectroscopic binaries.

The internal allocation of time within DGS will be adjusted as necessary to account for the actual population of satellites discovered in LSST and other surveys and to balance DGS priorities.

\subsubsection{The Cold Gas Survey (CGS)}

\begin{figure}[t!]
\centering
\includegraphics[width=\textwidth]{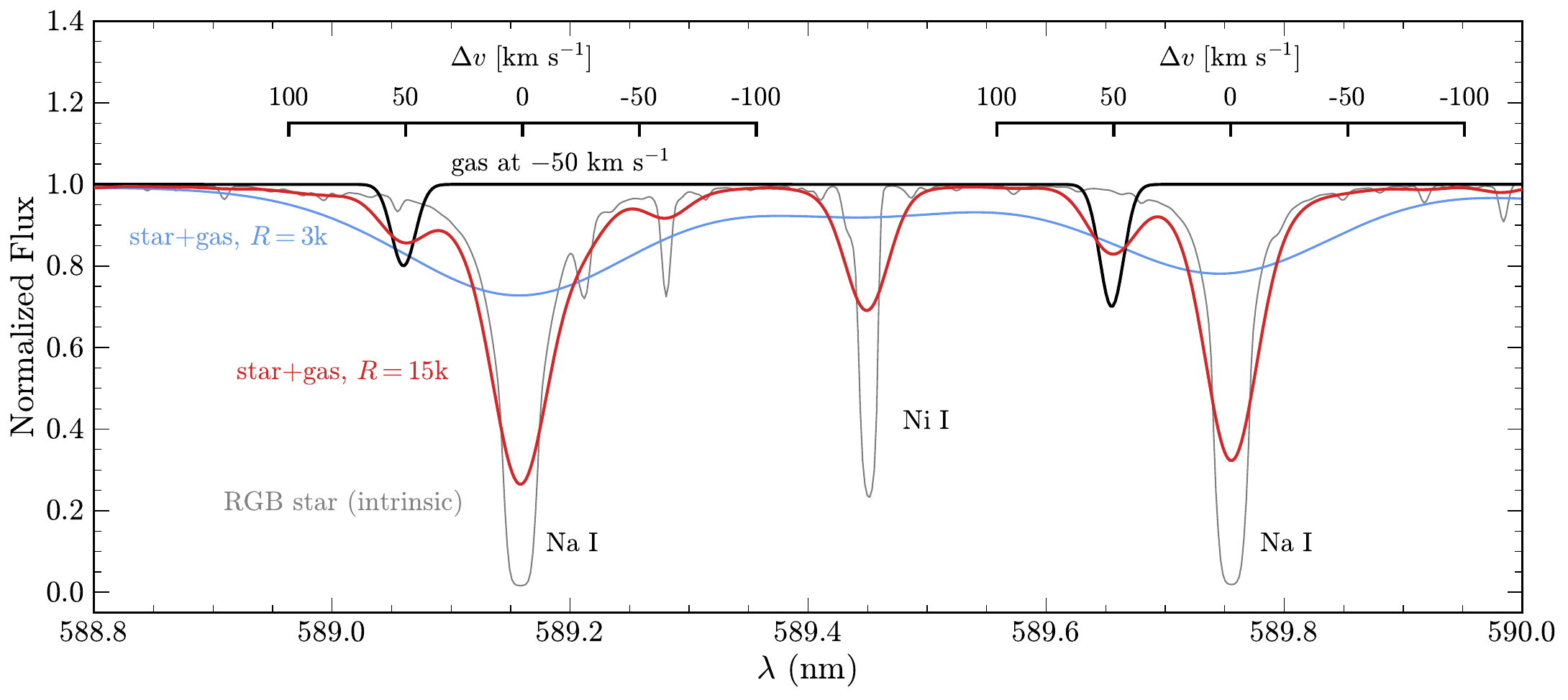}
    \caption{Simulated spectrum of an RGB star acting as a backlight for gas along the line of sight at a relative velocity of $-50\,\kms{}$, observed at $R = 3000$ (blue) and $R = 15000$ (red). 
    At the resolution of Viaspec, \ionn{Na}{i} absorption from cold gas can be disentangled from stellar absorption, revealing the density and velocity structure of the interstellar and circumgalactic medium.  Substantially higher spectral resolution is not advantageous owing to the broad wings of the photospheric \ionn{Na}{i} lines.}
\label{fig:nad_spec}
\end{figure}

Using hundreds of thousands of Milky Way halo stars as backlights, Via will map the distribution of cold gas in the CGM in three spatial and one velocity dimension. The \ionn{Na}{i} doublet at 589.0 nm and 589.6 nm is well suited for this purpose: it is strong, sensitive to cold neutral gas, typically optically thin in the CGM, and lies in a spectral region where distant giant stars are relatively bright. Via's high spectral resolution will enable robust separation of interstellar absorption from stellar sodium features at most velocities (see Figure~\ref{fig:nad_spec}). The higher fiber density will allow us to measure stars at many distances along each sightline simultaneously.  Rather than just determining a distance bracket on large individual clouds, we will be able to reconstruct the entire distribution of gas density and velocity.

\begin{figure}[t!]
\centering
\begin{minipage}[l]{0.6\textwidth} 
\includegraphics[width=1.\textwidth]{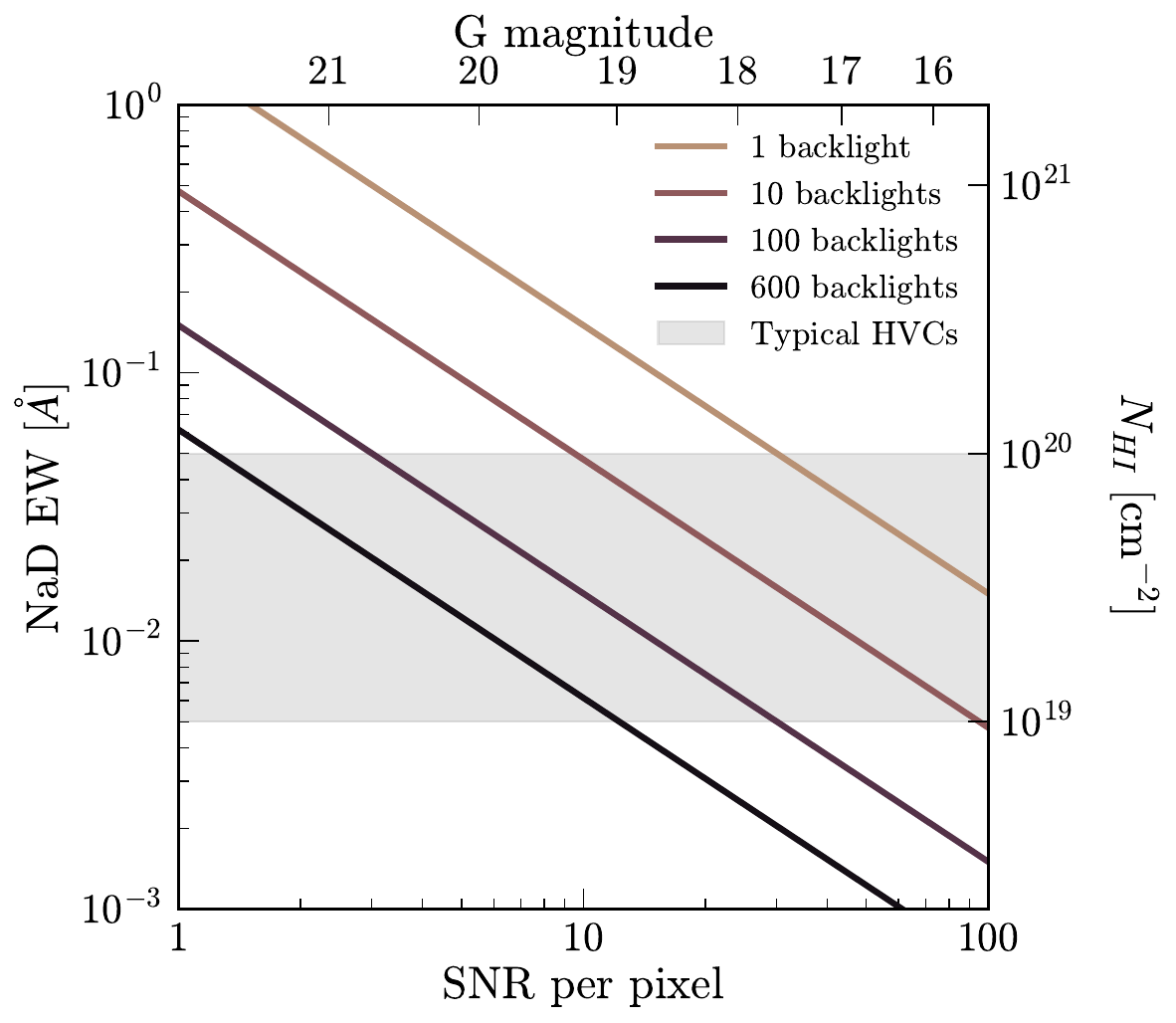}
\end{minipage}
\begin{minipage}[c]{0.35\textwidth}
  \caption{Relation between \ionn{Na}{i} 589.0/589.6nm EW (in \AA) and spectrum SNR per pixel for a \ionn{Na}{i} detection threshold of $3\sigma$.  \ionn{Na}{i} EW is converted to an approximate \ionn{H}{i} column density along the right axis \citep{Murga2015}.  Limiting $G$-band magnitude at a given SNR is shown along the top axis.}
\label{fig:nad}
\end{minipage}
\end{figure}

Sensitivity to cold gas will be determined by both the signal-to-noise ratio of individual spectra and the number of stars observed behind a cloud. We used a Fisher forecast to propagate the SNR per pixel near the sodium doublet to an equivalent width (EW) uncertainty, marginalizing over the unknown radial velocity of the absorber and assuming a Gaussian line profile set by the instrument resolution. We also estimated the \gaia\ $G$ magnitude required to reach this per-pixel SNR using mock spectra, assuming a 1-hr integration time and nominal conditions ($1.0\arcsec$ seeing and an airmass of 1.2). We converted the sodium equivalent width to an approximate neutral hydrogen column density using the relation from \citet{Murga2015}, although this conversion is highly uncertain and is likely variable cloud to cloud and across cloud substructures \citep{Peek:2019}. The results are shown in Figure \ref{fig:nad}. For typical clouds with $N_{\rm HI} \gtrsim 10^{19}\, \rm{cm}^{-2}$, Via will have moderate sensitivity for clouds backlit by only a few bright stars, but will easily detect clouds backlit by a substantial fraction of the 600 stars in a field.

For clouds within a few kpc where the density of backlights is high, we will reconstruct the 3D distribution of cold gas.  However, creating a 3D map from noisy, inhomogeneous, line-of-sight integral measurements is an inherently challenging task. Reconstructing a field with infinite degrees of freedom using finite data is ill-posed without some form of regularization, and quantifying correlated uncertainties on millions of parameters is computationally demanding. However, powerful tools to address these challenges have recently been developed in the context of high-resolution 3D dust mapping \citep{Edenhofer2024, Leike2019}, and we will adapt these techniques.  A key new feature of Via is that we will reconstruct the velocities along with the density, potentially aiding the reconstruction and enabling us to address a broader array of questions related to the nature and origin of the cold gas in the Milky Way.

\begin{figure}[t!]
\centering
\includegraphics[width=\textwidth]{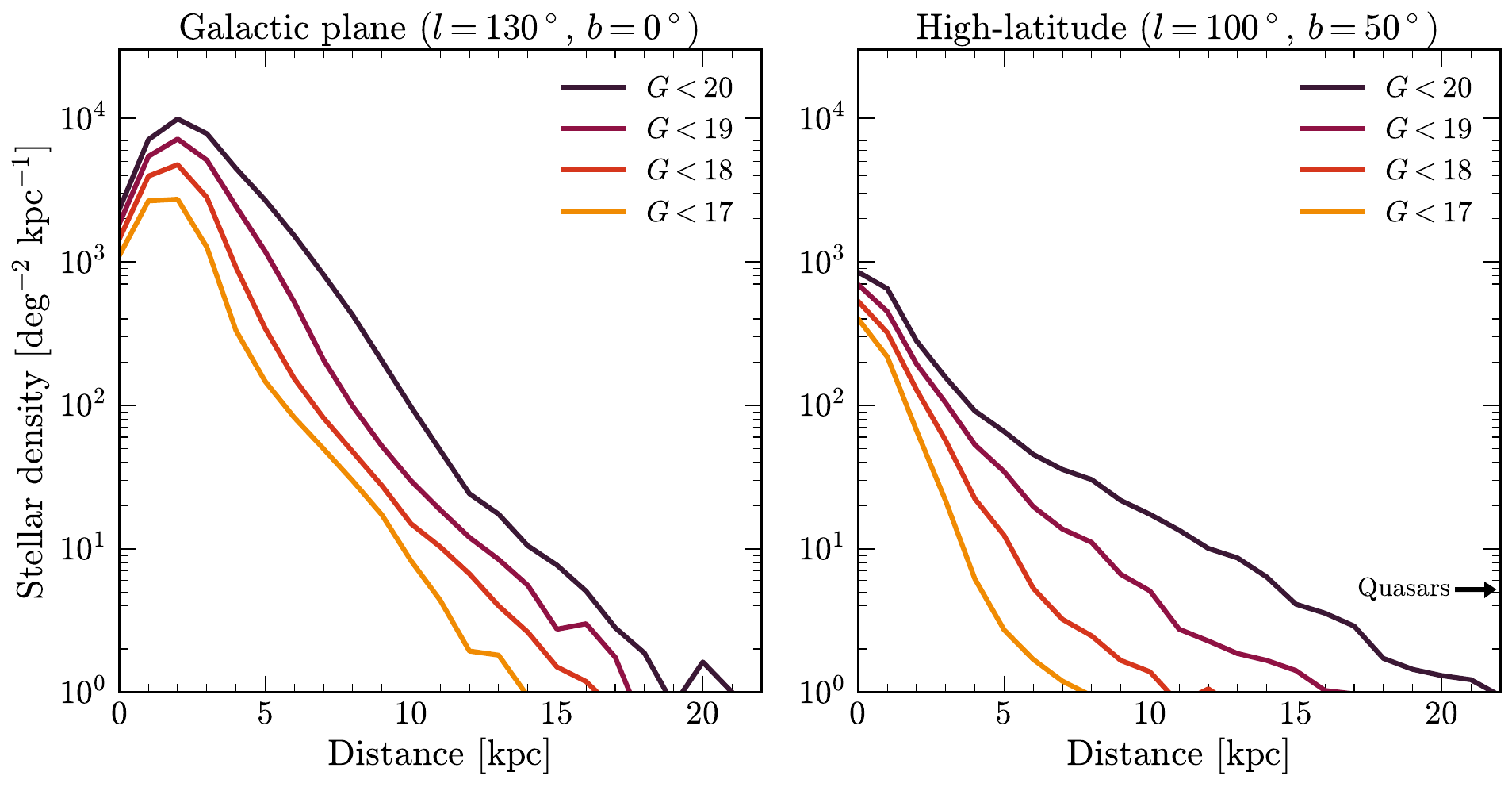}
    \caption{Distribution of stellar sources along two example sight-lines, according to the GUMS mock catalog \citep{Robin2012}. The Galactic plane sightline intersects Complex H, for which the only direct distance measurement for the gas is a $6$ kpc lower limit \citep{Smoker2011}. The high-latitude sightline intersects Complex C, which is currently bracketed at $10\pm 2.5$ kpc. We also expect multiple quasars per field providing an infinite distance backlight.}
\label{fig:nad_stars}
\end{figure}

At greater distances, the density of backlights drops; beyond 10 kpc we expect $\sim10$ stars brighter than $G=19$ in each Via FoV and at $>40$ kpc we expect only $\sim1$ star on average.  We will supplement stellar backlights with the expected tens of quasars in every field \citep{Storey-Fisher23}.  Quasars provide a total column density, important for measuring the amount of cold gas in the outermost regions of the halo.  The distribution of stellar sources along two example sight-lines is shown in Figure \ref{fig:nad_stars}.  Careful selection of fields with higher than typical density of distant stars (e.g., along the Sagittarius stream, where the stellar density is $>10\times$ higher than average) will increase the number of distant backlights.  While low stellar densities preclude detailed 3D reconstruction of the cold gas density field, simply detecting the existence of cold gas reservoirs in the outer Galaxy is important, as models disagree about the fraction of high-column-density cold gas beyond 10 kpc.  In a ``super-resolution'' simulation of a Milky Way galaxy and its surrounding CGM, \citet{Lucchini2026} find that 10\% of the stellar halo beyond 40 kpc has column densities exceeding $10^{19}$ cm$^{-2}$; that number drops to 1\% for $>10^{20}$ cm$^{-2}$ gas.

\begin{figure}[t!]
\centering
\begin{minipage}[l]{0.5\textwidth} 
\includegraphics[width=1.\textwidth]{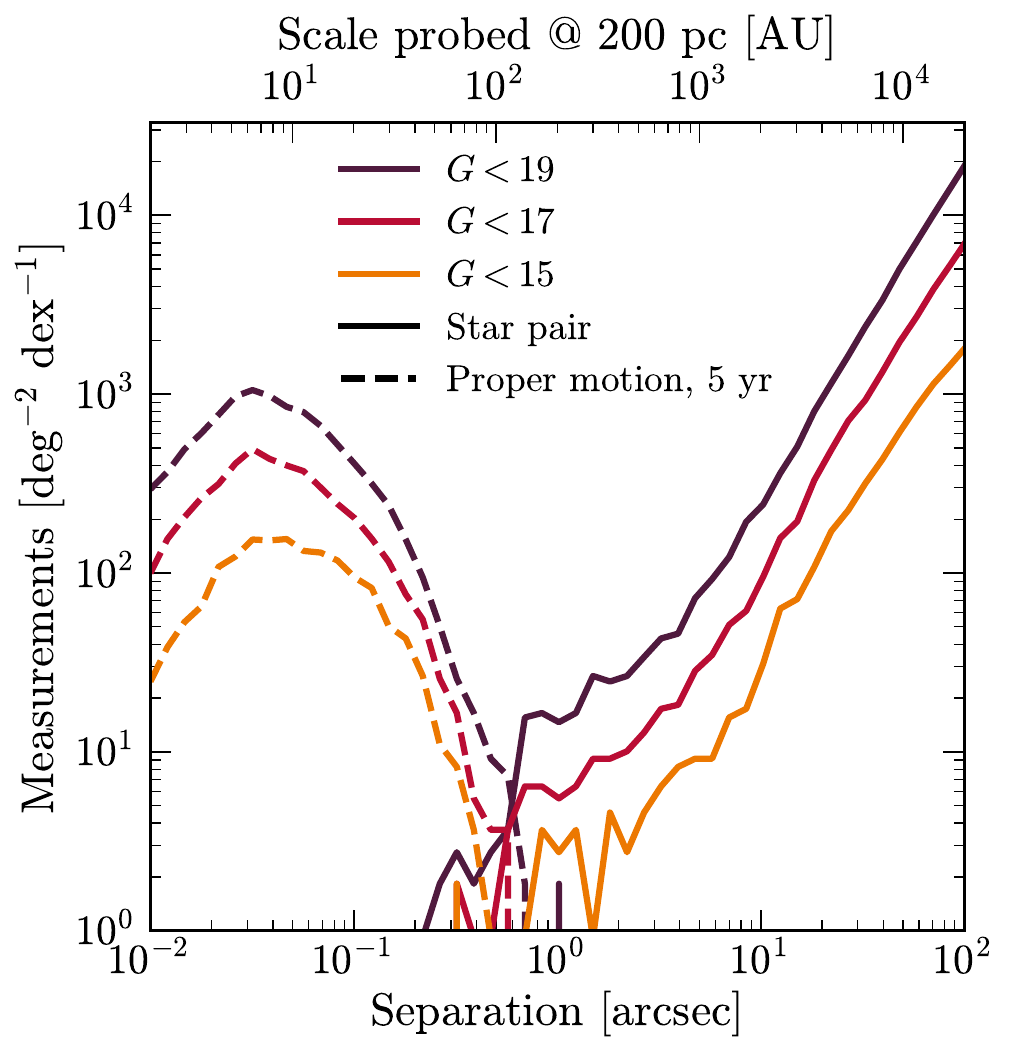}
\end{minipage}
\begin{minipage}[c]{0.45\textwidth}
    \caption{Scales probed by stars behind a notional gas cloud at $(l,b)=(198^\circ, 35^\circ)$ and a distance of 200 pc. Small-scale structure can be detected by measuring a pair of nearby stars simultaneously, or by measuring a single star with a high proper motion over the 5-year survey baseline.}
\label{fig:tsas}
\end{minipage}
\end{figure}

Via will make novel cold gas measurements in the Galactic disk as well. The interstellar medium (ISM) is structured over an enormous dynamic range, from kpc to sub-AU scales.  The origin of this structure is not known, but likely candidates include a continuation of the turbulent cascade and discrete structures such as supernova remnants \citep[see][for a review]{Stanimirovic18}.  Some of the earliest observations of this very small-scale structure came from observations of \ionn{Na}{i} EWs \citep{Munch53, Munch57}.  Measurements of binary star systems and stars with significant proper motions enable the detection of structure at the smallest (sub-AU) scales.  With \gaia\ astrometry we can now identify millions of such stars suitable for probing small-scale structure in the ISM.  Figure~\ref{fig:tsas} shows the range of physical scales probed within a fictitious gas cloud at a heliocentric distance of 200 pc.  Star pairs probe scales $>10^2$ AU, while high proper motion stars probe scales $<10^2$ AU when observed over a 5-year baseline.   This measurement could be made for dozens of nearby clouds, enabling the first comprehensive measurements of the small-scale structure in cold clouds.

This survey will target a variety of \ionn{H}{i}-selected gaseous features in the Milky Way halo (e.g., high velocity clouds).  All \ionn{H}{i} clouds with peak \ionn{H}{i} column density above $10^{19}\, {\rm cm}^{-2}$ will be observed with at least one pointing, requiring 100 ehr of survey time.  In bright time, the CGS will observe nearby clouds in the Galactic plane to obtain high-resolution 3D maps. The CGS can use data from the other key surveys to generate cold gas maps along unbiased lines of sight.  For example, these data will allow us to search for distant halo clouds at velocities consistent with disk rotation, currently undetected in \ionn{H}{i} due to confusion with Milky Way disk emission.  Finally, we will explore the use of distant galaxies as backlights to probe the CGM of nearby galaxies such as M31 and M82.

\subsubsection{The Transient Follow-up Survey (TFS)}

\begin{figure*}[t!]
    \centering
    \includegraphics[width=\textwidth]{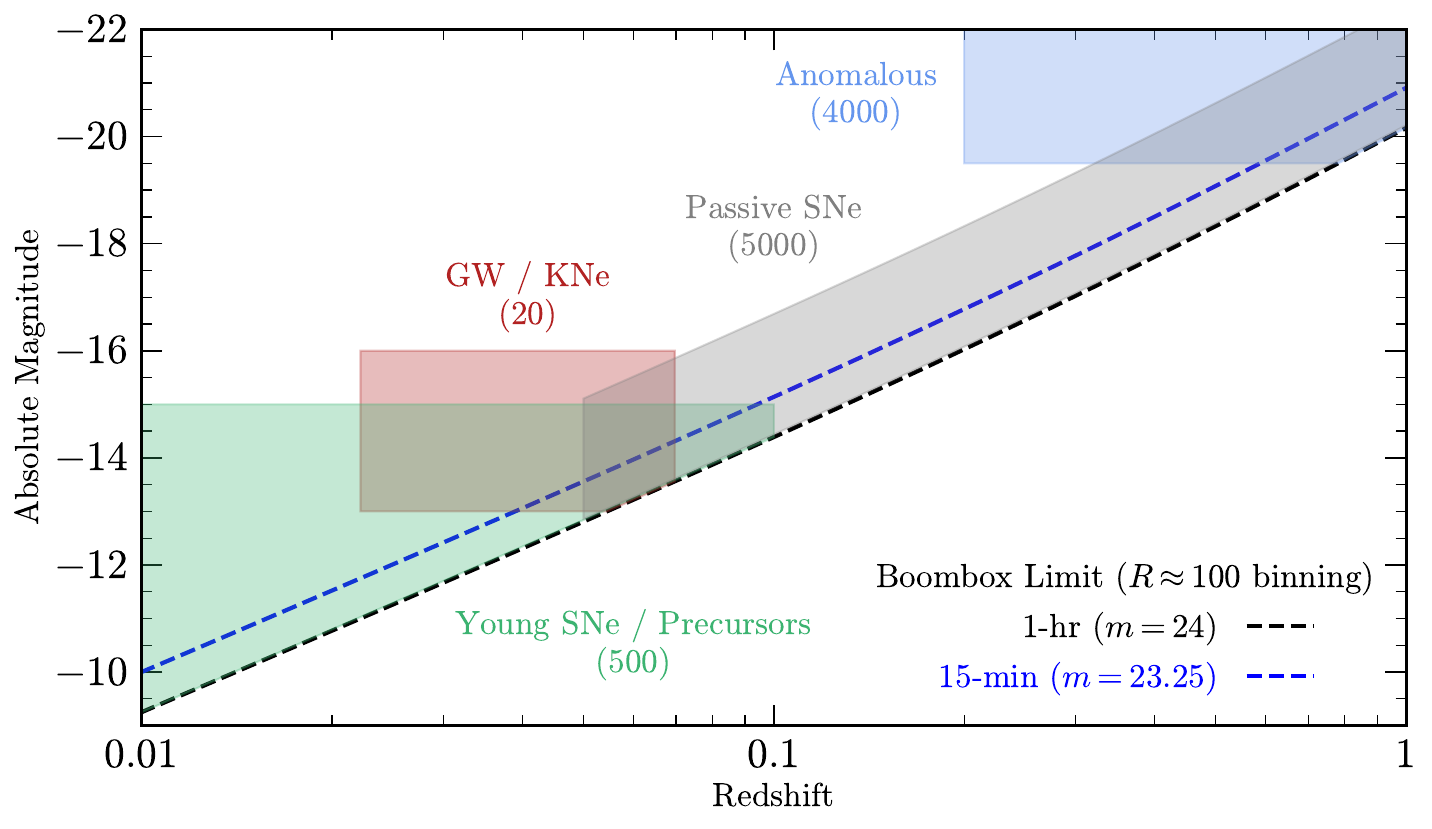}
    \caption{Four major target categories for the Boombox time-domain program.
    Each colored box represents the typical luminosity and redshift range of a target class: gravitational-wave counterparts and kilonovae (GW / KNe), young core-collapse supernovae and precursor events (Young SNe / Precursors), rare or anomalous transients (Anomalous), and passively selected active supernovae in Via fields (Passive SNe). 
    The dashed lines indicate the approximate limiting magnitude for Boombox spectroscopy when binned to $R \approx 100$ for transient classification. 
    The parenthetical numbers indicate the approximate number of spectroscopic targets expected over the duration of the Via campaign in each category. 
    At the native $R \approx 1000$ resolution of Boombox, these curves are about one magnitude shallower. 
    }
    \label{fig:transcience}
\end{figure*}

LSST will discover millions of astrophysical transients over its 10 year survey.  When Via is on-sky (either at MMT or Magellan), instantaneous follow-ups of transients are enabled for almost four months of the year.  Due to the high expected density of transients ($1$--$2$ supernovae brighter than $m_r < 24$ per sq. deg.), Boombox fibers will be devoted to transients and their host galaxies in every Via pointing.  The transient follow-up survey will also devote 400 ehr for targeted follow-up of the most interesting transients from LSST, including gravitational wave (GW) counterparts, young supernovae and precursor events, and unexpected phenomena identified by machine learning.  An interrupt-driven, target-of-opportunity mode will obtain rapid, deep ($m\sim24$ in 1 hr) Boombox spectroscopy of the rarest and most exciting transients.

GW followup, especially of candidates identified through LSST target-of-opportunity observations, will be a priority. When Via is on-sky, we can trigger follow up observations of LSST-identified GW candidates within an hour of a GW alert. Early spectroscopy (in the first hours to day) constrains the geometry and composition of the merger ejecta. Perhaps the most surprising aspect of the double NS merger GW170817 \citep{abbott2017gw170817} was the early ($\lesssim1$\,day) blue emission. This early blue emission may originate from low lanthanide abundance ejecta \citep[e.g.,][]{Drout2017,Villar2017}, or shock cooling emission from a post-merger wind or jet cocoon \citep[e.g.,][]{Nakar2017,Piro2018}.  Boombox spectra obtained $\lesssim12\,$hrs after the merger \citep{arcavi2018first}, would constrain the temperature and kinematics of the emitting material. Early GW170817 spectra showed hour-scale evolution \citep{Shappee2017} that Boombox will allow us to follow.  Depending on the mass ratio, equation of state, and other properties, differing quantities of r-process material with differing lanthanide ratios may be ejected from NS mergers \citep[e.g.][]{Rosswog2013}.  Detailed spectra of many NS merger counterparts are required to elucidate their impact on heavy element nucleosynthesis.  The electromagnetic counterparts to a NS--BH merger have not been observed so far. However, the lack of a neutron star primary leads us to expect weaker post-merger neutrino fluxes, a more lanthanide-rich composition, and a redder spectrum \citep[e.g.,][]{Kasen2015}.  Boombox's wide spectral coverage will be ideally suited to constrain any blue emission at early times that could come from lanthanide-poor or especially hot emission from ejecta during the merger. 

The combination of Rubin discovery and Boombox spectra will be important for the study of tidal disruption events (TDEs).  Measuring BH masses from a wide range of galaxies to higher redshifts allows us to follow BH growth and evolution \citep{Bricman2020}.  The formation and growth of low mass BHs is of particular interest. The lowest mass BHs have the most rapidly evolving light curves \citep[e.g.,][]{Mockler2019}, so Boombox's ability to quickly vet targets and characterize their early-phase spectra is key for understanding these BHs.  We are still trying to understand how stellar material circularizes to form a disk that feeds the BH \citep{Bonnerot2020}. We know disks are forming because late time observations have plateau emission in the optical and UV \citep{vanVelzen19, Mummery2020, Nicholl2024}, but the early phases are typically obscured by material that reprocesses any disk emission \citep{Piro2020}. Combining early spectra from Boombox with X-ray observations and three-dimensional, frequency-integrated and multi-group radiation hydrodynamic simulations \citep[e.g.,][]{Huang2025} will constrain the circularization process.

Fast blue optical transients are among the most exciting and mysterious transients \citep[FBOTs,][]{Drout2014,Pursiainen2018,Ho2023}. These transients have blue colors and short peaks ($\lesssim12\,$days), but they have otherwise heterogeneous properties, including radio, submillimeter, and X-ray emission, and spectral types (e.g., hydrogen versus no-hydrogen). This heterogeneity argues for multiple classes, such as the so-called luminous FBOTs with AT2018cow as the poster child \citep{Prentice2018,Perley2019}.  A wide range of progenitors have been proposed, including: 1) a TDE by an intermediate mass BH \citep{Perets2016,Kremer2021}, 2) collapse of a massive star to produce a BH \citep{Quataert19}, 3) magnetar formation \citep{Margutti19}, 4) electron capture of a merged white dwarf \citep{Lyutikov19}, 5) a shocked disk interaction buried within a supernova \citep{Margutti19}, 6) a common envelope with jets \citep{Soker19}, or even 7) a natal kick sending a compact object into its companion \citep{Tsuna2025}. Many of these scenarios have implications for other types of Boombox transients, such as gravitational wave counterparts and the lowest mass TDEs. Obtaining early, broad-band Boombox spectra is ideal for characterizing the fastest moving material from FBOTs.  These spectra, combined with radio and X-ray observations, will give a more complete picture of the ejecta energetics and geometry to constrain the many models. Boombox will also characterize the galaxy hosts of FBOTs. Many types of supernovae show correlations with galaxy type and environment, providing important clues about their progenitors. Early indications are that the FBOTs occur in less extreme environments than long gamma-ray bursts and Type Ic broad-line supernovae \citep[e.g.,][]{Wiseman2020}. More hosts must be identified to verify this connection, especially if FBOTs arise from a range of scenarios.

Growing evidence indicates that massive stars become increasingly violent at the end of their lives. The evidence includes flash ionized features from circumstellar material illuminated by early supernova emission \citep{Bruch2023}, bright shock cooling emission indicative of dense material recently lost by the massive star \citep{Piro2021}, and even outbursts that precede the main supernova explosion \citep{Ofek2014}. This activity is still not understood, but these studies have far-reaching implications for massive star evolution, the explosive transients they generate, and the types of compact objects they leave behind \citep{Smith2014}. Theories for the late-stage behavior include especially violent convection \citep{Smith2014b}, g-modes propagating energy into the shallow regions of the star \citep{Quataert2012,Fuller2017}, binary interactions that are enhanced as the massive star evolves off of the main sequence \citep{Chevalier2012,Wu2022}, or even energy injection in the stellar envelope that triggers eruptive mass loss \citep{Ko2022}. Another exciting idea is that large scale pulsations impact the outer radius of these massive stars \citep{Bronner2025, Laplace2026}.  The detection and study of these oscillations would provide a new probe of the stellar interior. 

Boombox will be adept at providing early spectra from these supernovae to constrain these models. Measuring the velocities of the flash-ionized features constrains the energetics of the mass loss, and the timing of the disappearance of the flash ionized features measures the radius of the material, constraining the history, and potentially the stages of burning, of the ejected material. The luminosity and temperature measured from the continuum emission during early phases is directly related to the radius of the densest circumstellar material. With early spectra, we probe multiple regions within the ejected material.  With more observations in recent years, these features have been detected in a variety of supernova types  \citep[e.g.,][]{Dong2025}, providing more scope for Boombox exploration. In a few cases, we have directly observed and characterized the pre-explosion activity rather than just inferring it from the supernovae \citep[e.g.,][]{JacobsonGalan2022}. The number of events available for study will increase dramatically with LSST data \citep{Tsuna2023}. Spectra from Boombox will be indispensable for studying these precursors: measuring the kinematics, constraining the outburst energies, and exploring the composition and its relation to the layers ejected from these stars.

Via will allow spectroscopic exploration of the rarest classes of extragalactic transients discovered by LSST. \textit{With anticipated spectroscopic limits comparable to the single-pointing limiting magnitude of LSST}, we will identify novel physics in real time. Via will rapidly identify and capture anomalous transients with sensitivity down to $m\sim24$. We will continue to develop data-driven methods for anomaly detection, combining light-curve, multi-wavelength, and host-galaxy information to quickly identify the most exciting events.  To estimate the rate of anomalous targets available to Via, we model a followup strategy using the expected population of Type I superluminous supernovae from the PLAsTiCC LSST simulation \citep{kessler2019models}. Selecting only events with $m>22.0$, we expect to observe $\sim3200$ luminous anomalies with Boombox over five years.

Target-of-opportunity observations will search for the ``unknown--unknown'' within the LSST datastreams: rare astrophysics that has not yet been observed, or observed only in extraordinarily small numbers. These include pair-instability supernovae, merger-driven supernovae, electron capture events and intermediate-mass black hole tidal disruptions. Recently, machine-learning–based anomaly detection has proven effective at identifying physically rare transients from wide-field alert streams \citep[e.g.,][]{villar2021deep, gagliano2023physics, aleo2024anomaly, gagliano2025evidence}. Given the immense volume of LSST alerts, even phenomena occurring at the one-in-ten-thousand level should be discovered on a monthly basis, provided that we have sufficient sensitivity and flexibility in spectroscopic follow up.

A broad, randomly selected spectroscopic sample of LSST supernovae and their host galaxies reaching the survey limit ($m\sim24$) is essential for unbiased population studies.  An unbiased sample enables relative rate estimates of supernova classes, calibrates photometric classifiers and allows us to understand the dependence of supernova properties on their host environments. Via will build this sample through a passive survey to observe at least one active transient and/or a host galaxy in every pointing of the main survey. In addition to classification of $2500$--$5000$ active (mostly high-$z$) supernovae, Via will target the host galaxies of known or previously discovered transients to provide redshifts and, when feasible, estimates of, e.g., star formation rate and metallicity. We expect $\sim25,000$ host galaxy spectra over the five-year survey. Other potential targets include sources with ambiguous variability, archival transient candidates lacking spectroscopic confirmation, or interesting systems flagged photometrically for further study. This strategy allows Via to build a rich statistical sample across a diverse range of transient-related phenomena, maximizing the impact of each exposure.  In Figure \ref{fig:transcience}, we summarize the key target classes of the \textit{active} transients described above. In target-of-opportunity mode, we will follow GW event counterparts (red); young supernovae and their precursor events (green); and human-vetted, anomalous targets driven by machine learning algorithms (blue). 

Via's time domain program will be complementary to other anticipated multi-fiber, spectroscopic followup campaigns.  The 4MOST Time Domain Extragalactic Survey (TiDES) is expected to commence in 2026 \citep{frohmaier2025tides} on the 4-m class VISTA telescope. TiDES' major goals are threefold: 1) to build a cosmological sample of Type Ia SNe, 2) to followup $\sim10,000$ non-Ia transients with $m<22.5$, and 3) to repeatedly observe $\sim1,000$ active galactic nuclei (AGN) for reverberation mapping studies. TiDES selection of active transients will \textit{not} (in currently reported plans) give any additional weighting to rare events. This strategy is highly complementary to our plans to heavily favor the most exotic phenomena and build a deeper set of ``passive'' supernova observations down to $m\sim24$ mag.

\subsection{Ancillary Science} \label{sec:ancillary}

A wide variety of ancillary science topics will be pursued in parallel to the four key projects using the spare fibers available in each pointing or with dedicated pointings.

\subsubsection{High-Value Ancillary Targets}

High-value ancillary targets have a compelling science case and a moderately low density on-sky ($\lesssim 10$ per sq.~deg).  While key project targets will be assigned priority classes 1 or 2 (\S\,\ref{s:key_projects}), high-value ancillary targets will be given priority classes 3 or 4 in the fiber-assignment procedure, with class 4 reserved for fainter and/or less probable members of the target class. Below we describe opportunities in cosmology and galaxy formation (Lyman-alpha forest, host galaxies of fast radio bursts, gravitational lenses, integrated kinematics of  low-mass galaxies), stars (white dwarfs, metal-poor stars, distant giants, astrometric binaries, exoplanet candidate hosts), comets, and more.

\noindent
{\bf Quasars and the Ly$\alpha$ forest}: The Lyman-alpha forest (Ly$\alpha$) flux power spectrum produces some of the most stringent constraints on the distribution of dark matter, and can constrain dark matter microphysics in ways that are complementary to Milky Way dwarf galaxies and streams \citep{Rauch1998, Irsic17, Garzilli21, Villasenor2023}. Observing the Lyman-alpha forest at high resolution also traces the thermal evolution of the intergalactic medium (IGM), for example revealing heating events triggered by the reionization of hydrogen and helium \citep[e.g.,][]{Hiss2018, Walther2019, Gaikwad2021, Villasenor2022}. In particular, the transition of intergalactic helium from singly to doubly ionized---\ionn{He}{ii} reionization---was driven primarily by hard ionizing radiation from quasars and left observational signatures in the IGM at $z=3-4$ \cite{Worseck2016, Worseck2019}. The temperature history of the IGM over this critical period is currently very poorly constrained, and the morphology of \ionn{He}{ii} reionization, e.g., the bubble size distribution and the spatial scatter in the thermal state of the IGM, is essentially unconstrained.

\begin{figure}
    \centering
    \includegraphics[width=\textwidth]{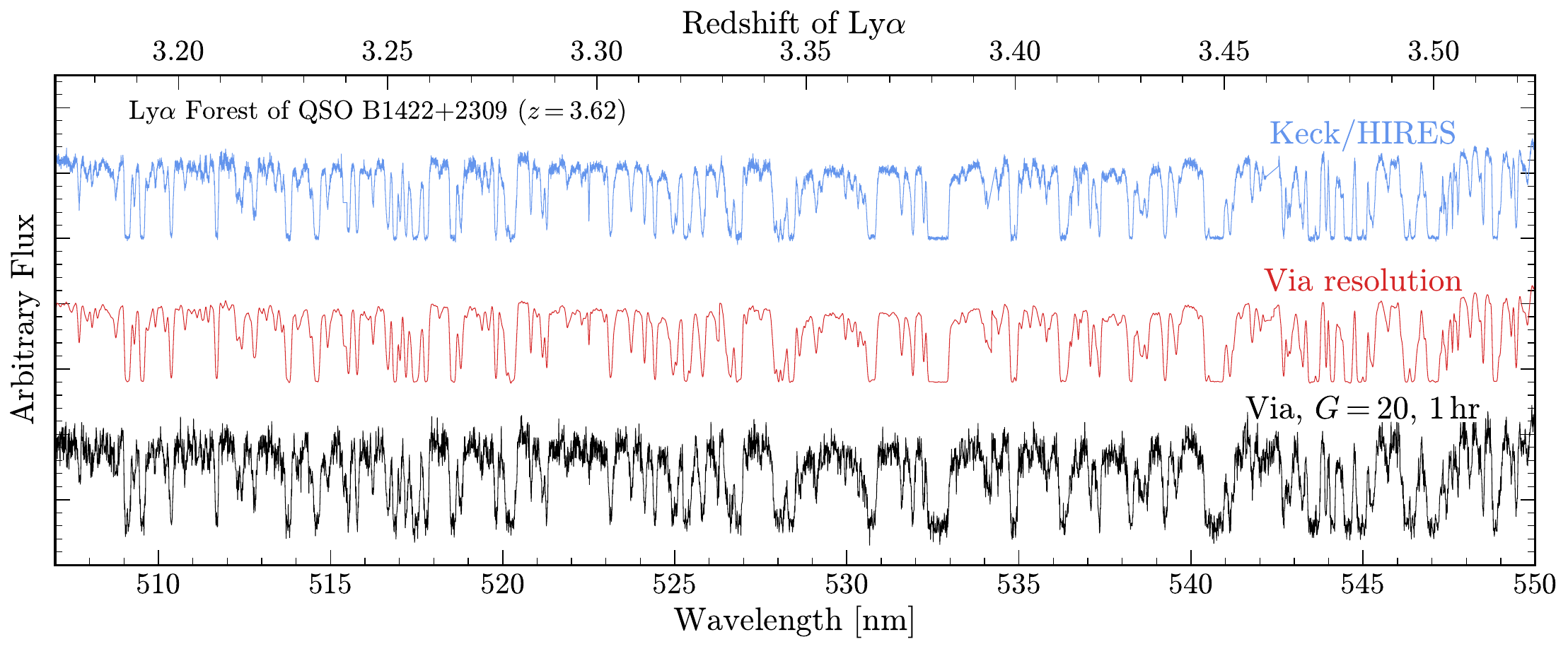}
    \caption{The Lyman-alpha forest of the $z = 3.62$ quasar B1422$+$2309, which covers half the Viaspec wavelength range.
    The top spectrum is an observation by the $R \approx 48{,}000$ Keck/HIRES instrument (top; \citealt{Songaila1996}). 
    The middle spectrum is smoothed to the Viaspec resolution, and the bottom spectrum is ``observed'' for 1 hour in dark time with the full Viaspec data simulator. 
    }
    \label{fig:lya_spectra}
\end{figure}

Existing high-resolution measurements are limited to less than 30 UV-bright quasars in the \ionn{He}{ii} Ly$\alpha$ forest \cite{Worseck2019} and several hundred high resolution objects that probe the HI regime, with a complex selection function that favors exceptionally luminous quasars and those in over-dense regions \citep{Karacayli22}. The Via wavelength range probes the Ly$\alpha$ forest with quasars beyond $z > 3.2$ (Figure~\ref{fig:lya_spectra}). An all-sky catalog of quasars based on \gaia\ DR3 now exists and contains $20$--$30$ quasars per sq. deg to $G=20.5$ \citep[Quaia,][]{Storey-Fisher23}. We expect that an even more robust catalog is possible with \gaia\ DR4, expected in time for the Via Survey.  Based on the Quaia catalog, we expect to fortuitously observe one quasar in this redshift range in every second Via pointing, enabling a well-defined sample of $\sim 1500$ high-resolution quasar spectra by the end of the survey, an order-of-magnitude increase over past surveys at high resolution \citep{OMeara2015, Lopez2016, Murphy2019}.

Via's combination of resolution, sample size, and full-sky access thus uniquely positions it to characterize \ionn{He}{ii}  reionization through four complementary measurements of the IGM's thermal state and spatial structure: (i) the global temperature evolution, pinning down the timing of when \ionn{He}{ii} reionization completes (ii) the slope of the temperature--density relation, which provides an independent timing constraint (iii) the line-of-sight variance of the temperature, a direct measurement of the patchiness of \ionn{He}{ii}  bubbles, and (iv) the transverse coherence of the Ly$\alpha$ forest across paired quasar sightlines, which constrains the characteristic bubble size.

Figure~\ref{fig:lya_spectra} shows the spectrum of QSO~B1422$+$2309 at $z = 3.62$ observed at $R \sim 48{,}000$ with the Keck/HIRES instrument, over approximately half the Via wavelength range.  We also show this spectrum degraded to the Via resolution of $R \approx 15{,}000$, and ``observed'' for $1$~hr through our data simulator at a representative magnitude of $G = 20$, achieving SNR~$\approx 10$ per pixel. Due to the typical broadness of Ly$\alpha$ features at these redshifts, the spectrum is minimally degraded at Via resolution and SNR, and the key features of the Ly$\alpha$ forest are preserved. 

\begin{figure}
    \centering
    \includegraphics[height=7cm]{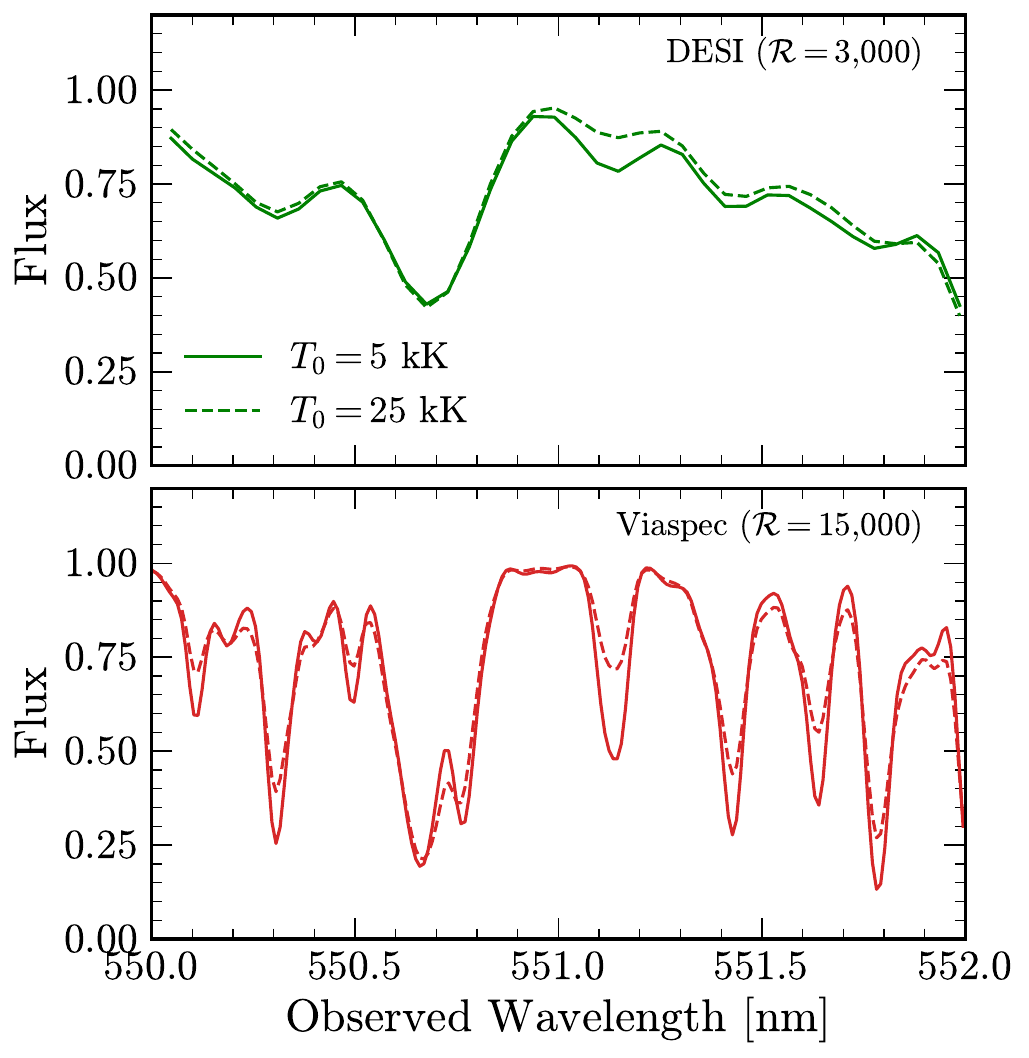}
    \includegraphics[height=7.1cm]{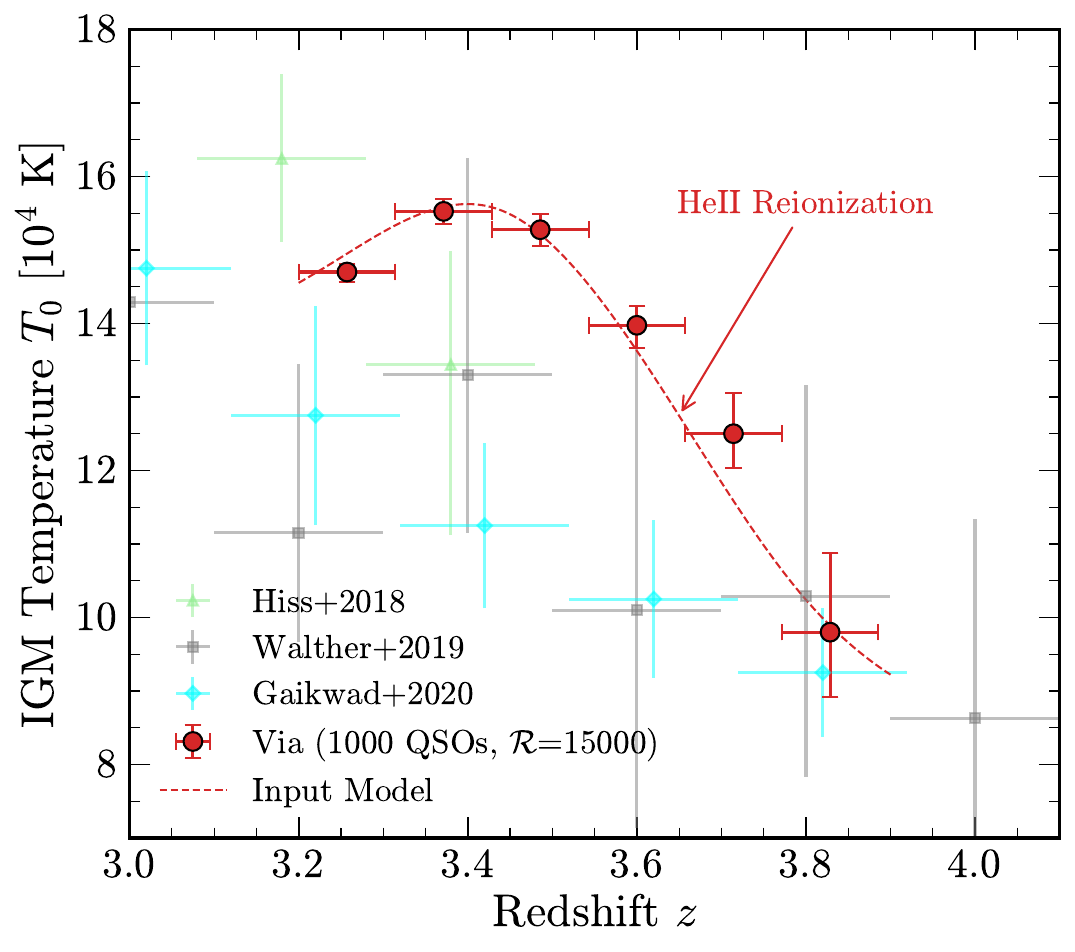}
    \caption{Left: Simulated Lyman-alpha forest spectra at $z = 3.1$ for two different IGM temperature parameters $T_0$, as observed by a low-resolution survey like DESI (top) and at Viaspec resolution (bottom). Right: Results from fitting a mock Via survey of 1000 quasars drawn from the Quaia quasar catalog. This mock survey recovers the evolution of the IGM's thermal state from $z = 3$--$4$, in this case recovering the temperature peak induced by the reionization of \ionn{He}{ii}.}
    \label{fig:lya}
\end{figure}

To quantify Via's sensitivity to the Ly$\alpha$ forest, we generate and fit mock spectra using our data simulator. The left panel of Figure \ref{fig:lya} illustrates the sensitivity of Viaspec spectra to the thermal state of the IGM. Power spectra are generated at two values of the IGM temperature at the mean density $T_0$ at $z = 3.1$ using the LaCE emulator \citep{Cabayol-Garcia2023}, and mock spectra are created using the lognormal formalism described in \cite{McDonald2006} and \cite{Karacayli2020}. We show corresponding spectra for low-resolution surveys like DESI in the top panel, and for Viaspec in the bottom panel. Viaspec can resolve the thermal state of individual IGM features, enabling a measurement of the IGM's thermal evolution over cosmic time. There are diminishing returns at resolutions higher than $\approx 10,000$, since the intrinsic velocity dispersion of the IGM becomes dominant \citep{Villasenor2022}. Some surveys have operated at even higher resolution using the Keck HIRES and VLT X-Shooter instruments, but these surveys are limited to a few hundred bright quasars in this redshift range \citep{OMeara2015, Lopez2016, Murphy2019}. 

To estimate the sensitivity of the Via survey to the IGM's temperature evolution, we adopt a $T_0(z)$ relation from \cite{Villasenor2022} and produce a mock survey of Viaspec spectra for $1000$ quasars drawn from the Quaia catalog at $z \gtrsim 3.2$. This is the expected number of quasars we might observe ``for free'' during the regular survey, given their source density on-sky. At each redshift and $T_0$ value, the intrinsic ``true'' power spectrum is generated with the LaCE emulator \citep{Cabayol-Garcia2023}. The mock Ly$\alpha$ forest spectra incorporate the Viaspec resolution, pixel sampling, and typical signal-to-noise ratios. We then estimate the Ly$\alpha$ flux power spectrum in redshift chunks across the survey, and fit these power spectra over the LaCE emulator grid to determine the $T_0$ measurement (and its uncertainty) as a function of redshift. 

The right panel of Figure~\ref{fig:lya} summarizes the results from this mock survey. Previous measurements using high-resolution spectra are shown for comparison \citep{Hiss2018, Walther2019, Gaikwad2021}. By conducting an unbiased high-resolution survey of over an order of magnitude more quasars in this redshift range, Via will produce leading measurements of the IGM's thermal evolution and its spatial structure over this crucial phase in cosmic time. By measuring the physical parameters that govern the IGM's evolution, this survey will also help disentangle IGM physics from the underlying physics of dark matter across all redshifts \citep[e.g.,][]{Irsic17, Villasenor2023}. 

Apart from this primary redshift-selected Ly$\alpha$ survey, Via will observe all bright candidate quasars as filler targets in each field. High-resolution spectra of quasars have several science applications. For example, quasars will serve as valuable backlights for the cold gas science case described in \S\ref{s:coldgas}, probing the cold gas content in the outskirts of the Galactic halo.

\begin{figure}[t!]
    \includegraphics[width=\linewidth]{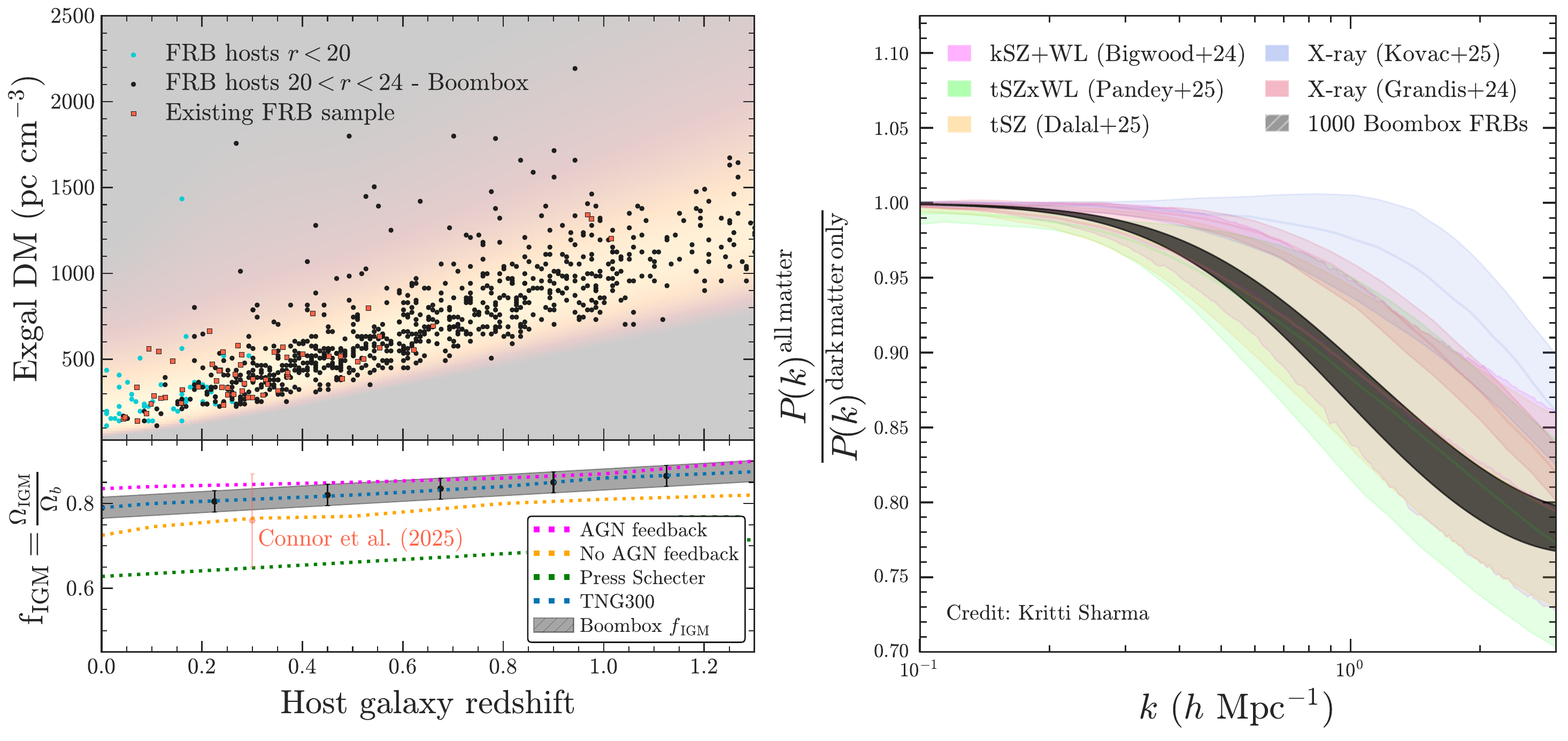}
    \caption{The impact of 1000 fast radio burst host galaxies measured by Boombox. Black points on the left figure shows a simulated distribution of extragalactic dispersion measures vs. redshift for FRB host galaxies measured by Boombox, with $20<r<24$ and $0<z<1.3$. The bottom panel shows the corresponding constraints on the IGM and its redshift evolution. The right figure shows constraints on the baryon power spectrum from the Boombox sample, compared with current constraints from other probes of cosmic gas.}
    \label{fig:frbfig}
\end{figure}

\noindent
{\bf Host Galaxies of Fast Radio Bursts}: Fast Radio Bursts (FRBs) are short-duration radio pulses from the distant Universe whose origin remains a key mystery in transient astrophysics. Upcoming radio telescopes such as the Deep Synoptic Array (DSA), CHORD, and the VLBI extension of CHIME will detect as many as 10,000 FRBs per year. Follow-up spectroscopy of FRB host galaxies is required to measure their redshifts, discover the nature of FRBs, and use them as a cosmological probe of baryons that are otherwise difficult to detect. FRBs are distributed randomly across the sky so Via's Boombox will be able to obtain redshifts of FRB host galaxies in any survey pointing. We will also place spare fibers on foreground galaxies whose CGM is intersected by the FRB, contributing to excess dispersion measure (DM) and radio scattering. By modeling the empirical host galaxy brightness distribution of FRBs, we estimate that Boombox could deliver 1,000 FRB host galaxies with $z \lesssim 1.3$ and $r < 24$, over five years. These redshifts will be critical to uncover the physical origin of FRBs and to apply FRB observations to cosmology and extragalactic astrophysics. In Figure~\ref{fig:frbfig} we show a simulated DM/redshift relation for 1000 Boombox FRB host galaxies (left) as well as the matter power spectrum normalized by the dark matter only power spectrum, with recent constraints from other cosmological probes (right). FRB constraints in the dark shaded region assume a mapping between line-of-sight dispersion measure variance and the baryon power spectrum, as proposed by \citet{Sharma2025}. Combined with the kSZ, a large sample of localized FRBs will break the optical-depth degeneracy in cosmology \citep{madhavacheril2019} and constrain feedback. Via's significant time on sky, sensitivity, and flexible fiber orientation will enable unique FRB science.

\noindent
{\bf 
Gravitational Lenses}:  Rubin Observatory's LSST and the Roman High Latitude survey will contain thousands of galaxy-scale lensed quasars, $\sim$100 lensed supernovae, and tens of thousands of galaxy--galaxy Einstein Rings. Most of the associated science benefits strongly from having well-measured lens and source redshifts, and two high priority cases additionally require the deflector stellar velocity dispersion as an independent constraint on the lens mass: 1) time-delay and distance-ratio cosmography for dark energy, and 2) dark matter substructure in massive lens galaxies. Assembling this sample is a large enterprise to which DESI,  the 4MOST Strong Lensing Spectroscopic Legacy Survey (4SLSLS; \citealt{collett2023}), Subaru/PFS \citep{takada2014}, and other instruments will contribute. At radio wavelengths, the DSA will add a largely independent population of $\mathcal{O}(10^4$--$10^5$ lenses, including time-domain AGN whose radio time delays and VLBI astrometry make them percent-level $H_0$ targets \citep{mccarty2025} for which Via need only supply the optical redshifts and dispersions. Via will observe all of these as low urgency, high-value ancillary targets with no cadence requirement:  a single dual-camera Boombox pointing yields the source redshift and, where the deflector is not outshone by the lensed images, the lens redshift, in one or two fiber placements, reaching fainter and higher-redshift deflectors than the 4m surveys. At $\sim 1$ lensed quasar per 10 sq deg and a few Einstein rings per sq deg, the survey footprint will deliver redshifts for hundreds of lensed AGN and thousands of ring detections; the rarer lensed SNe (less than 1 per 1000 sq deg, $\sim$ 10 discoveries expected per year) can drive dedicated pointings.

Via's distinctive contribution is one of scale, uniformity, and sky coverage. For the tens of thousands of AGN-free Einstein rings, Viaspec delivers clean single-aperture $R \sim 15,000$ stellar velocity dispersions in both hemispheres,  building a far larger and more homogeneous sample of deflector kinematics than targeted campaigns can reach, and anchoring the population-level mass modeling that underlies the substructure, stellar-mass--halo-mass, and cosmographic analyses. For lensed quasars a fiber cannot separate the faint deflector from the bright lensed images, so their deflector dispersions---and the resolved kinematics needed to break the internal mass-sheet degeneracy for precision cosmography---come from dedicated resolved spectroscopy elsewhere; for these systems Via instead contributes the source and lens redshifts and ties them into the larger sample. That same large single-aperture sample can also calibrate the velocity-dispersion scale against the smaller resolved-kinematics samples, controlling the leading systematic for time-delay cosmography---where any fractional error in the deflector velocity dispersion propagates into the inferred $H_0$ at twice its size.

\begin{figure}
    \centering
    \includegraphics[width=\linewidth]{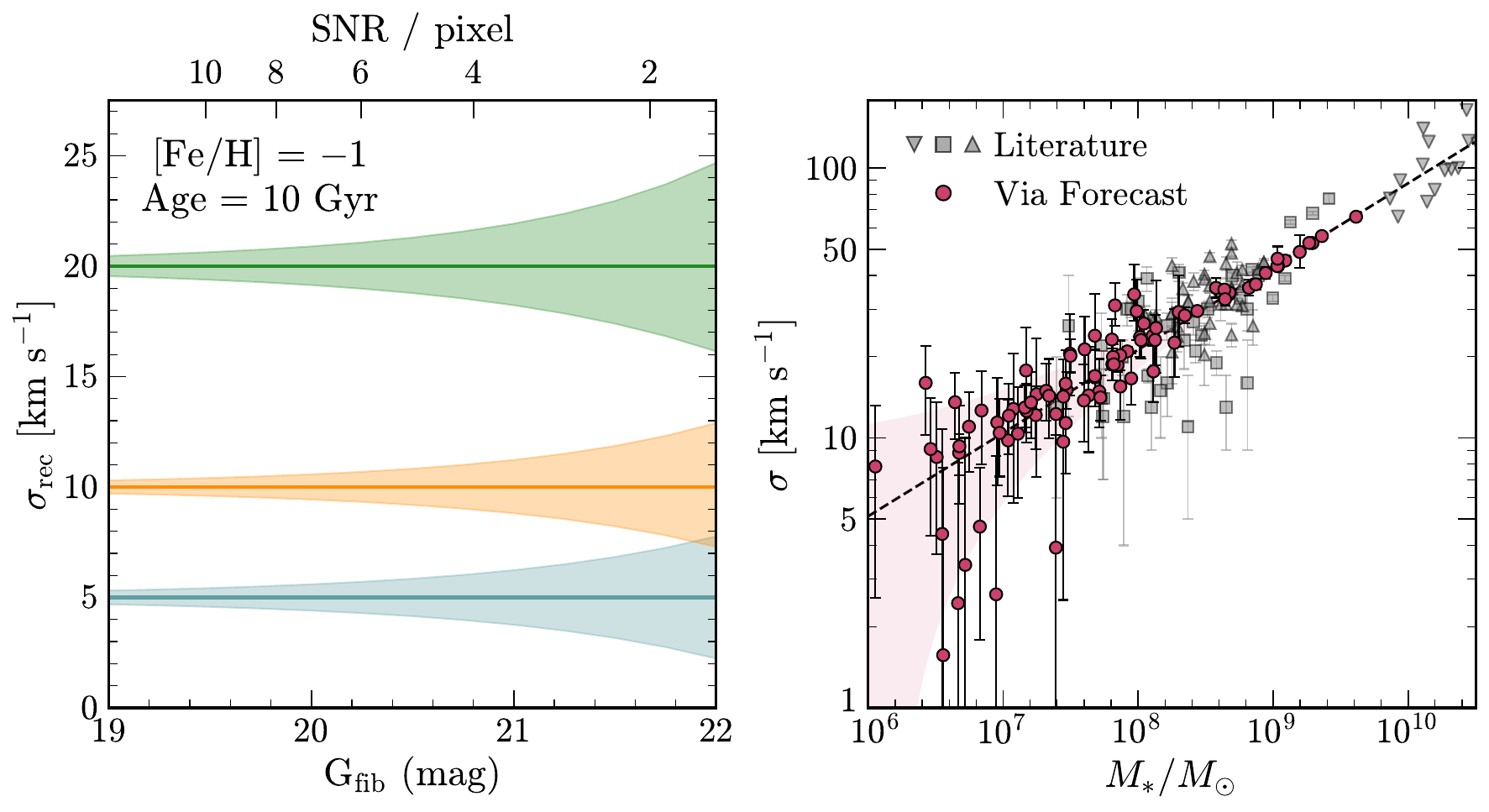}
    \caption{Left: Recovered velocity dispersion as a function of  fiber magnitude $G_{\rm fib}$ for galaxies with metallicity [Fe/H] = -1, age of $10$ Gyr, and input dispersions $\sigma = 5,10,$ and $20$~km\,s$^{-1}$. Shaded regions denote the 68\% confidence intervals of the recovered values. The upper axis shows the corresponding signal-to-noise ratio per pixel for a 1-hr exposure in dark time and nominal observing conditions. Right: Stellar velocity dispersion as a function of stellar mass. Pink points show the Via forecast for one third of the dwarf galaxies in the Fornax cluster from \citet{Venhola.Fornax.dwarfs.2018}. Grey points show existing literature measurements from Fornax \citep[squares;][]{SAMI.FORNAX.2022} and Virgo \citep[triangles;][]{Virgo.vdisp.2014} and the SAURON project \citep[inverted triangles;][]{SAURON.etal.2011}; the dashed line is a linear fit to these data. The Via forecasts assume no intrinsic scatter in this relation; the shaded pink region shows the median 68\% uncertainty in Via measurements, accounting for the fiber magnitude distribution. Via will extend these measurements to lower stellar masses and smaller dispersions than previous surveys, and more precisely measure the scatter at higher masses.}
    \label{fig:exgaldwarfs}
\end{figure}

\noindent
{\bf Integrated kinematics of low-mass galaxies}: Viaspec is an ideal instrument for the internal kinematics of dwarf galaxies at $z \lesssim 0.1$, measuring the broadening of absorption and emission lines in integrated-light spectra (Figure~\ref{fig:exgaldwarfs}). Velocity dispersions can be recovered to $\approx 0.5$ times the spectrograph velocity resolution when the LSF is well characterized \citep{Cappellari17}, as demonstrated for ultra-diffuse galaxies with MUSE \citep{Emsellem19}, reaching $\sigma \gtrsim 4\,\kms$---well below existing faint-object spectrographs. Drawing on the DESI DR1 dwarf catalog as the input list \citep{Manwadkar26}, there are $\sim 10\,\deg^{-2}$ galaxies with $M_\ast < 10^9\,\msun$ brighter than $G_{\rm fib} = 21$, and the final sample will include many thousands spanning isolated field systems to dense clusters. For the partially rotation-supported systems common at these masses, a single integrated aperture captures the total dynamical support $S_{0.5} \equiv (\tfrac{1}{2}V_{\rm rot}^2 + \sigma^2)^{1/2}$ \citep{Kassin07}, a clean virial-mass proxy across the rotation- to dispersion-supported transition. The primary goal is to measure the scatter in dynamical mass at fixed stellar mass and its dependence on environment, which constrains the stellar mass--halo mass relation and the efficiency of the baryon cycle. Scatter in dynamical mass and its environmental dependence are also sensitive to dark matter self-interactions, which drive a diverse, potentially bimodal distribution of central densities through gravothermal core formation and collapse \citep[e.g.,][]{Kaplinghat19, Nadler20}. The large sample will also uncover rare outliers with extremely low or high dark matter content, constraining alternative dark matter models.

This sample also enables a dynamical test of modified gravity. In chameleon-screened theories such as Hu--Sawicki $f(R)$, main-sequence stars are self-screened \citep{Hui09}, so a Viaspec stellar velocity dispersion is a Newtonian dynamical-mass anchor insensitive to screening, whereas the gas and dark matter in an \emph{unscreened} dwarf feel an enhanced coupling of up to $\Delta G/G = 1/3$ \citep{Jain11, Vikram13}. Comparing the screened stellar dispersion against the unscreened gas kinematics---\ionn{H}{i} line widths, or for star-forming systems the nebular lines in the \emph{same} Viaspec fiber---therefore probes a fifth force directly. The signal is cleanest for the most isolated, lowest-mass dwarfs ($M_\ast \lesssim 10^{7}\,\msun$) in the local volume ($D \lesssim 40$~Mpc, where \emph{Roman} can also provide surface-brightness fluctuation distances), which remain unscreened down to $f_{R0} \sim 10^{-8}$, the regime of the strongest current astrophysical bounds \citep{DesmondFerreira20}. A sample of a few tens of such well-characterized systems would yield an independent screening test with systematics orthogonal to the morphological constraints that dominate today.

\noindent
{\bf White dwarfs}: \gaia\ DR3 parallaxes have revolutionized the selection and study of white dwarfs (WDs), dense stellar remnants that represent the final evolutionary state of most stars.  There are $\sim 5$ \gaia-selected WDs per sq. deg. at $16<G<20$ \citep{GentileFusillo21}.  Between $20$--$50\%$ of WDs show metal pollution in their atmospheres \citep{Koester2014}.  Due to the rapid diffusion timescale of WDs, this implies the ongoing accretion of planetary debris \citep[see][for a review]{Farihi2016}.  Polluted WD spectra provide the gold-standard technique to measure the bulk composition of exoplanets.  A wide variety of metal lines including \ionn{Mg}{i}, \ionn{Fe}{i}, \ionn{Cr}{i}, \ionn{Ca}{i}, \ionn{Na}{i} will enable the detection and characterization of metal pollution in a large, unbiased survey of WDs.  Via will also characterize unpolluted WDs with helium-dominated atmospheres via the strong \ionn{He}{i} 587.6 nm line.  Since many Via fields will have repeat visits separated by long baselines, Via will also identify rare WD+WD binaries, including potential progenitors of Type Ia supernovae \citep{Maoz2014, Rebassa-Mansergas2019}.  The choice of whether to observe WDs with Viaspec or Boombox, or a combination of the two, requires further study; each spectrograph offers distinct advantages.

\noindent
{\bf Metal-poor stars}: Viaspec will obtain high-resolution spectra for photometrically selected metal-poor stars. As chemical fossils of the first stellar generations, metal-poor stars encode the earliest epochs of nucleosynthesis and Galactic assembly, and are a primary near-field probe of Population III enrichment and the build-up of the proto-Galaxy \citep{BeersChristlieb05, FrebelNorris15}. Viaspec's detailed abundances ([Fe/H], [$\alpha$/Fe], C) combined with precise multi-epoch radial velocities will yield a large, homogeneous sample to map early chemical evolution and---by joining chemistry to kinematics---separate the in-situ component of the Galaxy from its accreted building blocks \citep{Belokurov2022, Rix2022}.
\gaia\ DR3 has already delivered all-sky samples of metal-poor stars down to $G \approx 17$ and will extend to $G \approx 18.5$ in \gaia\ DR4 \citep{Andrae2023}.  In the southern hemisphere, we will target stars from the MAGIC survey, a narrow-band photometric survey that has $\approx10$ stars per Via FoV with $\varpi < 0.5$ mas to $G=20$ with estimated [Fe/H] $\lesssim -2.0$ \citep{Chiti26}.

\noindent
{\bf Distant giants}: 
Luminous red giants can be observed spectroscopically throughout the Milky Way's outer halo, where they bear on three questions central to near-field cosmology. First, they are premier dynamical tracers of the total mass and mass profile of the Galaxy at large radii---still uncertain at the factor-of-two level \citep{Deason2021, Bird2022}---which in turn sets the expected abundance of the subhalos and satellites that the SPS and DGS aim to measure. Second, their combined metallicities and velocities chemically tag accreted debris and map the density, anisotropy, and metallicity structure of the halo beyond 100 kpc, tracing the Galaxy's accretion history \citep{Conroy2019, Bird2021}. Third, distant halo stars reveal the Galaxy's live response to the infalling LMC---a reflex motion and trailing dark-matter wake \citep{PetersenPenarrubia2021, Conroy21} whose amplitude constrains the LMC mass, the Galactic potential, and even the nature of dark matter \citep{Foote2023}, and which reshapes the same stream population targeted by the SPS.
Upper-red-giant-branch stars can be accurately identified from broadband photometry \citep[e.g.,][]{Majewski03, Conroy18a, Chandra23a} and are detectable beyond 100 kpc at $G<18$. 
Building on the H3 Survey \citep{Conroy2019}, which mapped northern halo giants to $G\approx18$, Via will reach $G\approx
21$ across both hemispheres---essential, since the LMC wake and the halo's collective response are all-sky signals. We expect a few distant giants per Via FoV, yielding a homogeneous, all-sky sample of thousands with precise metallicities and radial velocities. Rare tracers such as RR Lyrae and blue horizontal branch (BHB) stars may also be observed with Boombox fibers, extending what is possible with surveys that only measure spectra for the bright end of these populations \citep{Medina2025, Bystrom2025, Feng2026}.

\noindent
{\bf \gaia\ astrometric binaries}: \gaia\ DR4, scheduled for release in December 2026, will deliver epoch-level astrometry for billions of sources.  This will enable the identification and characterization of millions of binary stars via the apparent motion of the photocenter of the unresolved binary. 
However, independent measurements are needed to validate the astrometric orbits, which are sometimes incorrect for no easily discernible reason \citep{Simon:2026}.
A few ground-based radial velocities (from e.g., Via) can confirm the \gaia-derived orbits.

One particularly exciting group of binary stars are those with compact object companions (white dwarfs, neutron stars, and black holes).
\gaia\ has already revealed a few black holes \citep{El-Badry:2023a, Chakrabarti:2023, El-Badry:2023b, Tanikawa:2023, Panuzzo:2024} and neutron star candidates \citep{El-Badry:2024}.
From these objects, there is evidence for a mass gap between neutron stars and black holes, as well as a strong relationship between low-metallicity and massive compact object formation. 
With at least an order of magnitude more detections to come in \gaia\ DR4, these emerging trends can be confirmed, and deliver novel constraints on the formation and evolution of compact object binaries.
Via can play an important role in the characterization of these binaries.
In addition to orbit confirmation mentioned above, high-quality spectroscopic follow-up is crucial for confirming the dark nature of the companion, ruling out a second luminous star, and characterizing the luminous stellar companion (e.g., surface gravity, mass, and metallicity). 

At the opposite mass range, \gaia\ is also sensitive to planets orbiting stars \citep{Sozzetti:2023,Stefansson:2025}.
Astrometric planet detection is particularly interesting because it is sensitive to much wider orbits than pure RV or transit searches.
In addition to the reasons listed in the previous paragraph, the velocity precision of Via will be very helpful for these binaries, as the RV semi-amplitude of these systems will typically be $\lesssim 1\, \kms$.

\noindent
{\bf  Convection, Pulsation, and Mass loss in Evolved Stars:}  Red giant and supergiant stars have extended envelopes with pulsations and vigorous convective flows, causing periodic and stochastic variations \citep{Trabucchi21}. Numerous puzzling properties of these stars can be studied with high-precision time series spectroscopy.

We have known for a century that many red supergiant stars display periodic or quasi-periodic variability on timescales longer than any known pulsation mode \citep{OConnell33, Payne-Gaposchkin54}.  Proposed mechanisms include unknown pulsation modes or binary companions  \citep{Wood04, Soszynski21, Goldberg24, MacLeod25}. Both mechanisms predict radial velocity variation, but with different phasing between the radial velocity and photometry \citep{Goldberg24}. High spectral resolution allows monitoring of photospheric and circumstellar lines to trace the physical origin of the variations \citep{Dupree2026}. 

Stochastic variability in giant stars is linked to the large-scale convective cells that define their envelopes. Convective motions are thought to sweep outward and shape a dense, multi-phase extended atmosphere around red supergiant stars \citep{Fuller24, Ma25}. Time series of high-resolution spectra can trace the emergence of plumes through atmospheric layers \citep{Kravchenko2021}. The properties of circumstellar matter relate to the observable appearance of early-time core-collapse supernovae light curves \citep{Tsuna25} and the emergence of dust-driven winds. Whether this mass is episodically launched in large bursts, perhaps like that observed in Betelgeuse’s “great dimming” event \citep{Dupree22}, or from a more-continuous percolation pertains directly to how these stars shed their envelopes and return mass to the interstellar medium. The power spectrum of variability is a powerful probe of the relative significance of stochastic variation on different timescales. \citet{Kiss06} first identified a distinct 1/frequency power spectrum in the photometric variability of red supergiant stars, with evidence for peaks at the fundamental mode and radial overtones of these stars. Extending this analysis to radial velocities of particular lines in giant star spectra will directly trace physical motions and allow a point of direct comparison for numerical simulations \citep{MacLeod25,Ma25}.

\begin{figure}[t!]
    \centering
    \includegraphics[width=\linewidth]{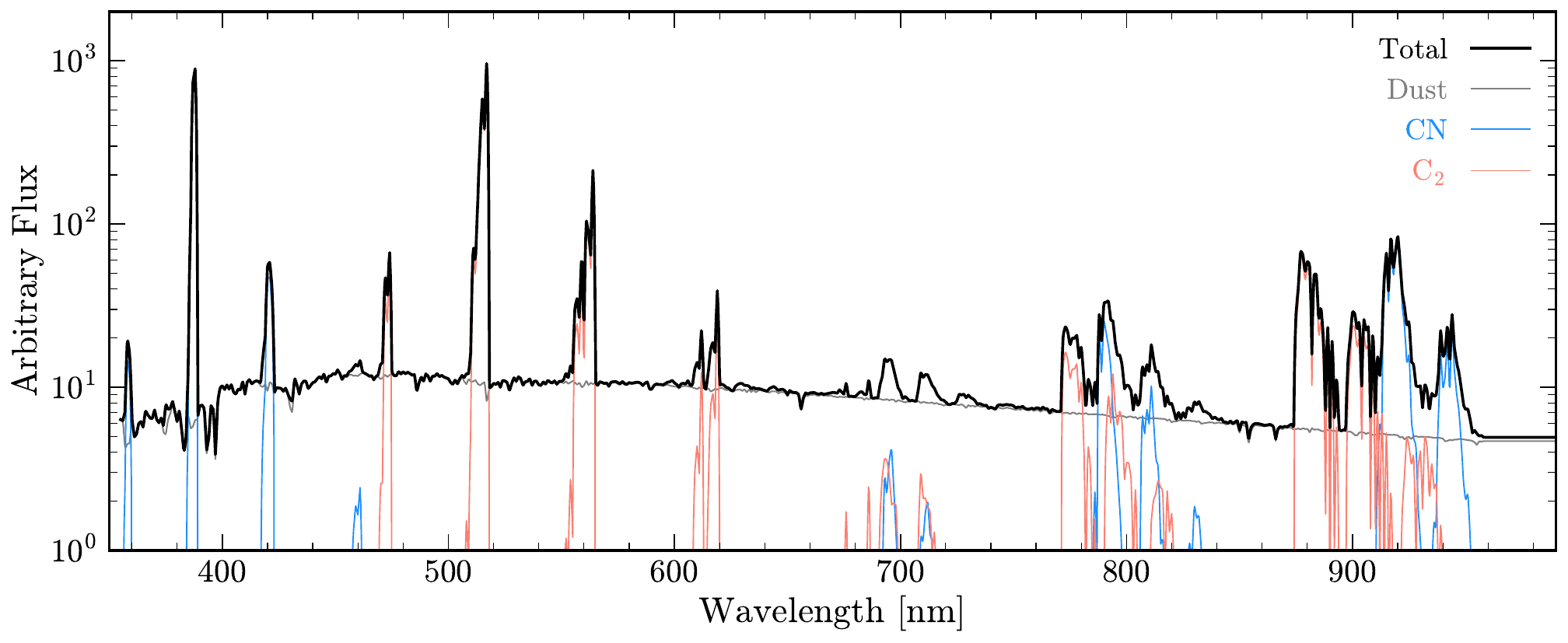}
    \caption{Simulated spectrum tuned to match Comet C/2014 Q2, generated with the NASA-GSFC Planetary Spectrum Generator \citep{Villanueva2018}. 
    Emission lines from three different carbon-bearing molecules are highlighted.}
    \label{fig:comet}
\end{figure}

\noindent
{\bf Bright Exoplanet Candidate Host Stars}:
Over the next few years, we expect the discovery of thousands of new exoplanets. Most new exoplanets around bright stars will come from the TESS \citep{Ricker2015} and PLATO transit missions \citep{Rauer2025}, and from \gaia\ DR4's astrometric planet search \citep{Perryman2014, Lammers2026}.  A subset of Via fibers will be allocated to high-value exoplanets and exoplanet candidates for rapid, homogeneous {\it reconnaissance} spectroscopy.  The RV requirement for triage of exoplanet candidates from these missions is modest: multi-epoch RVs at the $\kms$ level are enough to flag the dominant contaminants (SB1/SB2s, obvious EBs, hierarchical triples), to empirically calibrate false-positive rates, and to deliver uniform stellar parameters for the surviving systems.  Via performance estimates suggest that even in bright time, a 1\,hr pointing can deliver $\lesssim 1\kms$ uncertainties at $G \lesssim 17$, and correspondingly better precision for the typically brighter TESS, PLATO, and \gaia\ targets.  This puts Via squarely in  the regime needed for eclipsing and astrometric binary rejection.  The various exoplanet samples will benefit from Via's sky coverage.  \gaia\ will have an enhanced yield in the ecliptic-latitude bands where its scanning was densest.  The PLATO LOPS2 field \citep{Nascimbeni2025} is easily observed from Magellan/Clay.  The relevant TESS detections will be the large number of giant planet candidates around stars with 13$<$$T$$<$16 in the galactic plane \citep[e.g.,][]{Kunimoto2022}.  Via will help convert severely contaminated candidate lists into ``validated planet'' samples usable for demographics and for prioritizing scarce high-precision RV resources.

\noindent
{\bf Comets}: LSST will identify millions of rapidly moving objects in the solar system.  Within this class of objects, Via/Boombox observations of comets would be the most interesting, as comets display strong and broad emission features across the optical spectrum due to CN, C$_2$, C$_3$, CH, NH$_2$, \ionn{O}{i}, and \ionn{Na}{i}, as well as numerous molecular ions (see Figure \ref{fig:comet} and \citealt{Feldman04}).  Optical spectroscopy places important constraints on the abundances of these species.  Comets can be classified into two broad categories: short-period and long-period (LPC), with the conventional dividing line at a period of 200~yr.  LPCs have spent the past 4.5 Gyr in the Oort Cloud, and were only recently dynamically injected into the inner solar system.  The LPCs are therefore uniquely interesting, as they are relatively pristine relics of the early solar system.  LSST is predicted to discover $\sim10,000$ comets over 10 years \citep{Jones20LSST}, most of which will be visible for months and will have modest sky motions (a few arcseconds per hour).  There are $\sim1000$ currently known LPCs, and less than 100 with optical spectra.  The most mysterious rapidly moving bodies are the interstellar objects, of which only three have been discovered to-date.  Predictions are very uncertain, but LSST may discover 10 or more interstellar objects over the course of its survey \citep{Rice19, Jones20LSST, Hoover22, Marvceta23,2025arXiv250713409C}. We will observe comets in Via fields as high-value ancillary targets, increasing the sample of comet optical spectra tenfold and allowing the identification and characterization of different compositional classes of objects.  In addition, we will trigger, in ToO mode, on the dozen or so most exciting LSST rapidly moving objects.

\begin{figure}[t!]
    \centering \includegraphics[width=\textwidth]{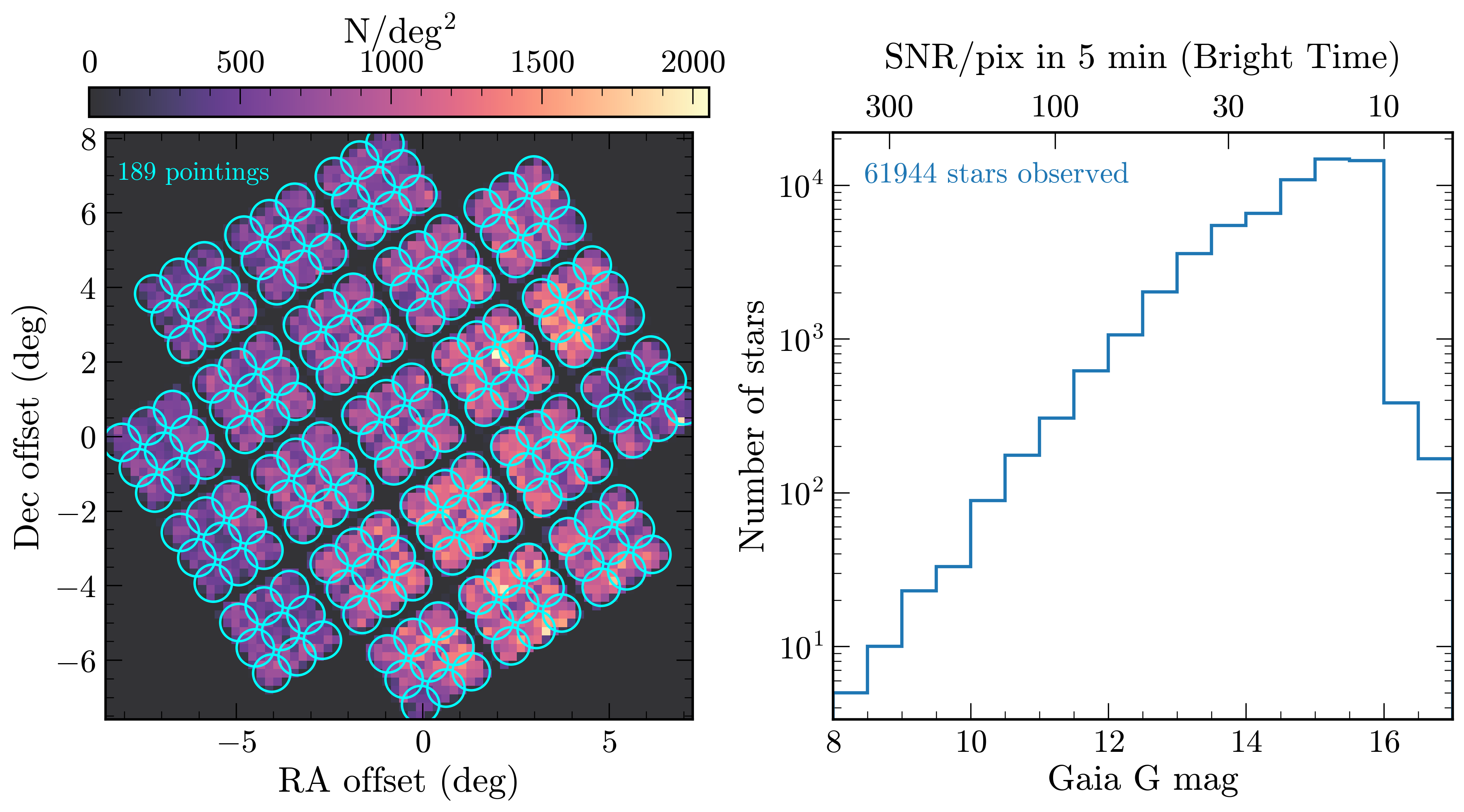}
    \caption{Simulated Via program to observe the Kepler field.  The entire field can be observed in 189 separate Via pointings (left panel), enabling uniform spectroscopic coverage over $\approx100$ sq. deg.  The right panel shows the number of stars observed as a function of $G$-band magnitude and SNR.  The latter was estimated from the Via ETC assuming 5m exposures in bright moon conditions.  Assuming 5m overheads, a single pass over the entire Kepler field can be completed in $3$--$4$ nights during bright time.}
    \label{fig:kepler}
\end{figure}

\noindent {\bf Brown dwarfs and hot Jupiters}: The mass distribution of brown dwarfs ($\approx13-80M_J$) orbiting stellar companions shows a clear deficit of objects in the range $35$--$55M_J$, known as the brown dwarf desert \citep[e.g.,][]{Ma14, Stevenson23}.  Existing sample sizes are small and are limited to bright stars with near solar metallicity.  Multi-epoch radial velocities from Via will be sensitive to the mass function of objects from $\approx10-100 M_J$. A $10M_J$ object orbiting at a 10 d period has a semi-amplitude of $1\kms$ when orbiting a $1\msun$ star.  Multi-epoch observations with Via will provide unique constraints on the brown dwarf desert and the metallicity dependence of the brown dwarf population.  Likewise, hot Jupiters, which have short periods ($\lesssim 10$ d) and relatively large semi-amplitudes ($\gtrsim 100\ms$), are found around $\approx1$\% of FGK dwarfs \citep{Wright12}.  Their population at low metallicity is poorly constrained, but would be informative for planet formation models \citep[e.g.,][]{Fischer05, Ghezzi18, Osborn20, Boley2021}.  Via will provide strong constraints on the frequency of hot Jupiters at low metallicity ([Fe/H]\,$\lesssim-1$).  The formation channels for brown dwarfs and hot Jupiters predict opposite trends with metallicity: the core accretion theory predicts a positive metallicity correlation with occurrence rate for brown dwarfs, while gravitational instability predicts a negative correlation for Jupiters.  This prediction has been tentatively verified \citep{Schlaufman18}.  Via observations of a large, uniform sample of systems containing brown dwarfs and Jupiter-mass planets will provide a strong test of this prediction.

\noindent
{\bf Diffuse interstellar bands}: Diffuse interstellar bands (DIBs) are common throughout the optical and NIR spectral range \citep{Herbig75, Hobbs08, Zasowski15, Tchernyshyov2018, Saydjari2023}. 
Their origins are uncertain but their strength appears to correlate with dust extinction column densities \citep{Herbig95, Saydjari2026}.  
The three strongest DIBs in the Viaspec wavelength range are the 578 nm, 579.5 nm, and 584.9 nm lines. 
These DIBs are relatively weak in the Via wavelength range, but will offer an important complement to the cold gas tomography provided by \ionn{Na}{i}.  
For example, in dense regions where column densities are high, combining information from DIBs and \ionn{Na}{i} will allow us to map the correlation between cold gas and dust masses.

\noindent
{\bf Magnetic activity}: Magnetic activity is a direct tracer of the stellar dynamo and of the rotation–activity–age relation used to date field stars, yet how the dynamo operates across metallicity and evolutionary state remains poorly mapped \citep{Wright2011, Mamajek2008}.  The \ionn{He}{i} D3 triplet at 587.6 nm is a nearly pure magnetic activity indicator \citep{Saar97}. This triplet is relatively weak even in active stars \citep{Schofer19}, but Via observations will deliver a large sample of active stars at a range of evolutionary phase and metallicity. Activity is also the dominant astrophysical source of radial-velocity ``jitter'' \citep{SaarDonahue1997}, so a homogeneous activity census from the \ionn{He}{i} D3 triplet will help separate stellar signals from genuine companions across Via's exoplanet and binary programs. 

\noindent
{\bf Carbon-rich stars}:  The C$_2$ Swan bands at 516.5 nm and 563.5 nm are broad molecular bands that can be very strong in stars with high carbon abundances.  Examples include carbon stars (defined to have C/O$>1$) and carbon-enhanced metal-poor (CEMP) stars, which have [C/Fe]$>0.7$.  The latter are common at low metallicities, comprising $10$--$30$\% of all stars at [Fe/H]\,$<-2$ \citep{Lucatello06, Yoon18}.  Carbon stars are the result of dredge-up phases during AGB evolution.  There are several plausible formation channels for CEMP stars, including binary mass transfer from AGB stars and formation from truly carbon-rich gas enriched from the first few generations of massive stars.  Via will deliver large samples of carbon-rich stars at a wide range of metallicities. The science payoff is twofold. At the lowest metallicities the ``CEMP-no'' subclass---lacking neutron-capture enhancement---preserves the nucleosynthetic fingerprint of the first stars, so the carbon-enhanced fraction and its abundance patterns constrain the explosion energies and mixing of Population III supernovae, and the carbon-driven cooling that may have enabled the first low-mass stars to form \citep{Yoon2016, Frebel2007}. Because CEMP-$s$ stars, by contrast, are almost always binaries that were enriched in carbon and $s$-process elements by a since-faded AGB companion, Via's multi-epoch radial velocities can reveal the orbital motion that separates binary CEMP-$s$ from intrinsic CEMP-no stars---chemically and dynamically classifying carbon-rich stars across metallicity \citep{Hansen2016}.

\begin{figure}[t!]
   \includegraphics[width=\textwidth]{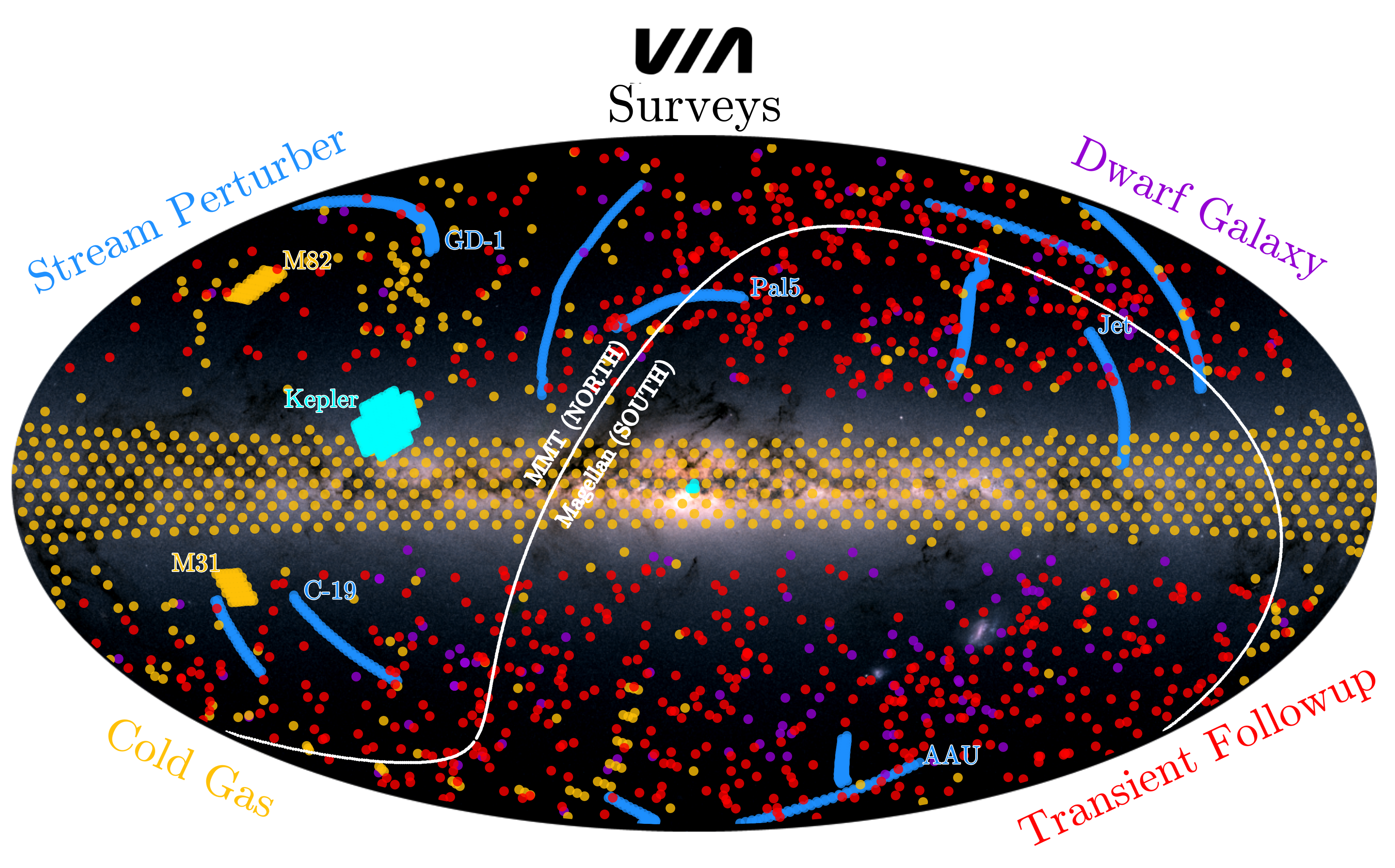}
    \caption{Simulated Via Survey in Galactic coordinates.  Individual pointings for the Via primary surveys are shown as colored circles (not to scale) and are overlaid on an all-sky image of the Galaxy. Several individual targets are labeled.  
    The white line indicates the equatorial plane.  This is a single realization of one possible survey configuration.}
    \label{fig:survey_geometry}
\end{figure}

\subsubsection{Dedicated Pointings}

\noindent
{\bf A Complete Census of the Kepler Field}: The Kepler Space Telescope exoplanet survey  \citep{Koch10} discovered most of the known exoplanets, and for decades will remain one of the most important datasets for studying the population of small planets in our galaxy.  A key limitation of the Kepler survey is uncertainty about the sample of stars searched for planets.  Better knowledge of the kinematics, binarity, and composition of the search sample is required to understand the survey demographics.  The SDSS-APOGEE survey made progress on this front by observing $\sim2,000$ stars in a portion of the Kepler field \citep{Zasowski17}. A multi-epoch Via program to characterize the entire 100 sq. deg. Kepler field will dramatically improve this situation.  Via measurements of precise radial velocities and stellar parameters for the sample of stars most sensitive to Earth-like planets will enable us to: (1)  disentangle and refine occurrence rates for planets orbiting single and binary stars, (2) determine planet occurrence rates as a function of kinematic and chemical thin/thick disk membership, and (3) provide mass constraints and measurements for thousands of previously unobserved planet candidates and eclipsing binaries.  Via will allow us to take full advantage of the Kepler sample and considerably refine our understanding of planet populations in our galaxy.  See Figure \ref{fig:kepler} for a simulated Via program in the Kepler field.

\noindent
{\bf Roman Transiting Exoplanets}:
Roman's Galactic Bulge Time Domain Survey (GBTDS) spans six 0.8$^\circ$$\times$0.4$^\circ$ fields near the Galactic bulge and center, and is well matched to the 1$^\circ$ diameter Via field of view \citep{ROTAC2025}.  A dedicated Via program observing Roman GBTDS transiting exoplanet candidates will directly address the core bottleneck for using this transiting-planet harvest: astrophysical false positives.  While the Roman GBTDS is expected to be sensitive to $\sim100,000$ transiting exoplanets \citep{Wilson2023}, disentangling these objects from eclipsing binaries containing star-sized planets is challenging.  Crowding will likely limit ground-based followup to sources with $R \lesssim 18$, a limit that is expected to yield roughly 600 transiting exoplanets over the full GBTDS survey.  Preliminary estimates suggest that these exoplanets will be hidden in a sample of a few thousand ``planet candidates'', including false positive eclipsing binaries.  Via will reach $\lesssim 1\kms$ precision for these candidates, sufficient to rule out eclipsing binaries with repeat observations.  A notional survey would tile each GBTDS field with multiple Via observations to overcome fiber-packing limits, obtaining 3--4 epochs per candidate for a brightness-limited sample.  Order-of-magnitude scaling suggests $\sim$9--12 nights would deliver statistical validation for $\approx$600 Roman transiting planets across all six WFI fields, while empirically calibrating the false-positive rate versus magnitude/crowding for the much larger but fainter transiting planet sample.  This calibration would enable occurrence-rate studies in a region of the Galaxy with chemistry, ages, and birth conditions distinct from Kepler and TESS fields.



\subsection{Lower Priority Classes}

Priority class 5 will consist of a more general selection of distant stars, approximately replicating the main priority class of the H3 Survey \citep{Conroy2019} but going fainter and taking advantage of \gaia\ DR4 astrometric parameters. We define a nominal baseline target selection for distant stars in \gaia\ DR4 using the following criteria: parallax $\varpi < 0.3$ mas and magnitude $17 < G < 21$. At $l = 90^\circ$, this selection produces $\sim 1600$ stars at $b = 30^\circ$ and $\sim 400$ stars at $b = 90^\circ$ within the Via FoV.  Finally, we will employ class 6 filler targets with $\varpi >0.3$ mas to ensure all fibers are assigned.


\subsection{Selection Function}

The selection function of a survey quantifies the relation between objects targeted for observation and the underlying parent sample.  An accurate selection function is essential for drawing any meaningful conclusions about the underlying population of stars.  To effectively and reproducibly track the selection function, priority class 5--6 (the main filler samples) will be drawn from homogeneous parent catalogs covering the entire sky and cross-matched to \gaia. Priority classes 1--4 can be defined from more specific catalogs depending on the primary survey goals. Priority classes will be assigned by well-defined rules that are tracked along with the parent catalog at the time of selection, prioritization, and fiber assignment. Supplemental catalogs may be added in special circumstances, but only for priority classes 1--4, and must be tracked along with the parent catalog in the same format.

\begin{figure}[t!]
    \centering
    \begin{minipage}[l]{0.5\textwidth}
        \includegraphics[width=\textwidth]{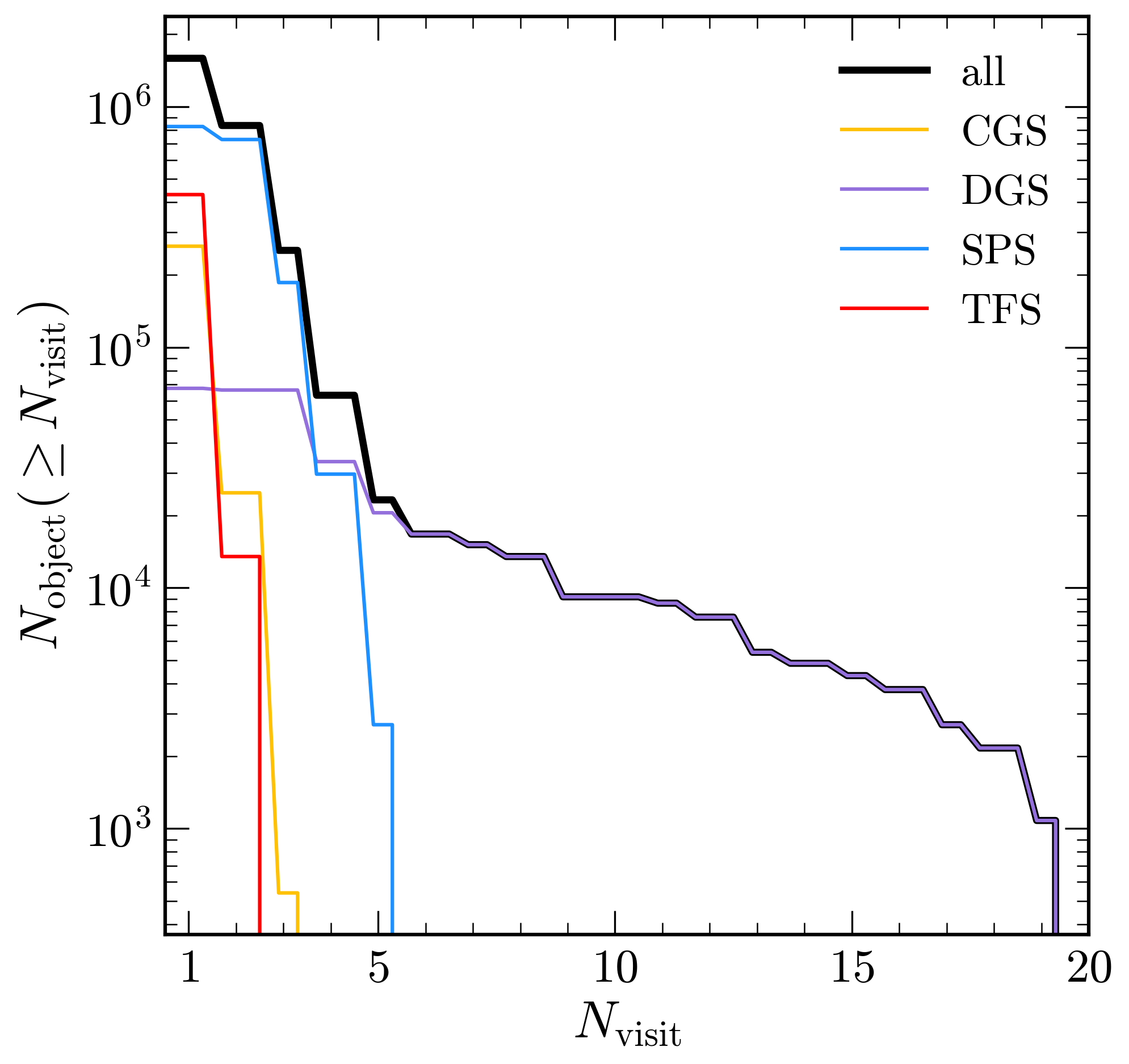}
    \end{minipage}
    \begin{minipage}[c]{0.4\textwidth}
      \caption{Cumulative distribution of the number of times a target will be observed by Via ($N_{\rm visit}$) under the nominal survey strategy.  Approximately $10^4$ objects will be visited at least $10$ times.  The separation between successive observations will range from days to years. CGS = cold gas survey, DGS = dwarf galaxy survey, SPS = stream perturbation survey, and TFS = transient followup survey.}
        \label{fig:nobsdist}
    \end{minipage}
\end{figure}

\subsection{Survey Simulator}

We are developing a survey simulator that includes the four key projects, multi-epoch observations, and site-specific seeing and weather conditions.   The simulator groups observations into month-long ``runs'' at each telescope spread throughout the year.

We define a baseline depth of one ``effective hour'' (1 ehr) as the exposure time required to obtain SNR$=5.5$ pix$^{-1}$ at $G=20$ mag in three 20 min exposures with nominal $1.0\arcsec$ seeing and airmass 1.2.  Exposure times can be significantly shorter than 1 hr in excellent seeing.  For short exposure times in excellent conditions we switch to a single exposure to minimize read noise.  Overheads include slew and settle time, guide star acquisition, wave front sensing (WFS), actuator configuration, metrology, and readout.  The WFS alignment process is expected to dominate Via overhead.  The current simulation conservatively assumes 5 min of overhead per field.  Magellan offers better average  seeing and weather than the MMT so the southern survey speed will be considerably faster.  LSST's footprint, however, will produce considerably more targets in the south.

Most fields will be observed at 1 ehr depth.  The stream and dwarf galaxy fields will receive at least one additional epoch of observations to remove binaries, with depths adjusted based on heliocentric distance. Many of the cold gas survey fields are relatively shallow and suitable for bright time. A subset of dwarf galaxy fields will reach much greater depth (as much as 10 ehr) through repeated observations required for binary rejection.

Figure \ref{fig:survey_geometry} shows a simulated Via Survey in Galactic coordinates.   Different colored symbols highlight the four key projects.  The transient followup survey (TFS) and dwarf galaxy survey (DGS) pointings are notional and concentrated in the LSST footprint.  The cold gas survey (CGS) pointings trace large over-densities in the \ionn{H}{i} column density, excepting the Magellanic Stream, and include several bright-time pointings within the plane.  The SPS pointings trace known streams.  The stream perturbation survey (SPS) pointings may be adjusted as more promising streams are identified.  The corresponding distribution of visits each target will receive is shown in Figure \ref{fig:nobsdist}.

\clearpage

\section{Modeling and Inference in Support of Via's Key Projects}
\label{sec:modelingmw}

This document has provided a broad overview of the scientific motivation for the Via Project and the instrument design required to meet these goals. 
In parallel with instrument and survey development, work to advance theory and inference is crucial preparation for interpreting data from Via and other contemporary spectroscopic surveys. 
This section summarizes modeling and inference work underway within our collaboration, and motivates future simulations and inference tools to enable new insights from our survey data. We discuss models of Milky Way substructure \S~\ref{subsection:mws_model},  the circumgalactic medium \S~\ref{subsection:cgm_model}, and the time domain \S~\ref{subsection:timedomain_model}.

\subsection{Modeling the Stellar Substructure in the Milky Way}
\label{subsection:mws_model}
Two of Via's key projects --- the Stream Perturbation Survey and the Dwarf Galaxy Survey --- require high fidelity models for Milky Way stellar substructure.  These models include accurate predictions for the small-scale dark matter structure in the Milky Way, the dynamical evolution of that dark structure over cosmic time, the mapping of stars to the underlying dark structure, and the dynamical interaction between the dark and luminous matter.  Here, we describe work underway in these areas. 

\subsubsection{\textsc{Compass}: a Computational Observatory for Milky Way Precision Inference with Accreted Stellar Substructure}
\label{sec:compass_pipeline}

Realizing the full potential of Via data requires robust comparisons to comprehensive models of the Milky Way's accreted stellar substructure.  We aim to connect the detection of perturbations in stellar streams (\S \ref{s:streams}, \S \ref{s:sps}), the kinematics of the smallest galaxies (\S \ref{s:dwarfs}, \S \ref{s:dgs}), and measurements of the Galactic potential to the physics of dark matter, the threshold of galaxy formation, and the Milky Way's accretion history. Via's radial velocity precision and depth will resolve subtle dynamical signatures beyond the predictive power of current models. 

A range of approaches for modeling the Milky Way's accreted stellar substructure have been proposed \citep[e.g.][]{BullockJohnston2005, Cooper2010, dreams2025}, and techniques for generating synthetic observations from these models are described by \cite[e.g.][]{Sharma2011, Lowing2015, Sanderson2020, Aurigaia2018, pyananke2024}. However, no existing framework captures accurate small-scale dark matter structure, its dynamical evolution over cosmic time, the mapping of stars onto that structure, and the interaction between dark and luminous matter at sufficient resolution and enough speed to generate large model ensembles. This motivates the development of \textsc{Compass}: a Computational Observatory for Milky Way Precision Inference with Accreted Stellar Substructure. \textsc{Compass} is a fast, modular, simulation-based forward-modeling pipeline that will generate suites of survey-realistic mock catalogs of stars, enabling robust inference and survey optimization for Via, and joint analysis with complementary maps of the Milky Way's stellar substructure from other surveys.

\begin{figure}[t]
   \centering \includegraphics[width=
   \linewidth,trim={4cm 3cm 0 0},clip]{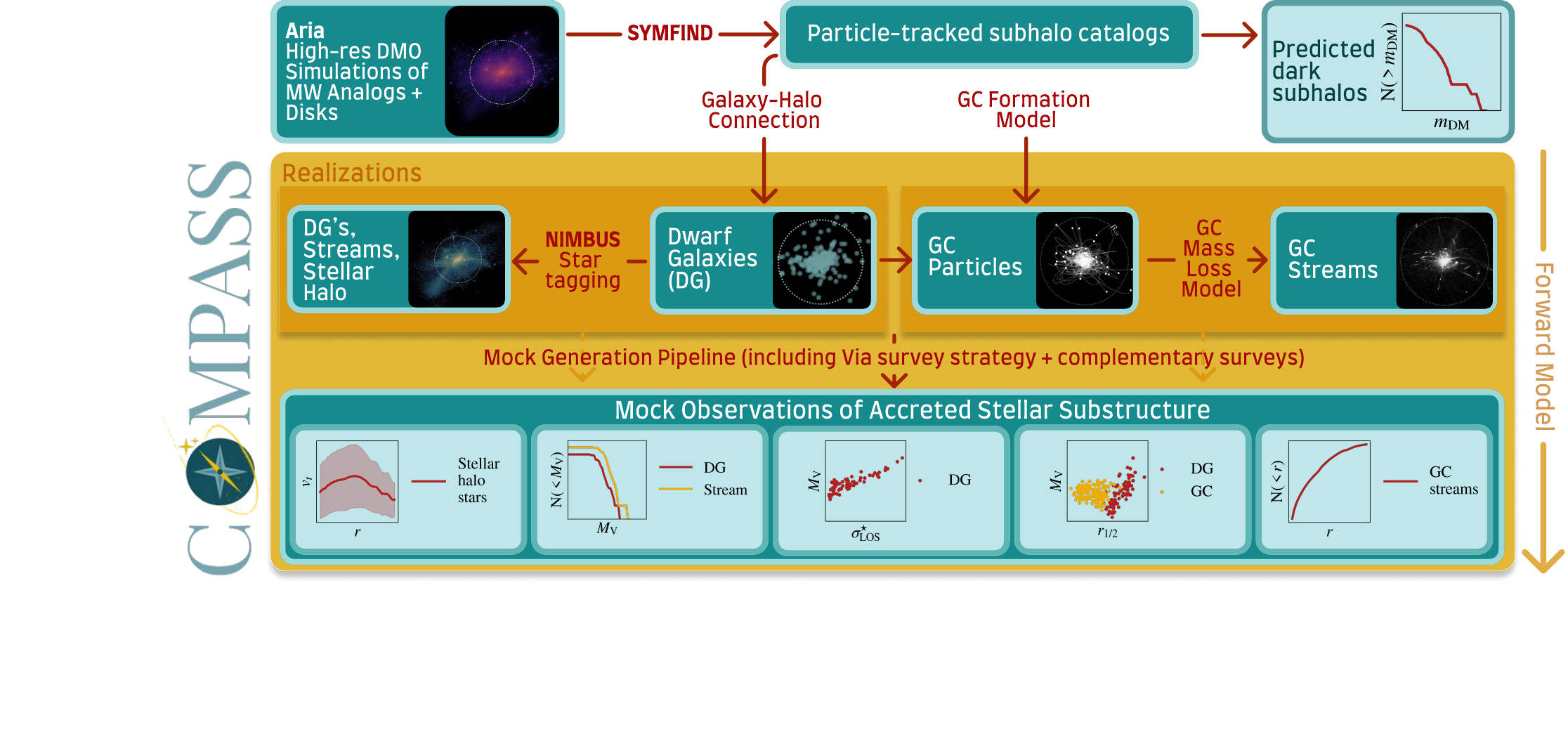}
   \caption{Schematic showing the \textsc{Compass} pipeline for modeling realizations of the Milky Way and its accreted stellar substructure. Teal boxes indicate simulation data products; red text indicates tools and tunable models that produce them. The yellow box indicates the readily repeatable modeling steps that can generate realizations over parameter spaces. Examples of comparisons of specific mock observables to data from specific surveys are included in the lower section.}
   \label{fig:compass_flowchart}
\end{figure}

Figure~\ref{fig:compass_flowchart} illustrates the organization of the \textsc{Compass} pipeline. The pipeline consists of several integrated modeling stages, each modular across a range of physical assumptions, allowing iteration.  The pipeline's first stage is a new zoom-in simulation suite, \textsc{Aria}. \textsc{Aria} is planned to comprise ten $10^9$-particle and forty $10^8$-particle simulations of MW-mass halos with embedded disk potentials \citep{Wang2025} and MW-like merger histories \citep{Buch2024}. This resolution is necessary to reliably follow the disruption of low-mass subhalos below the scale where galaxies form, a regime directly probed by Via's stream perturbation survey. A custom particle-tracking subhalo finder, \textsc{Symfind} \citep{Mansfield2024}, recovers subhalo populations that are missed by standard tools at small galactocentric radii.

Dark matter particles in each simulation are assigned stellar masses, metallicities, and ages using \textsc{Nimbus} (Mansfield et al in prep.), an energy-preserving star-tagging framework coupled to flexible galaxy--halo connection models. \textsc{Nimbus} allows tuning the galaxy--halo connection and galaxy tidal stripping, including galaxy sizes, stellar mass--halo mass relations, and disruption physics.  \textsc{Nimbus} can therefore generate diverse realizations of the ex-situ stellar halo, including intact satellite galaxies, dwarf stellar streams, and the smooth halo, from each simulation. This flexibility is essential for marginalizing over astrophysical uncertainties when comparing to Via data. Accreted star cluster populations are painted onto subhalos using flexible empirical star cluster--halo connection models and are evolved in the time-dependent gravitational potential of the host halo and its substructure. Star cluster disruption is modeled with a semi-analytic mass-loss prescription coupled to a particle-spray stream generation scheme, producing self-consistent stream populations whose properties depend on the assumed star cluster formation efficiency and mass-loss parameters.  The pipeline will be used to predict the perturber population expected for a given cosmological model, as well as the observable streams, dwarfs, and their dynamical properties.

The final stage of our pipeline produces star-by-star mock catalogs tailored to Via's selection function, including magnitude limits, fiber assignment, field placement, photometric and radial velocity uncertainties, and dust extinction. Mock catalogs are also generated for DESI, \gaia, Roman, and Rubin/LSST, enabling joint analyses across surveys with consistently treated observational effects.

With \textsc{Compass}, the modeling stages downstream of the simulations can be rapidly re-run over parameter spaces, generating large ensembles of mock observations from each simulation. 
The same underlying dark matter halo produces qualitatively different stellar halos and stream populations when galaxy sizes or star cluster mass-loss rates are varied. This flexibility is critical for three reasons.
First, inference methods can be rigorously tested on realistic mock data before they are applied to Via observations. Mass-modeling techniques, subhalo population analyses, and stream perturbation searches can all be benchmarked against ground truth, with systematic uncertainties quantified across the range of astrophysical models.
Second, \textsc{Compass} will be used to optimize Via's survey strategy. 
By forward-modeling target selection, exposure time, and field placement choices, the pipeline will help maximize Via's sensitivity to dark subhalo perturbations and dwarf galaxy kinematics.  This process directly informs the observing plan for the stream perturbation survey (\S\,\ref{s:sps}) and the dwarf galaxy survey (\S\,\ref{s:dgs}).
Third, \textsc{Compass}'s consistent treatment of multiple tracer populations (halo stars, streams, globular clusters, and intact and disrupting dwarf galaxies) enables joint analyses.  These analyses combine Via's high-precision radial velocities with complementary data including photometric and astrometric information from Rubin, Roman, and \gaia. The combination yields tighter constraints on the MW potential and dark matter physics than provided by any single tracer or survey.

\subsubsection{Modeling Subhalo Populations with \textsc{Galacticus}}

Whereas \textsc{Compass} is built on full $N$-body simulations, semi-analytic models offer a complementary, faster route at arbitrarily small mass scales. We are actively developing and using the semi-analytic model \textsc{Galacticus} \citep{Benson2012} to generate realizations of subhalo populations from which we compute the statistics of stream interactions (\citealt{Menker2026}; see also \citealt{Jiang2021, Adams2025}). 
A key strength of the \textsc{Galacticus} approach is the ability to generate constrained realizations of the Milky Way's mass assembly history. Recent work by \citet{Nadler2023} and others demonstrates the importance of including the Large Magellanic Cloud (LMC) merger (and potentially the Gaia--Sausage--Enceladus, GSE merger) into the host halo's evolution. By conditioning the merger trees on the presence of an LMC-like progenitor, \textsc{Galacticus} accounts for the significant ``infall'' subhalo population associated with the LMC, critical for interpreting the current state of Milky Way streams, their perturbers, and the population of satellites and anisotropy within the Milky Way. Ongoing work will extend this framework to multiple constraints (e.g., requiring both an LMC and GSE merger), as has been implemented in the \citet{Buch2024} $N$-body simulations.

Stellar streams are sensitive to very low mass dark matter substructure, well below the resolution limits of most existing cosmological simulations. \textsc{Galacticus} is able to evolve subhalo populations down to $10^5\, \mathrm{M}_\odot$ \citep{Menker2026} or lower. This capability enables quick calculation of the full subhalo mass function relevant to stream widths and density fluctuations, as well as the population that hosts the faintest observed galaxies.

The modular nature of the \textsc{Galacticus} power spectrum and transfer function physics also allows rapid exploration of alternative dark matter models. Whether investigating warm, self-interacting, or fuzzy dark matter, or modified power spectra, the model efficiently predicts how these cosmologies suppress or enhance small-scale structure \citep{Lonergan2025_DM}. This flexibility is essential for using stellar streams as a ``dark matter microscope'' to test physics beyond the $\Lambda$CDM paradigm.
To facilitate large-scale Bayesian inference, \textsc{Galacticus} and \textsc{Compass} can be used to train emulators using normalizing flows. As shown for \textsc{Galacticus} by \citet{Lonergan2025_Emulator}, these emulators can reproduce the statistical properties of subhalo populations across a wide parameter space with high fidelity. This allows the rapid generation of millions of subhalo realizations, enabling the marginalized likelihoods of dark matter parameters to be calculated in a fraction of the time required for traditional semi-analytic or $N$-body methods.

\subsubsection{Inferring Subhalo Properties from Via Observations}

\begin{figure}
    \centering
    \includegraphics[width=\linewidth]{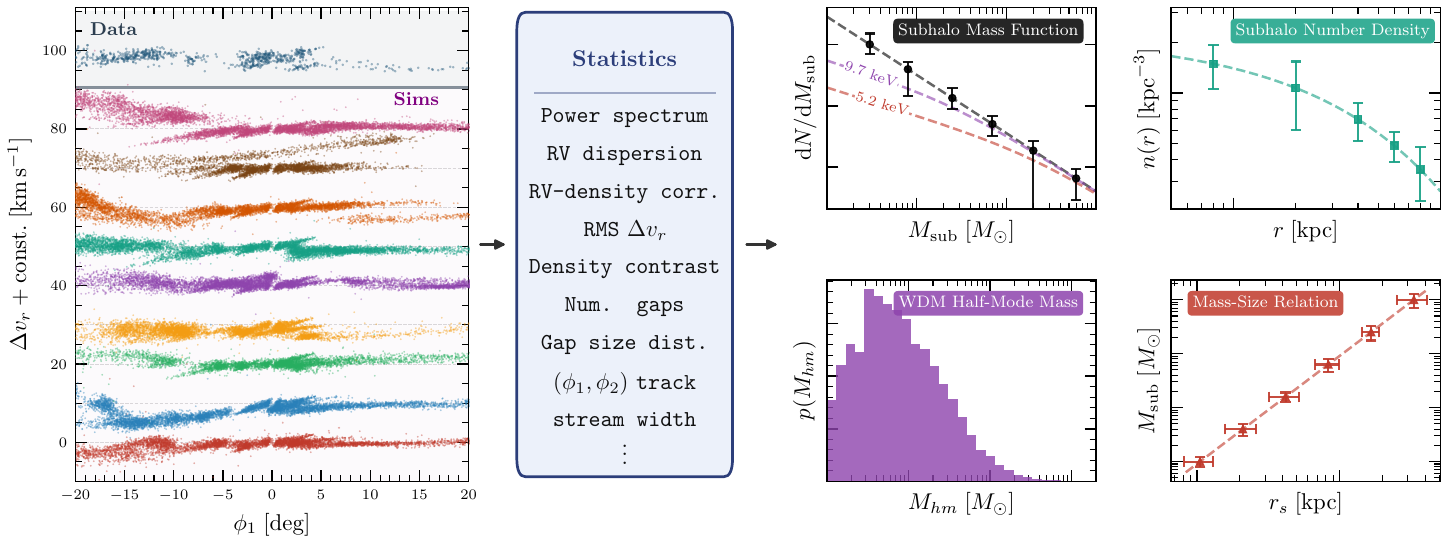}
    \caption{Left: The observed radial velocity structure of a stream (top, ``Data'') compared against numerous forward-modeled realizations (bottom, ``Sims'') under a specified dark matter model. The models rapidly capture the stochastic perturbations in radial velocity ($\Delta v_r$) along the stream coordinate ($\phi_1$) induced by subhalo impacts. Center: A suite of summary statistics computed for both the observed data and the simulated models. These metrics, such as the power spectrum of the velocity track, velocity dispersion, density contrast, and gap distributions, serve as the basis for comparing the models to the data within the inference framework. Right: Illustrative constraints on dark matter subhalos. By accepting models with summary statistics that closely match the data, constraints can be placed on the subhalo mass function, the subhalo radial number density, $n(r)$, the subhalo mass--size relation, and dark matter microphysics parameters such as the Warm Dark Matter (WDM) half-mode mass ($M_{hm}$).}
    \label{fig:streamillustration}
\end{figure}

Linking the observed radial velocity and density structure of a stream to the properties of dark matter subhalos requires fast and accurate forward modeling. We will use the GPU accelerated code \texttt{streamsculptor} \citep{nibauer:2025} to efficiently model the gravitational influence of subhalos on each stream. Our framework generates direct, explicit models of the stream's phase space distribution evolving with distinct dark matter subhalo populations. This framework gives us the flexibility to define different observables as needed by the data, ranging from the stream's density structure to its velocity track or width. Furthermore, the framework computes how these observables respond to variations in subhalo mass, concentration, and impact geometry within seconds. This efficiency is crucial, enabling us to generate millions of stream realizations and rapidly explore how different subhalo mass functions, mass--size relations, and impact geometries shape the observed stream.

Because the physical processes linking the dark matter subhalo population to the resulting stellar stream structure are stochastic and non-linear, deriving an analytical likelihood function to evaluate the data is intractable. Simulation-Based Inference (SBI; see, e.g., \citealt{banik:2021,Nibauer:2025heating}) allows us to overcome this issue. SBI is a family of likelihood-free statistical methods that bypasses the need for analytical likelihood functions. SBI instead relies on comparing observed data against an ensemble of forward-modeled simulations. Figure~\ref{fig:streamillustration} illustrates an example of the inference procedure. Given the observed velocity structure of a stream (left panel, data), we generate numerous forward-modeled realizations under a fixed dark matter model (labeled sims). We then compute a series of summary statistics (middle panel) for both the data and the models, such as the power spectrum of the stream’s velocity track and the local velocity dispersion. We use techniques like Approximate Bayesian Computation and Sequential Monte Carlo to accept models that produce sufficiently similar statistics, allowing us to construct an approximate posterior distribution over the dark matter model parameters.

By analyzing populations of both dwarf galaxies and stellar streams under this or related frameworks, we will be able to constrain the subhalo mass function, the radial distribution of subhalos, the mass--size relation, and dark matter microphysics parameters like the Warm Dark Matter (WDM) half-mode mass. Example posteriors on dark matter parameters (for illustration only, not an actual constraint), are provided in the right panels of Figure~\ref{fig:streamillustration}. From a population of streams spanning a wide range of galactocentric radii, precision radial velocities from Via will enable a stringent test of dark matter on small scales.

\subsection{Modeling the Circumgalactic Medium}
\label{subsection:cgm_model}

\subsubsection{From Via Spectra to the 3D Distribution of Cold Gas}

Via aims to characterize the three-dimensional distribution of cold gas using neutral sodium absorption lines in millions of stellar spectra. The combination of spectral resolution with high source density will enable maps of unprecedented detail, but the scale and scope of this science necessitates a novel approach to spectral reduction and inference. Several key challenges must be addressed to extract maximum value from Via data.

Stellar photospheric \ionn{Na}{i} lines are often blended with absorption from ISM and CGM gas, especially in the plane where a majority of our sample are cooler metal-rich stars (with stronger intrinsic absorption) and radial velocity overlap is common. We plan to employ the MADGICS algorithm \citep{Saydjari2023}, which uses a sparse, non-stationary, data-driven Gaussian process prior for stellar spectra. The main advantages of this approach are the ability to efficiently marginalize over stellar spectrum uncertainties and capture subtle deviations from theoretical models using the millions of stars observed by Via.  This approach is especially important for the \ionn{Na}{i} doublet, which suffers from significant non-LTE effects \citep{Lind11} that are not presently included in our default theoretical stellar libraries.

Interstellar sodium lines are integrals along the line of sight, and at Via's spectral resolution they have complex line profiles not well fit by a single Gaussian component. Adding sufficient flexibility while avoiding degeneracy with stellar lines on a per-star basis is difficult. Fortunately, ISM sodium absorption is highly correlated between stars at small angular separation and similar distances since they fundamentally probe the same gas. To take advantage of this correlation, we parameterize ISM and CGM sodium as density and velocity fields in 3D space, and then forward model to the observed spectrum using radiative transfer equations. Priors are naturally expressed in physical space, rather than per-star in spectral space.  Our 3D science products are directly output by the pipeline, and mock tests verify that this approach is computationally feasible and application to test data is currently underway. 

Stellar distance uncertainties propagate into sodium distance uncertainties, and can produce radial artifacts in the resulting map if not addressed. This is another effect that is difficult to account for on a per-star basis, but that is naturally incorporated into our forward-modeling framework. Stellar distance uncertainties can be fit or marginalized over during the sodium fit, using as a prior parallax or other distance constraints obtained by fits to spectra and photometry.

Continuous sodium density and velocity fields have vastly more degrees of freedom than the Via spectra constraining them, making inference an ill-posed inverse problem. We plan to regularize the map using a Gaussian process prior, following previous work in 3D dust mapping \citep{Edenhofer2024}. The main challenge is scaling the forward model and Bayesian inference to the number of parameters required. We will use the GraphGP algorithm \citep{Dodge2026} to efficiently represent the Gaussian process prior, capturing correlations on all scales with linear memory requirements and an efficient GPU implementation. We will draw posterior samples with the Geometric Variational Inference algorithm \citep{Frank2021}, which handles high-dimensional non-Gaussian distributions while still scaling to the hundreds of millions of parameters required for Via cold gas science.

\subsubsection{Hydrodynamic Simulations of the Circumgalactic Medium}

Empirical maps of the cold gas content throughout the Milky Way will shed new light on the cycling of matter into and out of the Galactic disk, as well as cycling between the various phases of gas within the Galaxy.  Additional insight into the physical origin and evolution of the cold gas can be provided by comparing the empirical maps to hydrodynamic simulations of the Galactic baryon cycle.  Unfortunately, the CGM has historically been challenging to simulate owing to the vast range of scales involved.  The CGM itself spans hundreds of kpc, while the smallest physical features in the CGM may extend down to pc scales (or smaller), requiring a dynamic range of $>10^6$ to properly resolve all scales.  Zoom-in simulations that focus on a small patch of the CGM can reach pc-scale resolution, but not in a cosmological setting, and not for an entire halo \citep[e.g.,][]{Gronke00, Tan24}, although see \citet{Ramesh26} for efforts to close the gap.  Conversely, cosmological simulations have resolution refinement schemes that tend to focus the highest resolution in the inner few kpc of the halo where the stars form and the feedback is launched, leaving the CGM at kpc-scale resolution \citep[see e.g., Fig. 2 in ][]{Lucchini2026}.

Recognizing this limitation, several groups have developed refinement techniques specifically designed to improve the resolution in the CGM \citep[e.g., FOGGIE, TEMPEST, GIBLE, SURGE, and ENGAWA:][]{vandeVoort2019, Hummels2019, Augustin2025, Ramesh2024, Lucchini2026}. However, even the latest simulations in this category only reach spatial resolution of $\approx 200$ pc in the CGM. For Via, we plan to combine these CGM-refinement simulations with subgrid models of cold gas \citep[e.g.,][]{Butsky24, Hummels24} in order to understand how the predicted distribution of cold gas depends on physical processes not directly captured in the simulations.  Ultimately these sub-grid models should be compared against zoom-in simulations of small patches of the CGM. The development of such models is an ongoing effort.

Another challenge/opportunity for CGM modeling is that the observations are sensitive to a wide range of uncertain physical processes, including turbulence driving and dissipation, thermal conduction, and cosmic ray propagation.  Models including cosmic rays are a particularly pressing case: they predict cooler temperature distributions as a result of the non-thermal pressure support \citep[e.g.,][and see \citet{FaucherGiguere2023} for a review]{Ji:2020, DeFelippis:2024, Girichidis:2024}.  Observations of the cool gas in the CGM have the strong potential to constrain these uncertain physical processes \citep[e.g.,][]{Butsky:2023, Lu:2026}. 

High-quality simulations of the thermodynamic structure of the CGM are only one step toward predicting the distribution of cold neutral sodium gas that Via will measure. The chemical evolution model, details of the feedback prescriptions and metal mixing, as well as the effects of radiative transfer can all impact the predicted distribution of neutral ions with low ionization energies such as sodium. We are developing tools to understand the impacts of each of these ingredients on the predicted sodium distribution from simulations. 

\subsection{Modeling the Transient Sky}
\label{subsection:timedomain_model}

LSST is expected to release billions of alerts annually, among which $\sim1$ million extragalactic, supernova-like transients will be hiding. From this data deluge, LSST Broker systems (e.g., Babamul/BOOM; \citealt{laz2025boom}) filter and tag active alerts with important metadata and classifications.  A major modeling effort is underway to identify and classify transients and to robustly select needles from the LSST haystack.

Customized algorithms such as Superphot+ \citep{de2024superphot} and LAISS \citep{aleo2024anomaly} allow identification of the most scientifically valuable transients as soon as LSST data are broadcast to the community. LAISS will be the workhorse populating the nightly Boombox queue with new, exotic explosions. LAISS relies on a supervised anomaly detection method, in which a forest-based classifier identifies transients as unusual through a combination of their host galaxy (e.g., apparent magnitudes, angular size, color) and light curve  (e.g., brightness, duration, skew) features.  LAISS works on data entirely generated from the LSST datastream.  We will incorporate unsupervised methods as well, including classic isolation forests \citep{villar2021deep} and a Distance Multi-Metric Anomaly Detection method \citep{chaini2025search}. These methods all produce a ranked list of ``most anomalous'' transients, from which we can select highly unusual and easily observable targets on a nightly basis. We have already developed versions of these anomaly detection algorithms for the pre-LSST Rubin datastream; the algorithms will be adapted and retrained as the LSST survey accumulates data. 
 
Precursor observations of SNe will require specialized filters beyond standard single-epoch alert selection. For example, filters based on ``clean'' (i.e., without precursor emission) archival DECam templates, such as \texttt{SLIDE} \citep{dong2025enabling}, can recover faint, long-lived pre-explosion emission (see e.g., \citealt{tsuna2024merger}) that may be partially imprinted in early LSST reference images and missed by the standard LSST alert stream. Precursor candidates for Via follow-up will be selected based on two key criteria. 1) the transient has been monitored by LSST for at least one month with a steadily increasing luminosity and its absolute magnitude falls within the range $-16 < M_r < -11$. 2) The transient has a projected offset within $\sim$30 kpc of a host galaxy located within 100 Mpc. Host galaxy distances will be taken from the GLADE+ catalog \citep{Dalya22}. This proactive precursor search paves the way for the first systematic, real-time spectroscopic studies of supernova progenitor systems in the final stages before explosion.

There will be a number of sources available for our ``passive'' followup program (e.g., transients, variables or host galaxies) that will be observed while other Via surveys drive the pointing. We will primarily target core-collapse supernovae, rare nuclear transients and/or SMBH activity, and host galaxies of said objects. We will use Superphot+ as our primary photometric classifier in selection of transient events, which has already been used on Rubin data \citep{Soto26}. For nuclear transients, particularly those in AGN, we will use data-driven methodology (an attentive neural process method) to flag anomalous behavior statistically inconsistent with a damped random walk process \citep{Shen_inprep}. 

We will model the active follow-up of all science cases above in an end-to-end simulation of the full Via survey, incorporating both the primary transient programs and the broader ecosystem of LSST, DSA, and other time-domain discoveries. Targets will be prioritized using a quantitative ranking scheme that accounts for scientific impact, rarity, target density, brightness, observational cost, and the sample sizes required for each science case. Rare events with exceptional discovery potential—including gravitational-wave counterparts, precursor eruptions, and highly anomalous LSST transients—will receive the highest priority, while more numerous populations such as AGN, FRB host galaxies, and transient host galaxies will populate the queue at lower priority and efficiently utilize otherwise unused fibers. This framework allows Via to balance rare, high-value events against target classes that require large samples for population studies. Our goal is to follow up $\gtrsim10{,}000$ transients across our primary science cases over the five-year survey.

\clearpage

\section{Comparison to Existing and Planned Instruments and Surveys}
\label{comp}

\begin{figure}[b!]
    \centering
    \begin{minipage}[c]{0.4\textwidth}
        \caption{Performance of Viaspec and the most comparable wide-field spectrograph on a 6m-class telescope, MMT/Hectochelle.   Points show observed data from the H3 Survey \citep{Conroy2019}, which were observed in bright time with 30\,min exposures.  Blue curves show predictions from the Data Simulator ($\S$\ref{sec:etc}) configured with Hectochelle parameters; the two seeing values bracket typical conditions at the MMT.   The red curve is the predicted Viaspec performance in identical observing conditions.
        }
      \label{fig:chelle_compare}
      \end{minipage}
    \begin{minipage}[r]{0.5\textwidth} 
    \includegraphics[width=\textwidth]{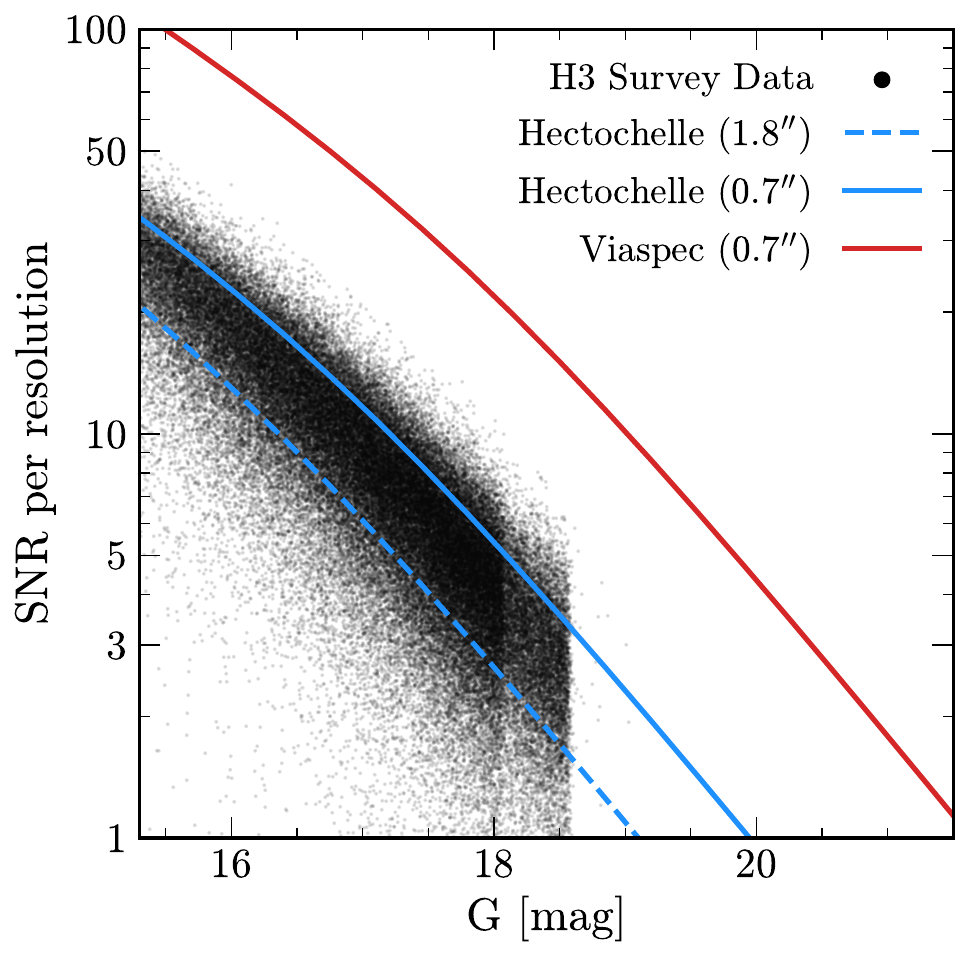}
    \end{minipage}
  \end{figure}

No existing or planned facility combines the radial velocity stability, depth, multiplexing, and field of view that Via requires. In this section we compare Via to existing and forthcoming instruments to demonstrate the unique capability space it occupies. As a high-resolution multi-object spectrograph on a 6.5m telescope, Viaspec is in some ways a successor to the Hectochelle instrument on the MMT \citep{Fabricant2005, Szentgyorgyi2011}. Although Viaspec has half the spectral resolution of Hectochelle, Viaspec surpasses the multiplex factor of Hectochelle by $2.4\times$, covers $7\times$ the spectral range, and has $5 \times$ the peak instrument throughput.  
To quantify the improvement, we have implemented Hectochelle instrument parameters into the Viaspec data simulator. 
Figure~\ref{fig:chelle_compare} shows the predicted performance of Hectochelle in bright time (blue curves), along with real data from the H3 Survey \citep{Conroy2019}.  
The red curve shows the predicted performance of Viaspec in identical observing conditions---at half the spectral resolution, Viaspec achieves comparable SNR two magnitudes fainter than Hectochelle.
We have also forecasted the statistical velocity precision of Viaspec and Hectochelle, using the data simulator described in $\S\ref{sec:etc}$---this allows a full assessment of the improved throughput and wider wavelength range. 
In dark time for metal-poor ([Fe/H]$= -2.0$) giants, Viaspec delivers a statistical RV uncertainty of $100\,\ms$ at $G \approx 17$, and $1\,\kms$ at $G \approx 20$ in $1$\,hr of exposure time. 
To deliver these RV uncertainties for stars at the same magnitudes, Hectochelle would require $6$\,hr and $30$\,hr of exposure time, respectively.  
These are statistical uncertainties only---in practice, the zero-point accuracy of Hectochelle is limited to $\approx 0.5\,\kms$, whereas Viaspec is designed to be stable at the $< 100\,\ms$ level. 

A number of highly multiplexed spectroscopic surveys and instruments are now operating or under construction.  (1) DESI at the 4m Mayall telescope has 5,000 fibers patrolling an 8 sq. deg FoV and 360--980~nm coverage at $R\sim$~2000--5000.  DESI's relatively low resolution precludes the high velocity precision required here.  (2) The SDSS-V, operating at 2.5m telescopes at APO and LCO, enables all-sky spectroscopy with 500 robotic positioners patrolling a 7 sq. deg. FoV.  The relatively small collecting area limits SDSS to bright targets.  (3) 4MOST at the 4m VISTA telescope and WEAVE at the 4.2m William Herschel telescope are beginning survey operations in the 2025--2026 timeframe.  Both have wide fields of view ($\sim2$ sq. deg) with $R\sim5,000$ and $R\sim20,000$ capabilities. The high-resolution modes have 800/940 fibers, but neither covers the critical \ionn{Na}{i} spectral region and both surveys plan high-resolution bright-time observations of relatively bright targets ($G<16$) for detailed abundance analysis.  Extensive high-resolution observations of faint targets in dark time are not currently planned in either of these surveys.  (4) PFS, operating at the 8m Subaru telescope, has 2400 fibers covering a 1.25 sq. deg FoV; it does not have a high-resolution mode and its maximum resolution is $R\sim5000$, precluding the high velocity precision that Via requires.  (5) MOONS at the 8m VLT has 1000 fibers patrolling a FoV $6\times$ smaller than Viaspec.  The high-resolution mode covers the CaT region at $R\sim9,000$ and the H-band at $R\sim18,000$.  The small FoV precludes large-area surveys with MOONS.

\begin{figure}[t!]
    \centering
    \begin{minipage}[l]{0.5\textwidth} 
    \includegraphics[width=\textwidth]{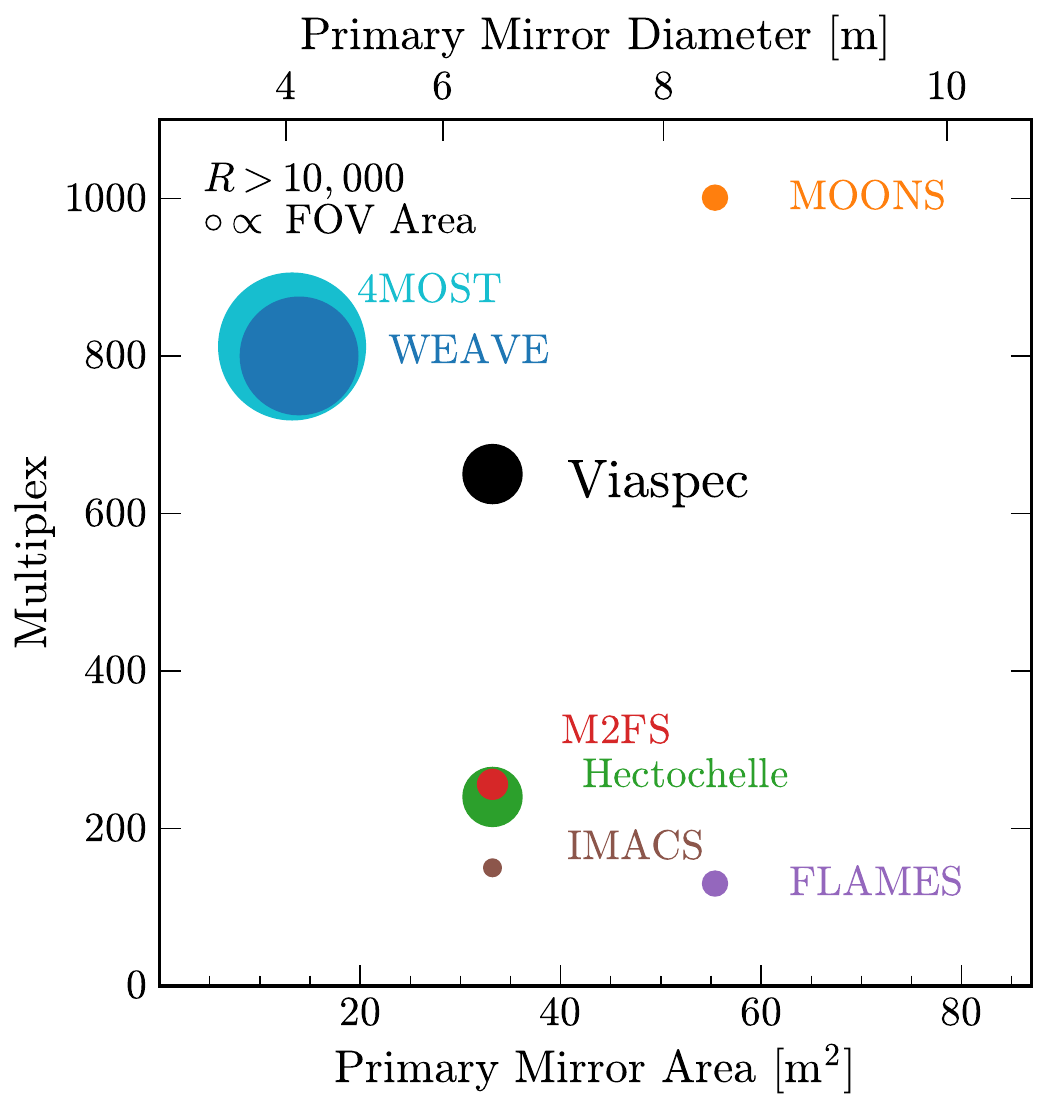}
    \end{minipage}
    \begin{minipage}[c]{0.4\textwidth}
      \caption{Current and forthcoming multi-object spectroscopic instruments with $R > 10,000$, compared to Viaspec along two axes: the primary mirror collecting area, and number of fibers in a single pointing. Marker sizes are proportional to the FoV area.}
    \label{fig:mos_compare}
    \end{minipage}
  \end{figure}

\begin{figure*}[t!]
    \centering
    \includegraphics[width=\textwidth]{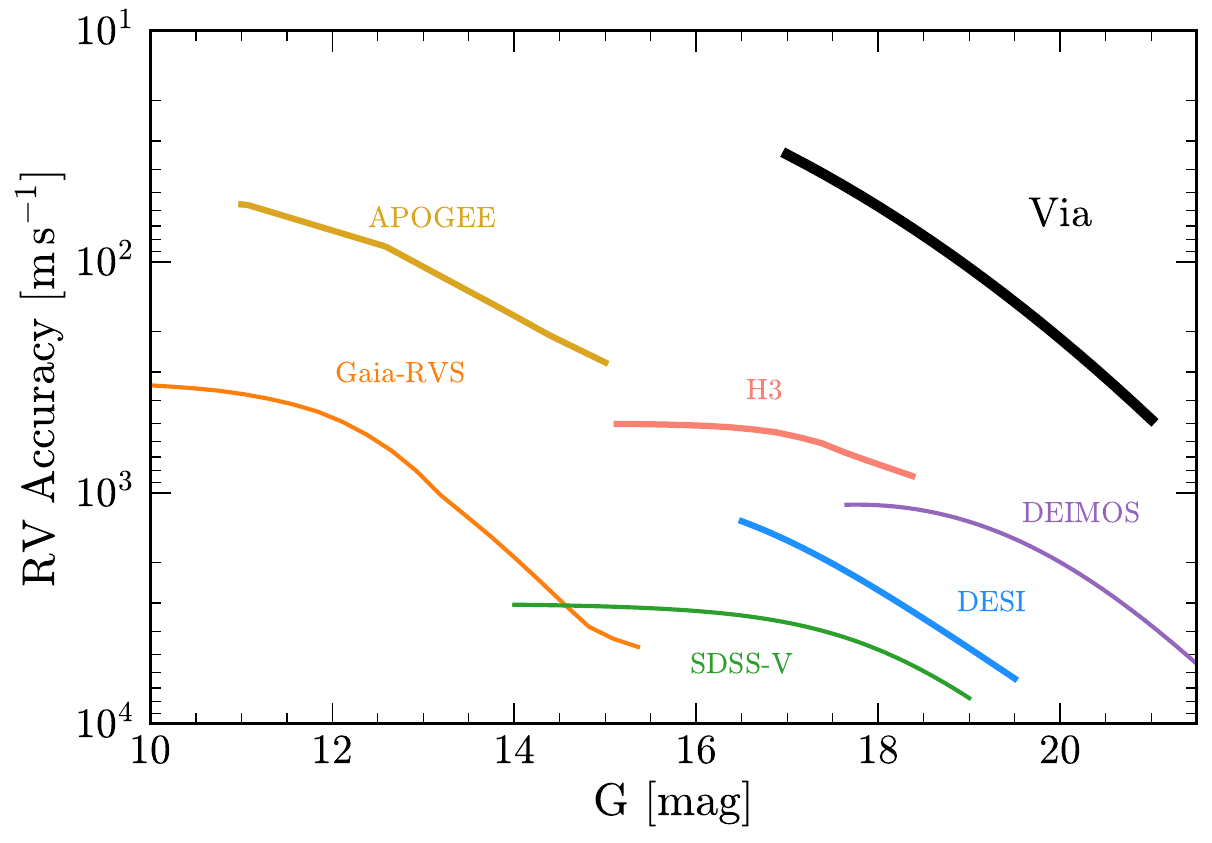}
    \caption{RV accuracy vs. $G$-band magnitude for a selection of previous, ongoing, and future surveys.  Here ``accuracy'' is defined as the quadrature sum of the magnitude-dependent RV precision, and the systematic RV stability floor of the instrument (typically determined via repeat observations).  The predicted Via precision for a [Fe/H]$=-1$ RGB star is shown. 
    For comparison, we show APOGEE \citep{Price-Whelan20, Saydjari2025}, DESI \citep{Cooper23}, Keck/DEIMOS \citep{geha2026b}, \gaia-RVS \citep{Katz23}, H3 \citep{Conroy2019}, and SDSS-V \citep{Kollmeier2026}. 
    The comparison here should be viewed as indicative rather than definitive, since estimates for other surveys are averaged over all metallicities and spectral types, and some surveys are based on collected data while others are forecasts.  Via stands out as by far the deepest survey to reach a velocity limit of $\sim100\ms$.}
    \label{fig:rv_comp}
\end{figure*}

Figure \ref{fig:mos_compare} compares the collecting area and multiplex of Viaspec to other existing or planned high-resolution spectrographs.  Viaspec has more than double the collecting area of wide-field survey instruments like VISTA/4MOST and WHT/WEAVE, and a significantly larger multiplex and field-of-view than workhorse dwarf galaxy follow-up instruments like Keck/DEIMOS and Magellan/M2FS.  The most comparable instrument is the VLT/MOONS, although Viaspec has six times the FoV.  Figure \ref{fig:rv_comp} compares the radial velocity accuracy vs. magnitude for several large spectroscopic surveys.  Via is the only large survey capable of reaching $\sim100 \ms$ velocity precision to $G\approx20$.  Although each of these instruments is highly capable in its own right, the combination of sensitivity, FoV, resolution, and institutional access to telescope time makes Viaspec uniquely situated to perform its primary survey experiment.

\clearpage

\begin{deluxetable*}{ll}
    \centering
    \label{tab:acr}
    \tablecaption{Acronyms and abbreviations used in this document}
        \tablehead{\colhead{} & \colhead{} }
    \startdata
    BH & black hole \\
(CC)SNe & (core collapse) supernovae \\
CDM & cold dark matter \\
CGM & circumgalactic medium \\
CGS & cold gas survey \\
DGS & dwarf galaxy survey \\
DIB & diffuse interstellar band \\
ehr & effective hour \\
ETC & exposure-time calculator \\
FBOT & fast blue optical transients \\
FoV & field of view \\
FPS & fiber-positioning system \\
FRB & fast radio burst \\
FRD & focal ratio degradation \\
GW & gravitational wave \\
IGM & intergalactic medium\\
ISM & interstellar medium \\
LSF & line spread function \\
LSST & Legacy Survey of Space and Time \\
MBH  & massive black hole\\
MMT & MMT Observatory\\
MSTO & main sequence turnoff \\
MW & Milky Way (galaxy) \\
PSF & point spread function \\
RGB & red giant branch \\
RV & radial velocity \\
SIDM & self-interacting dark matter\\
SLSNe & Super Luminous Supernovae\\
SNR & signal-to-noise ratio \\
SPS & stream perturbation survey \\
TDE & tidal disruption event \\
TFS & transient follow-up survey \\
ToO & target of opportunity \\
UFD & ultra-faint dwarf\\
Via & not an acronym! \\
WD & white dwarf \\
WDM & warm dark matter \\
    \enddata
\end{deluxetable*}

\clearpage

\bibliographystyle{aasjournal}
\bibliography{technical}

\end{document}